\documentclass[reprint,aps,prx,onecolumn,superscriptaddress]{revtex4-2}
\usepackage{hyperref}
\usepackage{amsmath,amsthm,amssymb,amsthm}
\usepackage{braket,physics}
\usepackage{enumitem}
\usepackage{tabularx}
\usepackage{color}
\usepackage{cancel}
\usepackage{tikz}
\usepackage[normalem]{ulem}
\usepackage{outlines}
\usepackage{tcolorbox}

\usepackage{graphicx}
\usepackage{subcaption}
\usepackage{ragged2e} 
\usepackage{appendix}

\usepackage{blkarray}
\DeclareMathOperator{\hmax}{h_\text{max}}
\DeclareMathOperator{\Spause}{s_\text{pause}}
\newcommand{\bseq}{\begin{subequations}}
\newcommand{\eseq}{\end{subequations}}

\begin{document}

\title{Magnetic Memory and Hysteresis from Quantum Transitions: \\ Theory and Experiments on Quantum Annealers}


\author{Frank Barrows}
\affiliation{{\em Q-MAFIA}, Los Alamos National Laboratory, USA}
\affiliation{Theoretical Division, Quantum \& Condensed Matter Physics, Los Alamos National Laboratory, USA}
\affiliation{Center for Nonlinear Studies, Los Alamos National Laboratory, USA}
\author{Elijah Pelofske}
\affiliation{{\em Q-MAFIA}, Los Alamos National Laboratory, USA}
\affiliation{Information Systems \& Modeling, Los Alamos National Laboratory, USA}
\author{Pratik Sathe}
\affiliation{{\em Q-MAFIA}, Los Alamos National Laboratory, USA}
\affiliation{Theoretical Division, Quantum \& Condensed Matter Physics, Los Alamos National Laboratory, USA}
\affiliation{Information Science and Technology Institute, Los Alamos National Laboratory, USA}
\author{\\ Francesco Caravelli}
\affiliation{{\em Q-MAFIA}, Los Alamos National Laboratory, USA}
\affiliation{Università di Pisa \& Scuola Normale Superiore, Pisa, IT}
\affiliation{Theoretical Division, Quantum \& Condensed Matter Physics, Los Alamos National Laboratory, USA}
\author{Cristiano Nisoli}
\affiliation{{\em Q-MAFIA}, Los Alamos National Laboratory, USA}
\affiliation{Theoretical Division, Quantum \& Condensed Matter Physics, Los Alamos National Laboratory, USA}
\affiliation{Center for Nonlinear Studies, Los Alamos National Laboratory, USA}

\begin{abstract}

Quantum annealing leverages quantum tunneling for non-local searches, thereby minimizing memory effects that typically arise from metastabilities. Nonetheless, recent work has demonstrated robust hysteresis in large-scale transverse-field Ising systems implemented on D-Wave’s analog quantum hardware. The quantum nature of these intriguing results remains to be understood at a deeper level. Here, we present a conceptual framework that explains the observed behavior by combining two-level Landau-Zener transitions via a first-order piecewise-constant propagator  with semiclassical domain-wall kinetics. We test this approach experimentally on a quantum annealer, where we observe clear coercivity even in one-dimensional rings with periodic boundary conditions comprising up to 4,906 qubits—regimes where classical hysteresis is forbidden, but quantum hysteresis is not. Our framework reproduces the measured kink densities, hysteresis loop shapes, and longitudinal sweep-rate scaling trends observed in data from three different D-Wave quantum annealers. In particular, it captures striking non-monotonic features and transiently negative susceptibilities, identifying them as genuine quantum memory effects. These results establish programmable quantum annealers as powerful testbeds for exploring memory-endowed non-equilibrium dynamics in quantum many-body systems.

\end{abstract}

\maketitle

\section{Introduction}
\label{section:Introduction}

Hysteresis curves have been used to study ferromagnetic materials and succinctly demonstrate magnetic memory since Warburg produced the first hysteresis loop for iron in 1880~\cite{SCHMOOL2018}. The understanding of  the relationship between magnetization and applied field, influenced the development of Curie's law in 1895~\cite{curie1895proprietes}, Langevin's description of paramagnetic and diamagnetic materials in 1905~\cite{Langevin_1905} and Weiss's domain law of ferromagnetism in 1907~\cite{Weiss_1907}. This has inspired analogous curves in ferroelectric, ferroelastic, ferrotoroidic, and memristive systems~\cite{Gnewuch_JSolStatChem_2019,Strukov2008}. 

Historically, hysteresis loops were determined by rotating permanent magnets around a magnetic sample~\cite{Searle_RSTA_1902}; contemporary techniques use a range of fluxmeters, pulsed field magnetometers, and SQUIDs~\cite{Vrijsen_RevSciInst_2013,Fiorillo_IEEE_2007,Sawicki_IOP_2011}, but the core protocol has not changed dramatically. 
Recently in Ref.~\cite{Pelofske_arxiv_2025}, we introduced a new experimental platform to study hysteresis, in systems of interacting spins embedded in quantum annealers that mimics those performed in a laboratory. Although quantum annealers, which are a type of analog quantum computer, were originally developed for non-local optimization~\cite{PhysRevE.58.5355, farhi2000quantumcomputationadiabaticevolution, Santoro_2002}, these devices have also served as platforms for exploring out-of-equilibrium quantum dynamics~\cite{King_PRX_2021, Morrell_PRA_2023} and classical critical phenomena~\cite{sathe2025}. More broadly, they have become a platform for controlled experiments on strongly correlated spin systems~\cite{qubit_spin_ice,lopez2023kagome,lopez2024quantum,lopez2023field}. 

Quantum annealers exploit quantum tunneling to escape metastable states, which would suggest they are \textit{not} natural platforms for studying hysteresis.
Nonetheless, our previous work demonstrated memory effects and showed how hysteretic features depend on the transverse field, and thus on quantum fluctuations. In that context, we observed Barkhausen noise in disordered models, as well as striking non-monotonic behaviors, which have been rarely, though occasionally~\cite{krempasky2023efficient,yoshimi2018current}, reported in experimental studies of spin glasses. These findings, however, remain without a clear theoretical interpretation. For instance, do these non-monotonicities reflect genuine quantum effects? How does the hysteresis  area depend on the local fields? Further, quantum hysteresis has been often approached via mean-field approximations or collective models, though these often neglect discrete quantum transitions~\cite{Wernsdorfer_Science_1999,Tejada_JMM_1999}. The question remains---what are the quantum mechanisms that give rise to collective magnetization reversal?

In this work, we develop a hybrid framework motivated by two-level Landau-Zener dynamics that combines low-order unitary evolution, with a semiclassical description of domain wall kinetics to study magnetic hysteresis under far-from-equilibrium longitudinal field driving in quantum spin systems. We focus on the transverse field Ising model because it can be implemented in various quantum annealers. We then check our findings with experiments on analog D-wave quantum computers and use native hardware controls~\cite{Pelofske_arxiv_2025} to probe hysteresis loops in one and two dimensional ferromagnets across thousands of qubits, varying the sweep duration to access rich dynamical behavior.

We observe reproducible, nontrivial features, such as non-monotonic magnetization reversals and hysteresis loop area variation, whose dependence on protocol parameters is captured by our framework and compared to mean field simulations, emergent classical Landau-Lifshitz-Gilbert dynamics, and time-dependent Schr\"odinger evolution. Our model links diabatic transitions and relaxation dynamics to quantum memory, reproducing key signatures of local entanglement and establishing quantum annealers as a powerful platform to study programmable quantum hysteresis in large scale spin systems.

The rest of the manuscript is organized as follows.
In Section~\ref{sec:QA_and_TFIM}, we describe the transverse field Ising model (TFIM) with a time-dependent longitudinal field that we study in this paper. 
In Section~\ref{sec:simulation_technique}, we present our simulation techniques based on a new theoretical model and compare experimental data with numerical simulations.
In Section~\ref{sec:expt_res} we present numerical and simulation results over a range or parameters.
Finally, in Section~\ref{sec:scaling_laws} we discuss scaling laws and emergent classical magnetism followed by a summary and conclusion in Section~\ref{sec:conclusion}.

\section{Quantum Annealers and the Transverse Field Ising Model}
\label{sec:QA_and_TFIM}

While our results can be generalized to a broader class of Hamiltonians, in this paper, we focus on the transverse field Ising model (TFIM) due to its simplicity and direct implementability on D-Wave quantum annealers. 
These devices consist of superconducting qubits that are connected by tunable couplers with a device-specific connectivity.
In the hysteresis protocol, the qubits are subjected to the time dependent Hamiltonian given by
\begin{subequations}
\begin{align}
    {\mathcal H} &= - \Gamma \sum_i\hat\sigma^x_i  -h(t) \sum_i\hat\sigma^z_i  -\sum_{\langle i,j\rangle}J_{ij}\hat\sigma^z_i\hat\sigma^z_j
    \label{eqn:QA_Hamiltonian_wo-h_gain_1}
    \\
    \begin{split}
    &= - \frac{A(s)}{2} \Big( \Gamma^\prime \sum_i \hat{\sigma}_{i}^{x} \Big) 
     - \frac{B(s)} {2} \Big(  \sum_i h^\prime_i(t) \hat{\sigma_i}^{z} + \sum_{\langle i,j\rangle} J^\prime_{i, j} \hat{\sigma_i}^{z} \hat{\sigma_j}^{z} \Big)    .
     \label{eqn:QA_Hamiltonian_wo-h_gain_2}
    \end{split}
\end{align} 
    \label{eqn:QA_Hamiltonian_wo-h_gain}
\end{subequations}

Eq.~\ref{eqn:QA_Hamiltonian_wo-h_gain_1} is the general TFIM during the hysteresis protocol, eq.~\ref{eqn:QA_Hamiltonian_wo-h_gain_2} is the experimental TFIM implemented in D-Wave quantum processing units (QPUs).   
Here, $J$, $\Gamma$, and $h(t)$ denote the spin–spin coupling, transverse field, and time-dependent longitudinal field, respectively. 
The unprimed quantities $J$, $\Gamma$, and $h(t)$ absorb the hardware-defined anneal schedule functions $A(s)$ and $B(s)$, where $s \in [0,1]$ parametrizes the anneal from the $\hat{x}$-basis ($s = 0$) to the $\hat{z}$-basis ($s = 1$). $A$ and $B$ are monotonically decreasing and increasing functions of $s$, respectively (see Appendix~\ref{section:methods}). At the end of the anneal,  $A=0$, with increasing $s$ corresponding to a monotonic increase of $J/\Gamma$. We will indicate when $A(s)$ and $B(s)$ are explicitly present or absorbed into $J$, $\Gamma$, and $h(t)$, except were otherwise stated we absorb $A(s)$ and $B(s)$ and use the un-primed notation in the text.


The experimental protocol implemented in QPUs different slightly from eq.~\ref{eqn:QA_Hamiltonian_wo-h_gain}, but is designed to closely approximate it. 
One key difference is the inclusion of fast quenches at both the beginning and end of the protocol.
While the full details are presented in Section\ref{sec:expt_res}, we note we operate the devices in the standard quantum annealing setting, wherein the conditions $s=0$ and $s=1$ at the beginning and the end of the protocol, respectively. We start $s$ at $0$, quickly quench it to an intermediate ``pause'' value $\Spause$, hold it at that value for the desired duration, and then quickly quench $s$ to $1$ to perform measurements. The quenches at the start and the end are implemented at the fastest available speed allowed by the hardware to reduce deviations from the target experiment, eq.~\eqref{eqn:QA_Hamiltonian_wo-h_gain}.

Prior theoretical studies on magnetic hysteresis in the TFIM have focused on transverse field sweeps \cite{Acharyya_JPhysA_1994,Banerjee_PRE_1995,Suzuki_Springer_2013,stinchombe1,stinchombe2}, while longitudinal field-driven hysteresis remains less explored.

The TFIM studied in this paper is related to but differs slightly from single molecule systems that display hysteresis. Single molecule systems consist of multiple magnetic atoms, spin flips are mediate by a global spin raising or lowering operator~\cite{Stoner_PhilTranRoySocA_1948,Wernsdorfer_PRB_2002,Zabala-Lekuona_CoorChemRev_2021}. This contrasts with the local transverse field in the TFIM where each spin couples to the transverse field. In the single molecule systems several intermediate states must be traversed in sequence to undergo magnetization reversal, producing a dynamical hysteresis. More broadly, purely dynamical hysteresis lack metastable states and arise when the driving timescale is comparable to or faster than intrinsic state‑transition rates.

Let us note a few properties of time-dependent TFIMs, and some of challenges in modeling hysteresis in these systems. 
Since the transverse field in the TFIM does not commute with the classical Ising Hamiltonian, the non-commuting coupling to fields gives rise to competition between quantum fluctuations and classical ordering tendencies driven by the longitudinal field. 
When driven dynamically, the TFIM evolves through metastable states, exhibiting rich physics influenced by parameter sweep rates and Hamiltonian topology~\cite{Dziarmaga_AdvPhys_2010,Polkovnikov_RevModPhys_2011}. 
Such dynamics are challenging to model, and techniques such as Suzuki-Trotter decomposition can introduce discrete errors that accumulate in the metastable magnetic states as fields vary in magnitude. Domain wall motion, pinning, and annihilation are low-energy phenomena but produce important features of hysteresis, e.g., Barkhausen jumps. Further, as longitudinal field is swept the energy gaps between eigenstates close, thus the dynamics display critical slowing. 
Standard scaling theories such as the Kibble–Zurek mechanism may not apply, especially in disordered or frustrated systems where universality is absent or ill-defined~\cite{Liu_PhysRevB_2014,Das_RevModPhys_2008}, particularly in finite size systems with non-zero $h$ that lacks both first- and second-order phase transitions; 
the longitudinal field confines excitations, the domain walls are not free to propagate but are instead bound together by an effective linear potential. 
Developing accurate models that can capture the role of quantum coherence, thermalization, and dissipation remains a frontier in both theoretical and experimental quantum dynamics. Such an understanding of the transient hardware dynamics in response to driving can enable further functionalities and state control \cite{Yamamura_PRX_2024,Pelofske_PhysRevResearch_2023}.

\section{Nonequilibrium Landau-Zener and Semiclassical Dynamics} \label{sec:simulation_technique}
In this section, we will present a new methodology to simulate hysteresis driven by a changing longitudinal field in a TFIM.
We will start with a few comments on the model and some of the challenges in Section~\ref{subsec:challenges}. 
The remaining three subsections form the core of our methodology.
The two main elements of our theory rely on insights gained from a Landau-Zener model (described in Section~\ref{subsec:diabatic}) and a piecewise-constant Magnus expansion (Section~\ref{subsec:piecewise}) and semi-classical domain wall kinetics (Section~\ref{subsec:domain_wall_kinetics}).
After describing them in detail in their corresponding subsections, we discuss how to combine them in Section~\ref{subsec:combine}.

\subsection{Challenges and Approach} \label{subsec:challenges}

The simplest explanation for hysteresis relies on the intuitive idea that in a ferromagnet, the magnetized state remains metastable under an oppositely applied field until the Zeeman energy becomes strong enough to destroy this metastability. 
This approach contains no reference to dynamics or timescale for metastability. 
More refined conceptualizations center on the role of domains and energetically unfavorable domain walls in the progressive and occasionally step-like mechanism of magnetization reversal. The Zeeman energy of the applied field drives the nucleation and growth of magnetic domains~\cite{Shinjo_Chap4_2009,Zhang_PRA_2018} thereby producing hysteresis~\cite{Graham_Wiley_chap9_2008,PhysRevB.59.4260}.
In a TFIM setting, the problem would be solved knowing the evolution of the system under time varying $h(t)$. 
Notably, analog quantum systems operate at finite temperature which influences dynamics and require the incorporation of environmental interactions to model them.
(D-Wave's quantum processing units (QPUs) are typically maintained at approximately 15 milliKelvin ($0.32$ GHz)~\cite{dwave-docs}.) 

We note that in the one-dimensional classical version Ising model without long range interactions, domain walls appear as zero dimensional kinks that possess no surface tension, which prevents long range magnetic ordering. 
Because the system lacks stable magnetization, it also exhibits no hysteresis~\cite{Schultz_RevModPhys_1964,Baxter}.
However, it can exhibit classical ``dynamical" hysteresis in finite size systems at zero temperature~\cite{Graham_PRE_2005}, which is trivially a transient lag between magnetization and field.
In the quantum version, the dynamics of the domain walls become important, as we show below.

There are multiple approaches to study the TFIM, we discuss limitations of several of these approaches here. 
While an exact solution for the one-dimensional TFIM can be obtained via the Jordan-Wigner (JW) transformation~\cite{Suzuki_Springer_2013} (even with a time-dependent transverse field~\cite{Suzuki_Springer_2013,Shukla_EDP_2020}), notoriously, no solution is known currently available in the presence of a static longitudinal field~\cite{Dutta2015}. 
Moreover, while a perturbative approach might seem natural, 
 the time-dependent longitudinal field $h(t)$ used in our hysteresis protocols is swept over a large range of values comparable to the bare TFIM, and thus cannot be treated perturbatively. 
In addition, implementations of near-equilibrium methods such as density matrix renormalization group (DMRG) fail to reproduce experimental results (see Supplementary Information~\ref{section:appendix_truncated_DMRG} for more details).
Finally, a density of states approach in the interaction picture produces sinusoidal hysteresis by treating the interaction term $J_{ij} \hat \sigma^z_i \hat \sigma^z_j$ as a perturbation, but misses essential features like domain growth and full magnetization reversal~\cite{Pelofske_arxiv_2025}.
In particular, it fails to capture steep magnetization reversals and negative susceptibility observed experimentally~\cite{Pelofske_arxiv_2025,krempasky2023efficient,yoshimi2018current}. 
While we focus on a single hysteresis cycle in this paper, it is in principle possible to probe multiple cycles. 
In the Supplementary Information~\ref{section:appendix_Floquet_states}, we present methods to construct time varying states under oscillating fields and examine the influence of $h(t)$.


\subsection{Diabatic Evolution} \label{subsec:diabatic}

To overcome these difficulties discussed above,  we introduce a hybrid model. 
We treat the system as an open quantum system in which the transverse field facilitates single spin-flip operations and open-system effects are captured by transitions and semiclassical kinetics, discussed below. This contrasts with the global spin reversal expected in adiabatic evolution from the aligned ground state. 
Here, we assume the spins are initially in the aligned along the $\hat{z}$ direction.
In our hardware implementation, a longitudinal field ramp is used to align the qubits in the $\hat z$ direction, as described in more detail later in Section~\ref{sec:expt_res}; see also, Ref.~\cite{Pelofske_arxiv_2025}.
As the longitudinal field changes, the system evolves due to the competing time scales of the changing applied field and the relaxation mechanisms. 
Spin flip excitations in the presence of a ramped longitudinal field can be understood in terms of the LZ model, which describes transitions between instantaneous eigenstates at avoided level crossings in asymptotic long time limit.
The LZ model and the conditions under which it is valued are discussed in Appendix~\ref{apd:QuantEvol}.  

We now rewrite the time dependent TFIM Hamiltonian in a form suitable for an LZ approach as
\begin{align}
    \mathcal{H}(t)=-\sum_{\langle i,j\rangle}J_{ij}\sigma^z_i\sigma^z_j -\sum_i\Gamma\sigma^x_i -h(0) \sum \sigma^z_i-\sum_i \dot{h} t \sum \sigma^z_i,
    \label{eqn:LZ_Ham}
\end{align}
where $\dot{h}$ is the longitudinal field ramp rate, i.e., the slope of the longitudinal field sweep, this form is valid for general hysteresis protocols. 
For an ideal two-level system in the TFIM, such as non-interacting qubits, $J=0$ (originally investigated by Majorana \cite{Majorana_IlNuovoCimento_1932}), the probability of a diabatic evolution   (p) or adiabatic transition, i.e., spin flip, (q) is
\begin{align}
p&=e^{-\frac{\pi\Gamma^2}{2\hbar \dot{h}}}    
, \qquad 
q =1-e^{-\frac{ \pi\Gamma^2}{2\hbar \dot{h}}}    .
\label{eqn:transition_prob}
\end{align} 
In what follows we work in the fixed computational basis. We refer to these $\sigma^z$-product states as diabatic states in the Landau–Zener sense, they are the states whose energies would cross linearly as the longitudinal field is swept if the transverse field  were absent. The corresponding adiabatic states are the instantaneous eigenstates of $\mathcal{H}(t)$, that smoothly connect through each avoided crossing.

Here, the formation of domain walls is governed by LZ-model non-adiabatic transition, driven by the rate of external field sweep, $\dot{h}$.  For complex systems, such as multistate systems, the LZ-model is only valid exactly if several conditions are satisfied \cite{Sinitsyn_PRL_2018,Li_PRA_2017,Malla_PRB_2021}. These are discussed in detail in Appendix~\ref{apd:QuantEvol}.  
The LZ model does not apply exactly to one-dimensional systems under a hysteresis protocol. Of course we are not interested in the exact LZ-model evolution, but instead understanding the excitations in our hysteresis protocol. Instead, we find a regime where the LZ model can approximate the quantum state transitions in a multi-state systems.

When a multilevel spectrum exhibits well separated avoided crossings, such that amplitudes generated at one crossing evolve without acquiring coherent phase relations that later interfere with those from another, the dynamics can be approximated as a sequence of independent two-level Landau–Zener transitions: this is the {\em independent level crossing approximation}. 
It is valid when the instantaneous energy eigenvalues are well separated such that there is a clear succession of avoided crossings both sequentially in time (or time varying longitudinal field) and in a hierarchy of energy levels, producing a clear \textit{time hierarchy} and \textit{energy hierarchy}. These hierarchies ensure the transition probability between states can be approximated by separate two-level scattering matrices. 
In the case of the one-dimensional TFIM, avoided crossings from a diabatic basis correspond to spin-flip operations and are separated by an energy $J=2$ (corresponding to domain nucleation or growth) and the system has a time hierarchy $\Delta t=J/ \dot{h}$. 
Thus the independent level crossing approximation and the LZ-model can capture the dominant transitions during the hysteresis protocol, but does not hold at some fast sweep rate.
Because the independent level crossing approximation does not capture interaction between excitations, non-equilibrium dynamics under a hysteresis protocol can be approximated with a mixed semiclassical and low-order exact quantum evolution framework. 
 
Previous work has implemented similar semiclassical dynamics~\cite{Sinitsyn_PRB_2015}, wherein the system consisted of two interacting spins and evolved under a linear ramped field, with quantum transitions occurring at pairwise avoided crossings. The diabatic energy lines produce a temporal sequence of well-separated pairwise avoided crossings, multiple coherent trajectories connect certain initial and final diabatic states, creating controllable interference of amplitudes. In contrast, our semiclassical model describes classical motion of domain walls and is interleaved with quantum evolution of the basis states, detailed below.

The independent level crossing approximation suppresses relaxation mechanisms essential to the emergence of memory in hysteresis. 
Diabatic transitions leave the magnetic system in an excited state due to the Zeeman energy under a reversed field. We note that open quantum systems effects and quench dynamics introduce classical and irreversible dynamics.  
As such mixed semi-classical treatment is motivated. 
Accounting for quench dynamics, i.e.,the rapid transition from the paused anneal point into the measurement basis, is non-trivial, but we find mixed semi-classical evolution and projecting to the $\hat{z}$-basis in numerical simulations is sufficient to reproduce experimental results, as demonstrated below.

In the simple case of multiple well-separated levels that satisfy integrability conditions, the evolution can be approximated as 
\begin{align}
    \mathcal{T} = \hat{T}_\text{sc}^{t_\infty,t_n}S^{n,n-1}\cdots \hat{T}_\text{sc}^{t_2,t_1}S^{0,1}\hat{T}_\text{sc}^{t_1,t_{-\infty}}
    \label{eqn:MixedModel}
\end{align}
with $\hat{T}_\text{sc}$ is the operator that implements and interleaves the semiclassical evolution (discussed in detail below)  and $S^{a,b}$ are scattering matrices which reproduce the LZ-model transition probabilities on successive times $t_i$. $\mathcal{T}$ is the superoperator from compounding $S$ and $\hat{T}$ acting on the diabatic basis.

While quantum evolution could be implemented by full Lie-Trotterization or high order Magnus expansion, in the next section we argue that first order Magnus expansion, i.e, a first-order piecewise-constant propagator, is justified by the independent level crossing. Further, a low-order propagator is needed to incorporate classical dynamics and account for decoherence due to open quantum system effects. Decoherence occurs on shorter timescales than the annealing times, such that the transition probability during a magnetization reversal is in fact dominated by short time interactions.

\subsection{Piecewise-Constant Approach} 
\label{subsec:piecewise}

In the $N$-qubit system, there are multiple avoided crossings between distinct sets of eigenvalues that occur at the same field value. 
Further, depending on the ramp rate, adjacent avoided crossings can influence transition rates. 
Under these influence, the transition amplitudes overlap and interfere, forcing a genuine multilevel treatment rather than a sequence of independent two-level scatterers.
To reproduce LZ-model dynamics while incorporating multiple level interactions, beyond the idealized LZ chain, we can model dynamics with a first-order Magnus expansion implemented as a short-time first-order piecewise-constant propagator of the full Hamiltonian $\mathcal{H}(t)$.  During a short slice $\Delta t \ll \tau_{\mathrm{LZ}}=\Delta E^{\text{min}}/\dot{h}$ we apply the propagator $\exp[-i\mathcal{H}(t_n)\Delta t/\hbar]$.  
Near the avoided crossing in finite time windows quantum transitions generate local entanglement. 
When the crossings are well separated this step reproduces the single–crossing LZ exponent to leading order, and  the probability of a quantum transition is well approximated by a LZ-model transition probability.  Thus, near every avoided crossing the dynamics factorizes as a single complex branch point controls the probability of leaving the adiabatic state. When levels overlap the same step retains every off–diagonal matrix element, thereby capturing multilevel interference. 

We propose, the evolution of well-separated levels, accounting for relaxation mechanisms and open-system processes, is defined by the time evolution
\begin{align}
    \mathcal{T} = \hat{T}_\text{sc}^{t_\infty,t_n}U(t_n,t_{n-1})\cdots \hat{T}_\text{sc}^{t_2,t_1}U(t_0,t_1)\hat{T}_\text{sc}^{t_1,t_{-\infty}},
    \label{eqn:MixedModel_2_main}
\end{align}
where we interleave quantum slices with a semiclassical drift–diffusion update for the domain-wall densities and first-order piecewise-constant propagator implemented with unitary operator $U(t_i,t_j)$.

Higher order multi-spin interactions and open-system effects like irreversible domain wall annihilation are not captured in our low-order quantum evolution. 
We can account for domain wall motion and interactions in excited states with a kinetic equation. This is described in the next section.

\subsection{Domain Wall Kinetics} \label{subsec:domain_wall_kinetics}

The semiclassical approximation applies when the system's wavefunction can be written as $\Psi=A(x,t)e^{\frac{-i}{\hbar}R(x,t)}$, and the phase and amplitude can be treated independently, as discussed in Appendix~\ref{apd:SemiClassEvol}. 
In our approach, by \textit{semiclassical-approximation} we simply mean that between well-separated avoided crossings the many-body state can be described by classical domain-wall particles, whose densities evolve via a kinetic transport equation;  the genuinely quantum ingredients enter only at the localized two-level crossings that create or move domain walls. This approach is justified because (i) the intrinsic TFIM frequencies $(J/\hbar)$ are much faster than the longitudinal-field sweep rate (rapid phase averaging), (ii) domain walls' effective mass (which  is $\propto \hbar^2 a^{-2} \Gamma^{-1}$, where $a$ is the lattice spacing~\cite{Suzuki_Springer_2013}) is ``heavy" $(\Gamma \ll J)$ so their wavepackets propagate with negligible quantum dispersion, and (iii) successive avoided crossings are well separated in time, suppressing coherent interference. Consequently, we keep probabilistic quantum transitions for spin-flip events and evolve the domain-wall density $n(x,t)$ classically in between.

The D-Wave hardware sets the TFIM energy scale,
and the frequency of quantum oscillations are much faster then the ramp rate of the longitudinal field. This decouples the semiclassical dynamics of the amplitude. The amplitude evolves as
\begin{align}
    \pdv{A(x,t)}{t} &=  -v_0 \pdv{A(x,t)}{x}.
\end{align}
 In Appendix \ref{apd:SemiClassEvol}, we derive that the amplitude evolves according to $\partial_t A(x,t)=\frac{1}{2\sqrt{n(x,t)}}\partial_t n(x,t)$, where $n(x,t)$ is the kink density.
Hence, the short-time dynamics of $  A(x,t) $ are governed by the redistribution of domain walls, as captured by the semiclassical kinetic equation. Thus, we use a kinetic equation of the domain wall density to update the amplitude of the quantum wavefunction.

Classically, domain wall motion results from torque induced by the external field, reflecting the gradient of the magnetization along the domain wall. 
We have two kinetic equations  for positive  ($n^{+}=n^{\uparrow\downarrow}$) and negative ($n^{-}=n^{\downarrow\uparrow}$) kinks.
Kinks interact through a local annihilation term.   
The velocity of the domain walls scales with the longitudinal field, $\mathbf{v}^\pm[h(t)]\equiv \pm v_0 h(t)$, determined by a Lieb-Robinson bound on domain wall measurement operations; in the case of large longitudinal field we can bound the velocity of the domain as $v(t)\leq v_0 h(t)$ \cite{Nachtergaele_ContMath_2010,Wang_PRX_2020}. 
Thus the velocity is linear in the longitudinal field and we define the kinetic equations for the domain walls as 
\begin{subequations}
\begin{align}
\frac{d n^+}{dt} &= -\mathbf{v}^+[h(t)]\frac{d n^+}{dt}-\Omega n^+ n^-  ,
\\
\frac{d n^-}{dt} &= -\mathbf{v}^-[h(t)]\frac{d n^-}{dt}-\Omega n^+ n^-  ,
\end{align} 
\end{subequations}
where $\Omega$ is the annihilation probability.
Evolution occurs between discrete states but $n^\pm$ is continuous valued, there is no domain wall diffusion. Below we discuss statistical methods we have developed to combine the quantum and semiclassical evolution, details of numerical simulation, and comparison with experimental data.

\subsection{Combined Model} \label{subsec:combine}

To incorporate semiclassical effects into the quantum dynamics, we update the diabatic basis amplitudes $\psi_i$ such that the wavefunction reflects the domain wall distribution from the semiclassical evolution. The wavefunction is normalized after the update. In this update we find a filtered state $\psi(t+\delta t)$ which is closest to an initial state $\psi(t)$ that has domain wall densities similar to the semiclassically evolved domain wall densities.  We  update the states  as 
\begin{align}
    \psi_i(t+\delta t) &=\psi_i(t)\prod_a \left(\frac{n_a^t}{n_a^\psi}\right)^{\alpha \hat{n}^i_a} .
\end{align}
This is akin to an iterative proportional fitting (IPF), the marginals of a target matrix are the semiclassically evolved domain wall densities, and the kernel is the domain wall densities. This  updates the matrix of domain wall density as
\begin{subequations}
\begin{align}
    \bra{\psi(x, t)} \hat{n}\ket{\psi(x, t)} &= \begin{pmatrix}
        n^i_a & n^i_{a+1} & n^i_{a+2} & \cdots 
        \\
        n^{i+1}_a & n^{i+1}_{a+1} & n^{i+1}_{a+2} & \cdots 
        \\
        n^{i+2}_a & n^{i+2}_{a+1} & n^{i+2}_{a+2} & \cdots 
        \\
        \vdots & \vdots & \vdots & 
    \end{pmatrix} 
    \\
    \bra{\psi(x, t+\delta t )} \hat{n}\ket{\psi( x, t+\delta t )} &= \begin{pmatrix}
        n^i_a \Pi_b \left(\frac{n_b^t}{n_b^\psi}\right)^{2\alpha \hat{n}^i_b}
        & n^i_{a+1} \Pi_b \left(\frac{n_b^t}{n_b^\psi}\right)^{2\alpha \hat{n}^i_b}
        & n^i_{a+2} \Pi_b \left(\frac{n_b^t}{n_b^\psi}\right)^{2\alpha \hat{n}^i_b}
        & \cdots 
        \\
        n^{i+1}_a \Pi_b \left(\frac{n_b^t}{n_b^\psi}\right)^{2\alpha \hat{n}^i_b}
        & n^{i+1}_{a+1} \Pi_b \left(\frac{n_b^t}{n_b^\psi}\right)^{2\alpha \hat{n}^i_b}
        & n^{i+1}_{a+2} \Pi_b \left(\frac{n_b^t}{n_b^\psi}\right)^{2\alpha \hat{n}^i_b}
        & \cdots 
        \\
        n^{i+2}_a \Pi_b \left(\frac{n_b^t}{n_b^\psi}\right)^{2\alpha \hat{n}^i_b}
        & n^{i+2}_{a+1} \Pi_b \left(\frac{n_b^t}{n_b^\psi}\right)^{2\alpha \hat{n}^i_b}
        & n^{i+2}_{a+2} \Pi_b \left(\frac{n_b^t}{n_b^\psi}\right)^{2\alpha \hat{n}^i_b}
        & \cdots 
        \\
        \vdots & \vdots & \vdots & 
    \end{pmatrix} .
\end{align}
\end{subequations} 
Ideally, IPF at a site $a$ would be ${n^t_a}/{\sum_i n^{i\psi}_a}$. 
 As the system evolves in the diabatic basis,  spatially varying amplitudes cannot be updated directly. Instead we perform a scalar update of the diabatic eigenbasis.
As our semiclassical kinetic equation is written in terms of the total domain wall density $n=\sum_i n^i$ this is the appropriate form of the IPF update in terms of $n$, the marginal of the domain wall density matrix. A heurestic weighting factor {$\alpha = \alpha_0 + \frac{\kappa}{2N}\sum_{a}\vert n_a^t(t) - n_a^\psi(t)\vert$} control the strength of the update, which allows the system to evolve in time and iteratively trend towards to semiclassical domain wall densities. When there is a greater mean difference in the number of domain walls the update is stronger. $\alpha_0$ and $\kappa$ are free parameters that are tuned for agreement with experiments.
This update minimizes the Kullback–Leibler divergence $D_{KL}(p^t || p^\psi)$ between target and initial probability distributions, implemented with a Lagrangian function with constraints on the probability and the domain wall density:
\begin{align}
\mathcal{L}=\sum p^t_i\ln{\left(\frac{p_i^t}{p_i^\psi}\right)}+ \lambda_0 \sum_i(p^t_i-1)-\sum_a\lambda_a\left(\sum_i p^t_i \hat{n}^i_a - n^t_a \right)  .
\end{align} 
The stationary point with respect to $p_i^t$ yields $p_i^t=c p_i^\psi e^{\sum_a \lambda_a \hat{n}^i_a} $. 
Here $p_i$ is the probability of being in a diabatic basis $\psi_i$, and $c$ is the normalization factor.  For single domain wall states,  $\lambda_a = \ln{\left(\frac{n^{i,t}_a}{n^{i,\psi}_a}\right)}$ enforces  $\sum_i p^t_a \hat{n}^i_a=n_a$ as desired. As the semiclassical equations are written in terms of total domain wall density $n$, in practice we approximate this as $\lambda_a=\alpha\ln{\left(\frac{n^t_a}{n^\psi_a}\right)}=\alpha\ln\left(\frac{\sum_i p_i^t\hat{n}^i_a}{\sum_i p_i^\psi \hat{n}^i_a}\right)$, reproducing the update above.

We implement numerical time evolution of the hysteresis protocol.
We discretize the total evolution time into steps of size $\Delta t$, “freezing” the Hamiltonian at the left endpoint of each interval ($t_n$) and updating the state via $U(t_{n+1},t_n) = \exp[-i\mathcal{H}(t_n)\Delta t/\hbar]. $
 We evolve the state with an exact diagonalization of the Hamiltonian frozen on short slices $\Delta t$, i.e, a first-order piecewise-constant propagator.  
Choosing $\Delta t\ll \tau_{\mathrm{LZ}}$, ($\Delta t/\tau_{\mathrm{LZ}}<10^{-2}$ in numerical simulations below), ensures that each piecewise‐constant propagator incurs only an $\mathcal{O}(\Delta t^2)$ local error. In practice, with $\Delta t / \tau_{\mathrm{LZ}} < 10^{-2}$, these errors remain orders of magnitude smaller than any physically relevant signal.

We perform numerical simulations via exact diagonalization to calculate the quantum transition probability. We see the avoided crossing to the first excited state occurs around $|h(t)|=2J $, independent of system size, as seen in experiments below. This exactly reproduces excitations from the independent level crossing approximation.  
In the Supplementary Information~\ref{section:appendix_eigenvalue_crossing}, we plot the longitudinal field value at which the avoided crossing gap is smallest as a function of system size and $\Gamma$. We see that $h_\text{crossing}$ is non-monotonic as a function of $\Gamma$ and system size, $N$, and generally $h_\text{crossing}$ is close to $2J$. 
 This is a rich area to explore, as tuning control parameters in analog quantum devices may allow one to adjust or navigate eigenspectrum bifurcations, offering a route to steer quantum dynamics beyond standard annealing protocols~\cite{Armen_PhysRevA_2006,Doherty_PRA_2000}.

For the numerical realization we diagonalize the $2^{N} \times2^{N}$ Hamiltonian at every time point, at each discrete time $t_n = n \Delta t$.
 The semiclassical domain‐wall density update is performed via a forward (explicit) Euler scheme, stability is ensured by negligible diffusion via the diffusive Courant condition.

\begin{figure}[t!] 
    \centering
    \includegraphics[width=.75\linewidth]{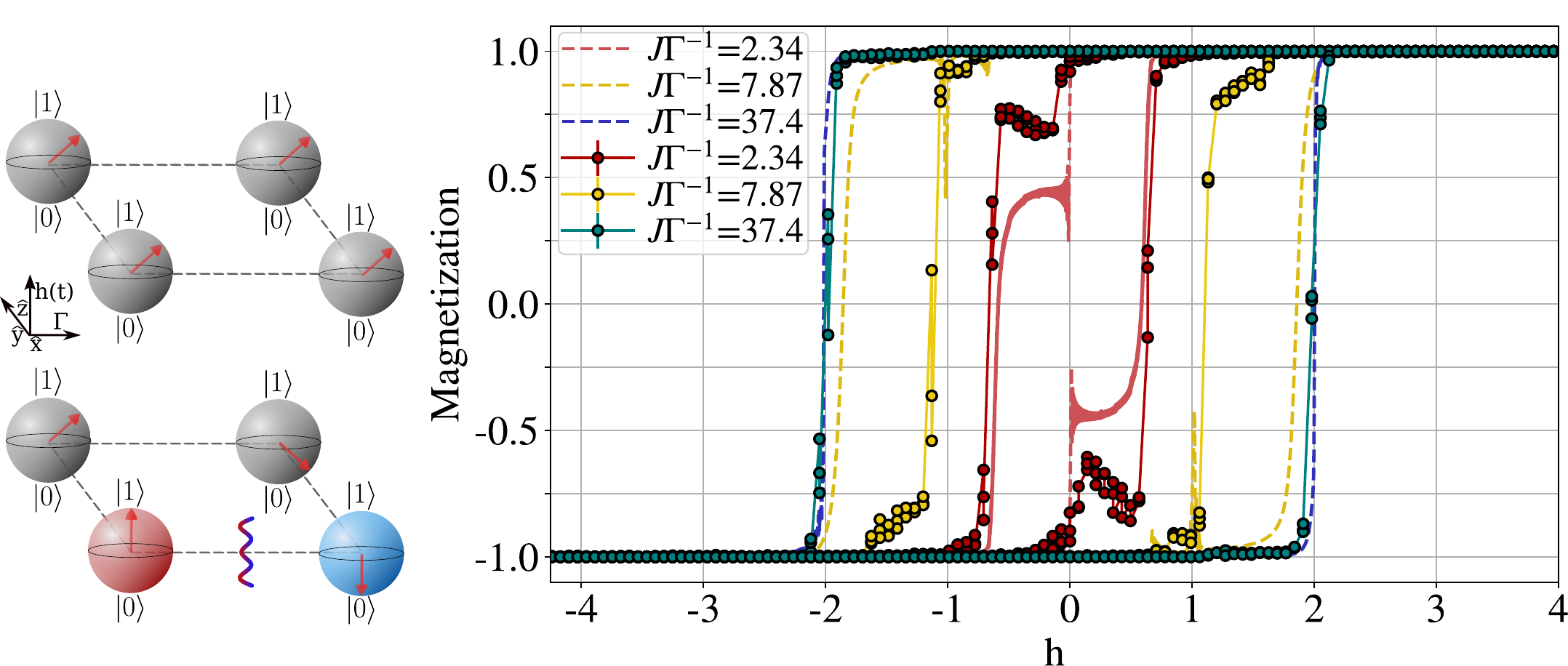}
    \caption{ 
    (Left) Schematics of a ring or plaquette of four spins in the transverse field Ising model. 
    Individual spins are represented as Bloch spheres, spins are coupled along the dashed line. Individual spins can be anti-aligned (red and blue spins),  forming a domain wall or kink. 
    (Right) Comparison between the experimental D-Wave magnetic hysteresis data and the numerical simulation developed in this study, on a $4$-spin ferromagnetic model with periodic boundary conditions. Experiment: solid lines, numerical simulation: dashed lines. Close correspondence for 4 spins at $ J \Gamma^{-1}\in\{2.34,7.87,37.4\}$ ($\Spause\in\{0.3,0.4,0.5\}$) with $11.2 \mu s$ annealing time are observed between simulation and experiments. Non-monotonicity is observed at similar $h$ values when $J\Gamma^{-1}=2.34$, though magnetization values differ. Similar reversal curves are seen for $J\Gamma^{-1}=2.34$ and $J\Gamma^{-1}=37.4$. At $J \Gamma^{-1}=7.87$ differences appear, we see a non-monotonicity at the same field values $h$ in both experiments and numerical simulation (e.g., magnetization reversal from $m=-0.5$ to $m=0.1$ during the backward sweep) but the simulation does not undergo complete reversal but instead re-magnetizes, and undergoes complete reversal at larger applied field values. Thus while capturing the beginning of the non-monotonicity there are some higher order reversal mechanisms not captured in the simulation, either through choice of parameters of just lacking from the model. The fully saturated regions of the hysteresis curves are not shown here for visual clarity of the reversal region. 
    }
    \label{fig:4spin_ExpSim}
\end{figure}

Fig.~\ref{fig:4spin_ExpSim} shows a  comparison of experimental and  simulated hysteresis in small, numerically tractable systems, and a corresponding schematic.
Using the mixed semiclassical and low order quantum framework, transitions and semiclassical evolution are treated as two distinct steps, as implemented in eq.~\eqref{eqn:MixedModel_2_main}, in comparison to D-Wave QPU experimental data that was run on the analog quantum processor \texttt{Advantage2\_prototype2.6}. Hysteresis protocol implemented in hardware is similar to previous work~\cite{Pelofske_arxiv_2025}, and is detailed in Appendix~\ref{section:methods}.

The first order quantum transitions transfer amplitude between states near an avoided crossing, and the system evolves semiclassically between quantum transitions. 
The combined algorithm provides a practical surrogate for the full quench dynamics targeted in this work.
This mixed model accounts for relaxation and excitation interactions. In the absence of this semiclassical channel the protocol stalls, the system remains trapped in excited manifolds and never completes a global magnetization reversal, as demonstrated in the Supplementary Information~\ref{section:appendix_simulation_without_relaxation_semiclassical_motion_of_domain_walls}. In the purely unitary evolution of the TFIM under a rapid longitudinal sweep, the magnetization can become trapped on one instantaneous eigenbranch due to exponentially small transition probabilities. Although no classical barrier exists, these ‘diabatic metastable states’ act like metastable wells on the timescale of the experiment.

\section{Experimental Results} \label{sec:expt_res}
\begin{figure}[t!] 
    \centering
    \includegraphics[width=\linewidth]{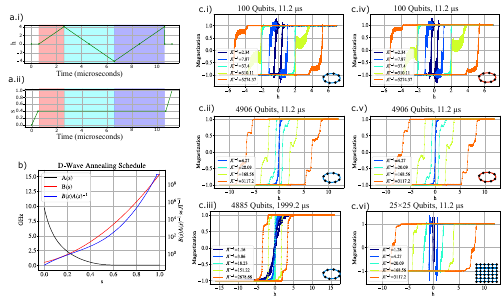}
    \caption{Hysteresis protocols and magnetic hysteresis loops with standard deviation for one dimensional and two dimensional ferromagnetic systems.   (a) Hysteresis protocol, (a.i) longitudinal field protocol with initial ramp (red) backward (cyan) and forward (dark blue) sweeps, a hysteresis loop corresponds to the cyan and blue sections. (a.ii) Anneal schedule, a quick quench fixes the ratio of transverse and longitudinal field, which is held constant throughout the hysteresis protocol, a $0.5 \mu s$ quench to the $\hat{z}$-basis is required to perform measurements. (b) Ratio of the transverse Hamiltonian (A(s)) and longitudinal Hamiltonian (B(s)) strength for a anneal schedule value ($\Spause$), for the \texttt{Advantage2\_prototype2.6} device. (c) Hysteresis loops shown as average magnetization, average magnetization as a function of applied longitudinal field $h$, standard deviation shown with error bars, inset schematic shows system and interaction, ferromagnetic bonds in blue and anti-ferromagnetic bonds gauge transformed in red. Panels (c.i), (c.ii), and (c.iii) show simulations of one dimensional periodic ferromagnetic  rings, with parameters ($100$ qubits, $11.2\mu s$ anneal time, $2000$ samples ), ($4906$ qubits, $11.2\mu s$ anneal time, $2000$ samples ),  and ($4885$ qubits, $1999.2\mu s$ anneal time, $2000$ samples ), respectively.
    Panels (c.iv), (c.v) show simulations of one dimensional periodic anti-ferromagnetic gauge transformed rings, with parameters ($100$ qubits, $11.2\mu s$ anneal time, $2000$ samples ) and ($4906$ qubits, $11.2\mu s$ anneal time, $2000$ samples ), respectively. (c.vi) Two-dimensional ferromagnetic square lattice, ($25 \times 25$ qubits, $11.2\mu s$ anneal time, $2000$ samples ). (c.i) and (c.iv) were obtained on the  \texttt{Advantage2\_prototype2.6} device, (c.ii) and (c.v) were obtained on the  \texttt{Advantage\_system4.1} device. (c.iii) was obtained on the  \texttt{Advantage\_system6.4} device and (c.vi) was obtained on the  \texttt{Advantage\_system4.1} device.}
    \label{fig:Hysteresis}
\end{figure}

We implement a hysteresis sweep with a time varying longitudinal field $h(t)$ in D-Wave quantum processing units (QPUs). 
 Fig.~\ref{fig:Hysteresis} displays the hysteresis protocol, we ramp up to a fixed $s$ value where we pause the system, $s_\text{pause}$ and then perform three linear longitudinal field ramps. 
Fig.~\ref{fig:Hysteresis} (a.i) and (a.ii) displays the longitudinal field and the anneal protocol, $s$, respectively. The system is first magnetized along the $\hat{z}$-axis in the initial ramp (red region), and then longitudinal field is swept from $|\hmax|$ to $-|\hmax|$ (cyan) and then back to $|\hmax|$ (dark blue). $\Spause$ defines the relative strength between the $\hat{x}$- and $\hat{z}$- basis Hamiltonians throughout the sweep. The sample is measured in the $\hat{z}$-basis via a $0.5 \mu s$ quench to $s=1$. Each measurement is performed on an independently prepared state, the hysteresis protocol is performed using progressively longer simulations during the application of the alternating longitudinal field up until a pre-defined total simulation time. Measurements at various times during the protocol are combined to form the full hysteresis loop in the form of the observable of average magnetization. Each point in the loop corresponds to 2000 measurements.  
In Fig.~\ref{fig:Hysteresis} (b), shows the anneal functions $A(s)$ and $B(s)$. The blue line in Fig.~\ref{fig:Hysteresis} (b) is the ratio $B(s)/A(s)$, or alternatively $J/\Gamma$ as $|J|=1$ and $|\Gamma|=1$, we will make clear when this is not the case.

Fig.~\ref{fig:Hysteresis} (c) shows hysteresis loops corresponding to the backward and forward sweeps, cyan and dark blue regions in (a). Prior work has demonstrated hysteresis in D-Wave systems for $s_\text{pause}$ values in the range $\sim 0.2\leq \Spause\leq 0.9$~\cite{Pelofske_arxiv_2025}. In Fig.~\ref{fig:Hysteresis} (c), average and standard deviation of magnetization versus applied field are shown for $s_\text{pause}\in\{0.3,0.4,0.5,0.6,0.7\}$, labeled with corresponding $J/\Gamma$ ratios, which vary slightly by device. We excluded $\Spause=0.3$ for the \texttt{Advantage\_system4.1} and \texttt{Advantage\_system6.4} devices at $11.2\, \mu s$ anneal time due to anomalous results that will be investigated in a separate work. Magnetization is aligned with the applied field after the ramp rate, hysteresis loops start from an aligned configuration. Fig.~\ref{fig:Hysteresis} (c.i), (c.ii) and (c.iii) are one-dimensional periodic uniform ferromagnetic systems ($J_{ij}=J>0$), and $h(t)$ is a global parameter applied to each spin in eq.~\eqref{eqn:QA_Hamiltonian_wo-h_gain}. The corresponding system size and anneal times for (c.i) are $N= 100$ and $11.2\mu s$, for (c.ii) are $N=4906$ and $11.2\mu s$, and for (c.iii) are $N=4885$ and $1999.2\mu s$. The hysteresis loop shrinks with longer anneal time. Panels (c.iv) and (c.v) show hysteresis simulations using the antiferromagnetic gauge transform method, with similar parameters as (c.i) and (c.ii), ($N=100$, anneal time of $11.2\mu s$ and $N=4906$ and anneal time of $11.2\mu s$ respectively). In the antiferromagnetic gauge, we set $J=-1$ and $h_i(t)$ is an alternating parameter with $\pm h(t)$ for even and odd sites, respectively. We transform back to the ferromagnetic gauge (flipping the orientation of every other spin) to plot the magnetization. We find good comparison in the antiferromagnetic gauge as the ferromagnetic gauge. Working with superconducting flux qubits there is the potential for drift over time of strongly ferromagnetically coupled qubits~\cite{Harris_PhysRevB_2010, King_NatComm_2021}, working in the antiferromagnetic gauge can mitigate this drift. If drift contributed to observed behavior, we would expect noticeable differences in the ferromagnetic and antiferromagnetic gauge.

Hysteresis for a two-dimensional ferromagnetic square lattice is shown in Fig.~\ref{fig:Hysteresis} (c.vi), $N=25\times 25$ and $t_\text{anneal}=11.2\mu s$. The hysteresis in the two-dimensional square lattice demonstrates similar properties to the one-dimensional ferromagnetic rings. The reversal occurs at larger field magnitudes than in the one-dimensional case (compare with (c.ii)), consistent with stronger magnetic ordering.

One dimensional quantum systems do display hysteresis, as there is a barrier to the magnetization reversal of the formation of domain walls above thermal or quantum fluctuations, in contrast to classical one-dimensional systems. Note that for all values of $\Spause$ investigated, $A(s)$ and $B(s)$ are much larger than the D-Wave hardware temperature ($0.32$ GHz), thermal fluctuations are on the order domain wall energy for small $s \sim 0 $, and the strength of transverse field is comparable to the thermal fluctuations for $\Spause\geq 0.5$.   Furthermore, the experimentally observed magnetization reversal does not occur intermediately once initialized. 
Instead magnetization reversal occurs gradually with finite susceptibility. Further, the one-dimensional systems show non-monotonicity, contrasting with classical ferromagnetic systems and the two-dimensional quantum systems. 
 As expected from eq.~\eqref{eqn:QA_Hamiltonian_wo-h_gain}, for one-dimensional systems magnetization reversal is centered around $|h|=2$ ($J=1$). 
For instance, in (c.i) reversal begins at $|h|=2J$ for $J\Gamma^{-1}=3.74$ ($\Spause=0.3$) and $J\Gamma^{-1}=310.11$ ($\Spause=0.6$). The reversal can occur at larger and smaller longitudinal field values for larger and smaller $J\Gamma^{-1}$ values, respectively. The strength of the transverse field determines the susceptibility of the qubit ferromagnet system, equivalent to the hardness of magnetic materials. The stronger the transverse field in the system, the more susceptible the system is to undergo spin reversals, making it easier to reverse a magnetic material and shrinking the area of the hysteresis loop.
In addition, by comparing the hysteresis loops in (c.i), (c.ii) and (c.iii), we see that the magnetic coercivity and saturation field are also influenced by the annealing time. Anneal time also influences coercivity; longer times allow more spin flips, shifting reversal to lower $|h|$ values.

The complete set of one and two dimensional system ferromagnetic hysteresis cycle simulations run on the three D-Wave processors is shown in Supplementary Information~\ref{section:appendix_additional_experiment_DWave_magnetic_hysteresis_loops}.

\begin{figure}[t!]
    \includegraphics[width=1\linewidth]{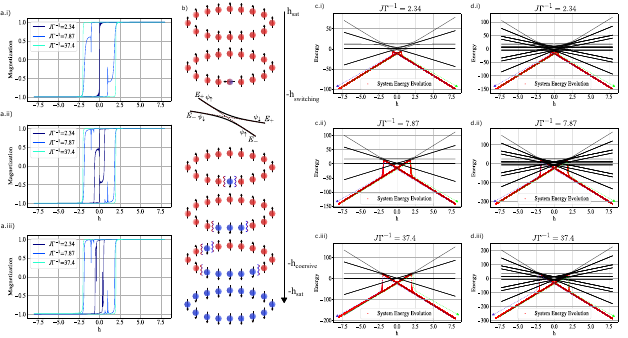}
    \caption{(a) Simulated hysteresis with a combined first-order piecewise-constant propagator and semiclassical domain wall kinetics.  (a.i), (a.ii), and (a.iii) correspond to ferromagnetic one-dimensional ring systems with $N=3,4,6$ qubits, respectively, simulated with a total anneal time of $1\text{E}3~\mu$s and step size $\Delta t = 100$ ps. Hysteresis is shown for  $\Spause \in \{0.3,0.4,0.5\}$ ($J\Gamma^{-1}\in \{2.34, 7.87, 37.4\}$). (b) Schematic of magnetic reversal via domain wall formation triggered when the longitudinal field reaches a switching value $(-h_\text{switching})$. An avoided crossing for a two-level system illustrates the competition between adiabatic evolution ($E_- \rightarrow E_- $ and $\psi_\uparrow\rightarrow\psi_\downarrow$ )and  diabatic evolution  ($E_- \rightarrow E_+ $ and $\psi_\uparrow\rightarrow\psi_\uparrow$ ).
    (c) and (d) Average energy (red dots) projected on the adiabatic eigenbasis (black lines) during the hysteresis protocol. Backward and forward sweep are indicated with purple and green arrows, respectively. The average energy obtains an excited state when the field changes sign, eventually the system relaxes to the ground state. (c.i), (c.ii) and (c.iii) $N=4$ qubits, and (d.i), (d.ii) and (d.iii) $N=6$ qubit, for $J\Gamma^{-1}\in \{2.34, 7.87,37.4\}$, respectively.  } 
    \label{fig:LZ-hyst}
\end{figure}

In Fig.~\ref{fig:LZ-hyst} hysteresis plots from numerical simulations are shown. The simulations yield hysteresis loops providing a qualitative agreement with the experimental results (e.g., in Fig.\ref{fig:Hysteresis}), including regions of negative susceptibility. Fig (a.i)-(a.iii) are simulated with  different  number of qubits in a ring, ($T_\text{total}=1E3 \, \mu s$, $\Delta t = 100 \, ps$ and $N \in \{3,4,6\}$), respectively. We note that $\Spause\leq 0.5$ and $\Spause > 0.5$ define qualitatively different regimes requiring distinct simulation treatments. Hysteresis plots from numerical simulations at $\Spause\in\{0.6,0.7\}$ ($J\Gamma^{-1} \in \{310.11, 5274.37\}$) are provided in the Supplementary Information~\ref{section:appendix_Weak_Gamma_hysteresis_simulations}. Fig.~\ref{fig:LZ-hyst}(b) gives a schematic of domain wall formation in a one-dimensional periodic TFIM, illustrating the avoided crossing.

In Fig.~\ref{fig:LZ-hyst} (c) and (d)  we plot the TFIM energy projected onto the adiabatic eigenbasis throughout the hysteresis loop.  (c.i) - (c.iii) $N=4$ and (d.i)-(d.iii) $N=6$, correspond to $\Spause\in\{0.3,0.4,0.5\}$ ($J\Gamma^{-1}\in \{2.34, 7.87,37.4\}$),  respectively. The backward and forward sweeps are indicated with purple and green arrows, respectively. The bowtie structure formed by the average energy during backward and forward sweeps shows the distinct evolution, indicating the presence of hysteresis and excited states. Semiclassical domain wall motion allows the system to relax to the aligned state.

In the weak transverse field regime, $\Spause\in\{0.6,0.7\}$ ($J\Gamma^{-1}\in \{310.11,5274.37\}$), complete hysteresis loops and full magnetization reversal are observed in numerical simulations. However, these loops do not qualitatively match experimental data (see Supplementary Information~\ref{section:appendix_Weak_Gamma_hysteresis_simulations}).
This suggests that the independent level crossing approximation breaks down when the transverse field is sufficiently weak and our simulation is neglecting some relevant higher-order interactions. In experiments, fluctuations can assist magnetization reversal as the transverse field is small compared to the hardware device temperature. Thus within this weak transverse field limit we don't expect the simulations to capture the magnetization reversal seen in experiment. 

\begin{figure}[t!]
    \centering
    \includegraphics[width=.99\linewidth]{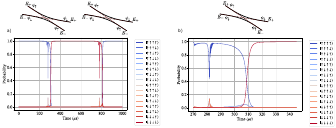}
    \caption{Eigenstate evolution as a function of time during the simulated hysteresis protocol. (a) For a ferromagnetically coupled 4-qubit ring, the system is initialized in an aligned state, blue, and undergoes magnetization reversal, indicated with the the schematic avoided crossing. (b) Zoom-in of the first magnetization reversal in (a), highlighting transient population of a nearly reversed eigenstate, responsible for the non-monotonicity.}
    \label{fig:LZ-eig}
\end{figure}
The diabatic eigenbasis state population, shown in Fig.~\ref{fig:LZ-eig}, provide insight into the non-monotonicity in magnetization reversal, e.g., in Fig.~\ref{fig:LZ-hyst} (a.iii). Fig.~\ref{fig:LZ-eig} (a) and (b) show the eigenbasis population for $\Spause=0.4$, transient occupation of excited states occurs just prior to full magnetization reversal. These excited states first appear as a fluctuation before the system undergoes full magnetization reversal. These excited states decay through  semiclassical domain wall motion and further diabatic transitions. As domain wall nucleation and domain growth is energetically unfavorable in higher dimensions, these fluctuations are exponentially suppressed outside one-dimensional systems.

Classical frustrated systems can display such non-monotonicity due to the re-entrant phases. Spurious coupling between qubits in the D-Wave hardware could enable such frustration. Existing models of such frustrated low-dimensional systems would indicate that spurious coupling should be of comparable magnitude as the ferromagnetic coupling, e.g., half the strength~\cite{Suzuki_Spring_chap4_2013}. Previous work has characterized the spurious coupling in D-Wave hardware are typically on the order of $0.1$ the ferromagnetic coupling strength~\cite{tuysuz_arxiv_2025}.

\section{Domain Wall Scaling and the Emergent Landau-Lifshitz-Gilbert Equation}\label{sec:scaling_laws}

\begin{figure}[t!]
    \centering
    \includegraphics[width=\linewidth]{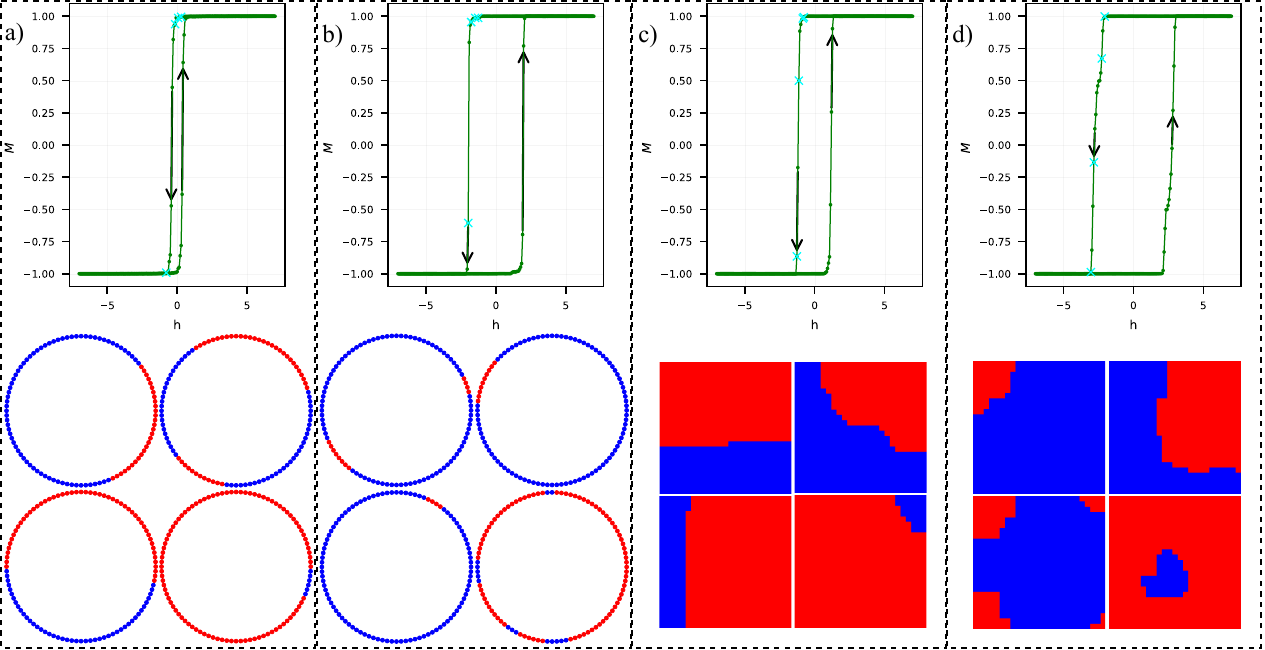}
    \caption{Representative single spin configurations from various points along the one-dimensional and two-dimensional ferromagnetic hysteresis cycles, run on \texttt{Advantage2\_prototype2.6}. Top row shows the mean magnetization $M_z$ hysteresis cycles as a function of the longitudinal field $h$, with four marked points as cyan $\times$'s. Bottom rows show four single spin-configurations, specifically selected because they show magnetic domains, from those four points in the hysteresis cycle (spin configurations are read top left to bottom right). The four real space configurations correspond to marked magnetization values read right to left. Panels (a) and (b) are one-dimensional $100$-spin ferromagnetic systems with periodic boundary conditions. Panels (c) and (d) show data from a two-dimensional $25\times 25$-spin ferromagnetic system with open boundary conditions. Panels (a) and (c) show simulations at $\Spause=0.3$ (stronger transverse field), and panels (b) and (d) show simulations at $\Spause=0.5$ (weaker transverse field). The spin configurations are all selected as representative not fully polarized configurations so as to illustrate the types of magnetic domains seen in the measured samples. The arrows overlayed on the hysteresis cycles denote both the direction of the longitudinal field sweep and the time progression of the protocol. Blue nodes/pixels denotes spin down ($-1$) and red denotes spin up ($+1$). }
    \label{fig:real_space_configs}
\end{figure}
In Fig.~\ref{fig:real_space_configs} we display selected magnetization configurations of individual measurements found on the quantum annealing hardware. These examples provide insight into the magnetic domains during the hysteresis protocol. 
In systems of $\mathbb{Z}_2$ symmetry, order corresponds to symmetry breaking between the two equivalent orientations. Under applied field, the longitudinal Zeeman energy of the field nucleates and grows domains of magnetization aligned to the field (see also average spin structure factors in Supplementary Information~\ref{section:appendix_2D_spin_structure_factor}). Examples of these magnetic domains are shown in Fig.~\ref{fig:real_space_configs}. These configurations in Fig.~\ref{fig:real_space_configs} were selected to be representative, in particular as the first samples found from the QPU readouts that are not fully polarized (note in particular many of the samples especially for $s=0.3$ simulations are fully polarized). Domains grow and combine to form larger domains as the magnetization reversal progresses.

 In a classical system at equilibrium, the number of domain walls $n_d$ depends only on temperature, scaling as $\ln( n_d) \sim-2\beta J$ at low wall density. In our out-of-equilibrium, quantum case, instead, the number of domain walls is influenced by the relative strength of the transverse and longitudinal field, as well as the rate at which the longitudinal field changes. In fact, as we will now show, stronger transverse field leads to fewer domain walls, as does a faster longitudinal field ramp rate. This is readily observed by comparing the domain walls in Fig.~\ref{fig:real_space_configs} poanels (a) and (b), corresponding $\Spause=0.3$ and $\Spause=0.5$, respectively. The majority of real space configurations corresponding to $\Spause=0.3$ are fully polarized.  We demonstrate now that our quantum evolution approach predicts scaling laws that are confirmed by  experimental data.

\begin{figure}[t!]
    \centering
    \includegraphics[width=0.999\linewidth]{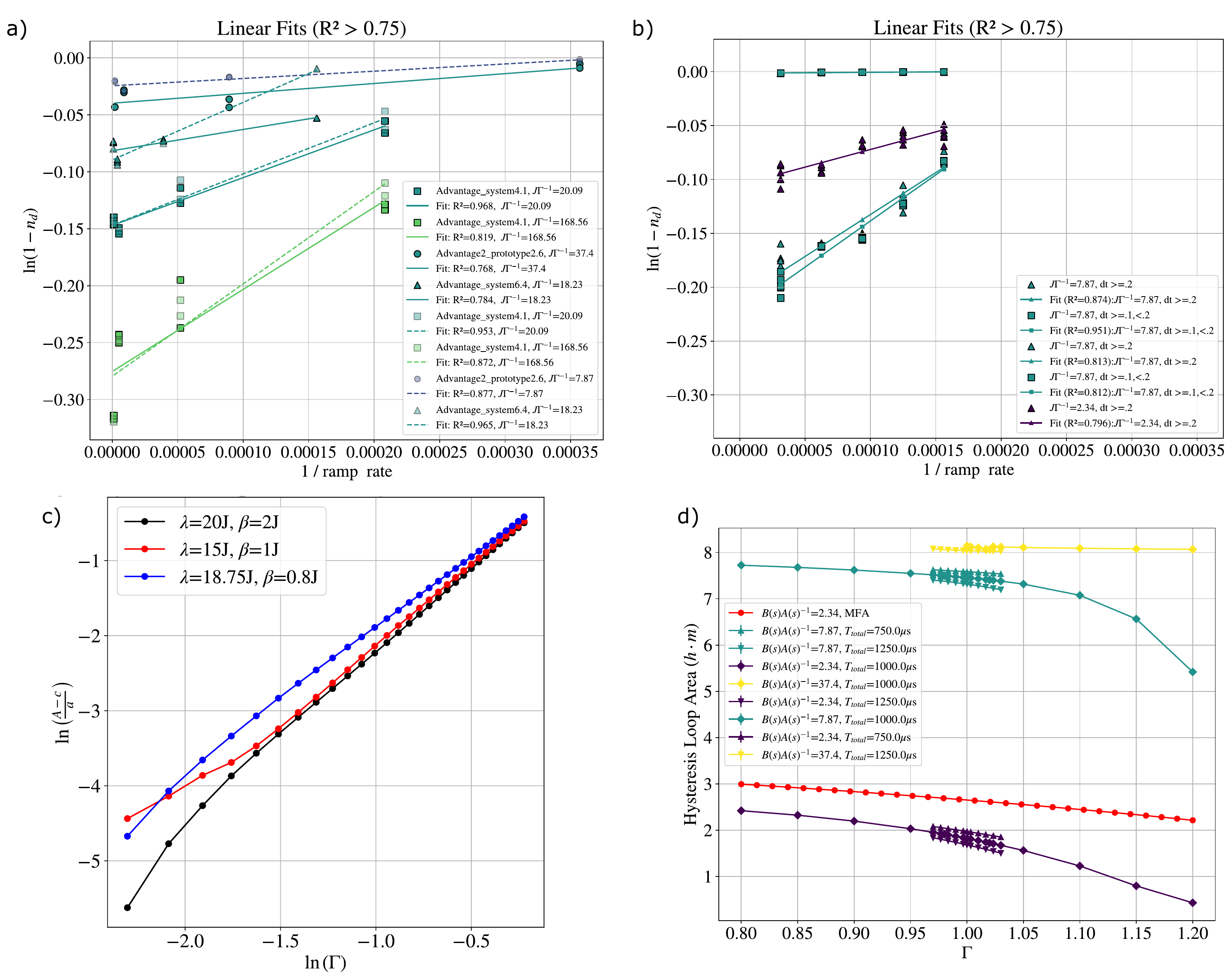}
    \caption{ (a) Linear fit of defects density $\ln(1-n_d)$ vs. inverse ramp rate ($1/\dot{h}$ in units of $\text{sec}/h_z$) for experimental (D-Wave) data on 1D spin systems with periodic boundary conditions. Fits with $R^2>0.75$ shown. Color indicates $J\Gamma^{-1}$ ratio, marker indicates device, solid/dashed lines indicate backward/forward sweeps, respectively. 
    (b) Linear fit for simulated data. Color corresponds to $J\Gamma^{-1}$ ratio, marker indicates time step size, $dt$. 
    (c) Plot of $\ln\left(\frac{A-c}{a}\right)$  vs. $\ln(\Gamma)$, showing power-law scaling loop area determined by the mean field approximation. Constants $a$ and $c$ are determined for each simulation parameters, $\lambda$ is a phenomenological relaxation rate,  $\beta$ is the thermodynamic inverse  temperature. 
    (d) Hysteresis loop area vs.  the scaled strength of the transverse field. Color indicates the ratio of the transverse and longitudinal Hamiltonians ($B(s)A(s)^{-1}$); markers indicate the anneal time in simulation. Mean field predictions as a function of $\Gamma$ (red dots). }
    \label{fig:Scaling}
\end{figure}

If we assume the independent level crossing approximation, where the first-order piecewise-constant propagator describes the quantum evolution, from the LZ-model we expect the number of kinks to relate to the adiabatic evolution probability, $q$ in eq.~\eqref{eqn:transition_prob}. For both experiments and simulations, we identify the scaling of kink density to the inverse of the ramp rate, $\dot{h}$, which is proportional to the maximum applied longitudinal field and inverse of the total annealing time. This can be expressed as 
\begin{align}
    \ln(1-n_d)= \frac{ \left[A(s)\Gamma^\prime\right]^2 5 T_\text{sweep}}{\hbar B(s)4|h^\prime_\text{max}|} ,
\end{align}
where $h_\text{max}$ is maximum strength of the longitudinal field, $T_\text{sweep}$ is the total (non-zero) longitudinal field sweep time in Fig.~\ref{fig:Hysteresis} (a), $\frac{2}{5}T_\text{sweep}$ corresponds to the duration of one forward or backward ramp, and $n_d$ is the maximum kink density during magnetization reversal. In simulation, $T_\text{sweep}$ is chosen to match experimental D-Wave values, as discussed above.  Thus, $\ln(1-n_d)$ is inversely proportional to the ramp rate. In Fig.~\ref{fig:Scaling} (a) and (b), we plot the linear fits  of $\ln(1 - n_d)$ versus $1/\dot{h}$ experimental and simulated data, respectively. The trends show strong agreement between experiments and simulation, with $R^2$ values for all experimental and simulation parameters provided in Supplementary Information~\ref{section:appendix_defect_density_as_a_function_of_longitudinal_field_ramp_rate}. 

The slopes given in Fig.~\ref{fig:Scaling} are comparable to theoretical predictions for the LZ-model for $0.4 \leq \Spause \leq 0.6$. 
Thus, within a range of $J\Gamma^{-1}$ ratios, the independent level approximation captures the kink density scaling with ramp rate observed in experiments. Notably, the correspondence between experimental (a) and simulated (b) data supports the validity of this approximation.


We now compare the hysteresis loop area with results from a mean field approximation (MFA), extending prior approaches to hysteresis in the TFIM \cite{Banerjee_PRE_1995,Acharyya_JPhysA_1994}. Beyond linking to this foundational work, this framework also provides a connection to emergent Landau-Lifshitz-Gilbert Equation (LLG) dynamics and semiclassical model of hysteresis. 
Governed by LLG dynamics, the precession of magnetization $\vec{m}$ in an effective field $\vec{H}_\text{eff}$ is described by the equation
\begin{align}
    \dot{\vec{m}}=-\gamma\vec{m}\times\vec{H}_\text{eff}-\lambda \vec{m}\times(\vec{m}\times\vec{H}),
\end{align}
where $\gamma$ is the gyromagnetic ratio and $\alpha$ the dampening parameter. This is a similar form to the Bloch equations for isolated non-interacting two-level systems~\cite{HELLER_PLA_1986}.

Hysteresis in the TFIM under an oscillating transverse field have been previously studied with mean field methods \cite{Acharyya_JPhysA_1994,stinchombe1,stinchombe2,Banerjee_PRE_1995}. 
The foundational work by Banerjee \textit{et al.} derived phenomenological equations from a mean field approximation for a TFIM coupled to a thermal bath. This model was derived without restriction on the bond coordination in the TFIM. 
We define the mean field as $\tilde{h}(t)=  J_{ij} \langle\sigma^z\rangle+h(t)$, yielding the Hamiltonian  $\mathcal{H}=-\sum_i\tilde{h}(t)\sigma^z_i+\Gamma \sigma^x_i$.  
We generalize this to the case of a time-dependent longitudinal field, as detailed in Appendix~\ref{apd:appendix_mean_field_approximation}. The expectation value of the magnetization is  $ m^\mu(t)=\Tr(\rho(t)\sigma^\mu)$; the $\hat{y}$-axis rotation operator $S_y=\exp{\frac{-i}{2}\sigma^y\arctan(\Gamma/\tilde{h}(t_0))}$ transforms the magnetization to the rotating frame, giving $m^\mu(t)=\Tr(\tilde{\rho(t)}S_y^{-1}\sigma^\mu S_y)$. Here $\tilde{\rho}$ is  the rotated density of states. In the weak $\Gamma$ limit, the dynamics are given by
\begin{subequations}
\begin{align}
    \frac{dm^z}{dt}&=-2\lambda\left[m^z-\tanh{(\beta h^\prime_0)}\right] -2\Gamma m^y
    \\
    \frac{dm^x}{dt}&=-\lambda m^x \cos{(h^\prime_0 t)}+2\tilde{h}(t) m^y
    \\
    \frac{dm^y}{dt}&=-\lambda \left[ m^y \cos{(h^\prime_0 t)} - m^x\sin{(h^\prime_0 t)}\right] -2\tilde{h}(t) m^x +2\Gamma m^z
\end{align}
\label{eqn:MF-LLG}
\end{subequations}
with $\beta=\frac{1}{k_B T}$,  $h_0=\sqrt{\tilde{h}(t)^2+\Gamma^2}$, $h^\prime_0=\text{sign}(\tilde{h}(t) )h_0$ and  $\lambda$ is the power spectral density of bath fluctuations (defined in the Supplementary Information~\ref{apd:appendix_mean_field_approximation}). 
 Eqs.~\ref{eqn:MF-LLG} resembles the LLG equation, with dampening arising from system-bath coupling. In the Supplementary Information~\ref{apd:appendix_mean_field_approximation} we derive $\frac{d\vec{m}}{dt}$ beyond the weak $\Gamma$ limit.

The hysteresis loop area, $A=\oint  \vec{m}\cdot d\vec{h}$ quantifies the energy dissipated during magnetization reversal. In this mean field model, the area scales with the transverse field via a power law,
\begin{align}
    A=a\Gamma^\alpha+c  , \qquad \alpha\sim 2 ,
\end{align}
as confirmed in simulations, shown in Fig.~\ref{fig:Scaling}(c).  We plot the scaling of hysteresis loop area in the MFA under various $\lambda$ and $\beta$, plotting $\ln((A - c)/a)$ vs.\ $\ln(\Gamma)$ to isolate the leading $\Gamma$-dependence. 
The strength of the transverse field can be related to the classical hardness of a magnetic material. Hysteresis loop area scaling with the transverse field in the MFA aligns with earlier studies using a time-dependent transverse fields, and no longitudinal field  \cite{Acharyya_JPhysA_1994}.

In Fig.~\ref{fig:Scaling} (d), we plot the hysteresis loop area vs. the transverse field for both simulations and the MFA. We incorporate the D-Wave parameters $A(s)$ and $B(s)$ in $\Gamma$, and $h(t)$ and  $J$, respectively, to better compare  simulations with the MFA. Here we set $B(s)A(s)^{-1}= 2.34$. As can be seen, hysteresis loop area decreases faster in simulation than in the MFA. 
 We find that the MFA and mixed semiclassical-first order quantum simulations yield comparable area scaling at small $\Gamma$, but diverge as $\Gamma$ increases. This indicates that while the MFA captures essential features of the TFIM under weak transverse fields and strong bath coupling, it fails to capture the hysteresis scaling observed in simulations and experiments. A complete description requires treating the full nonequilibrium quantum dynamics that are not captured in a MFA.

\section{Discussion}
\label{sec:conclusion}

We have shown that magnetic hysteresis in quantum annealers exhibits features that defy classical expectations, particularly in one-dimensional systems where classical theory predicts no hysteresis. Our results strongly suggests that finite coercivity, non-monotonic magnetization reversals, and transient negative susceptibilities are all hallmarks of non-equilibrium quantum dynamics
To explain these observations, we introduced a hybrid framework that interleaves a first-order piecewise-constant quantum propagator with semiclassical domain wall kinetics. This model captures both the diabatic excitations driven by longitudinal field sweeps and the relaxation pathways governed by domain wall motion. Numerical simulations based on this approach reproduce key experimental signatures across a wide range of parameters and system sizes.

Notably, our study reveals that magnetization reversal is governed by the interplay of transverse field strength, sweep rate, and domain wall energetics. 
We have shown that the qualitative and quantitative features of the experimentally observed hysteresis can be understood through the competition of adiabatic and diabatic evolution during longitudinal field sweeps.  The competition between the longitudinal field ramp rate and the energy gaps between eigenstates results in Landau-Zener transitions, which populate excited states and delay magnetization reversal. The observed bow-tie structure in the energy evolution during field sweeps (Fig.~\ref{fig:LZ-hyst}) is a direct signature of this diabatic evolution. Thus this model describes the stochastic evolution under a driving longitudinal field through quantum transitions but does not require coherent evolution. Instead quantum transitions occur through local entanglement which is interleaved with semiclassical evolution. These diabatic transitions alone are insufficient to describe the full hysteresis behavior. We introduced a semiclassical model to account for relaxation processes via domain wall motion, enabling magnetization to relax toward the aligned ground state. Interleaving the quantum and semiclassical dynamics required developing statistical matching techniques. The resulting hybrid low-order quantum–semiclassical model captures both the excitation and relaxation mechanisms contributing to hysteresis.

In one-dimensional systems, reversal consistently initiates around $|h| \sim 2J$, independent of system size, supporting a local domain nucleation picture. Magnetization reversal is influenced by anneal time and the relative strength of the transverse field.  Increased dwell time near an avoided crossing, or stronger transverse field, increases the probability of a spin flip, producing smaller hysteresis loops and magnetization reversal at applied fields of $|h_z|\leq 2J$. 
The observed negative susceptibilities and non-monotonic reversals are linked to transient occupation of excited states—features that are absent in two-dimensional systems where domain growth is more energetically costly. This behavior supports a picture in which quantum transitions are driven by local entanglement rather than sustained coherent evolution, and memory emerges from the interplay of discrete excitations and relaxation processes.

Comparison with mean-field models and density matrix renormalization group techniques highlights the importance of accounting for non-equilibrium dynamics. While mean-field approximations capture broad trends like loop area scaling, they fail to reproduce discrete transitions and domain-level memory effects. Our hybrid framework bridges this gap, revealing the essential role of quantum coherence interleaved with semiclassical relaxation.

These findings establish quantum annealers as programmable platforms for exploring quantum memory, emergent irreversibility, and driven transitions in many-body systems. The framework presented here can be extended to other quantum systems and protocols, offering a path toward understanding quantum dynamics under continuous measurement, decoherence, and control. 

Future work will explore applications to molecular magnets and other systems with intrinsic internal structure. Accurate modeling of non-equilibrium spin processes and emergent LLG-dynamics would advance simulations of realistic magnetic materials.  The control afforded by programmable annealers may reveal discrete steps, coherent signatures, and new types of quantum hysteresis.

\section*{Acknowledgments}
\label{sec:acknowledgments}
We thank Andrew King, Minseong Lee, and Vivien Zapf for helpful discussions. This work was supported by the U.S. Department of Energy through the Los Alamos National Laboratory. Los Alamos National Laboratory is operated by Triad National Security, LLC, for the National Nuclear Security Administration of U.S. Department of Energy (Contract No. 89233218CNA000001). The research presented in this article was supported by the Laboratory Directed Research and Development program of Los Alamos National Laboratory under project number 20240032DR. This research used resources provided by the Los Alamos National Laboratory Institutional Computing Program.

\appendix

\section{Quantum Annealing Hardware Methods}
\label{section:methods}

The sampling-based magnetic hysteresis protocol is implemented on D-Wave programmable quantum annealers using the methods described in ref.~\cite{Pelofske_arxiv_2025}, with a few small differences that are described below. D-Wave quantum annealing processors are based on superconducting flux qubits\cite{johnson2011quantum, Lanting_2014, Bunyk_2014}; these devices operate at approximately $15$ milliKelvin. The Hamiltonian that the D-Wave processor implements, dropping the $^\prime$ notation used in the main text, is given by 

\begin{equation}
    {\mathcal H} = - \frac{A(s)}{2} \Big( \sum_i \hat{\sigma}_{i}^{x} \Big) - \frac{B(s)} {2} \Big( g(t) \sum_i h_i \hat{\sigma}_i^{z} + \sum_{\langle i,j\rangle} J_{i, j} \hat{\sigma}_i^{z} \hat{\sigma_j}^{z} \Big)   ,
    \label{equation:QA_Hamiltonian_h_gain}
\end{equation}

where $\hat{\sigma}_{i}^{x}$ defines the transverse field driving Hamiltonian that facilitates state transitions, and the terms that are user-programmable are the local fields $h_i$ and the quadratic interaction terms $J_{i, j}$. The time-dependent function $g(t)$ defines a time-dependent schedule, which is typically known as the ``h-gain schedule''\footnote{We interchangeably use h-gain field and longitudinal field to refer to the same thing}, which acts as a global multiplier on all of the programmed local fields. The other critical component of this Hamiltonian that we program for the hysteresis sweep protocol is the \emph{anneal schedule} -- typically the default anneal schedule is a linear ramp, but on D-Wave hardware we can also program a time-dependent schedule as a function of the parameter $s$ which is defined to be within $[0, 1]$ and is typically known as the anneal fraction. Here we use the hardware convention where $h_i$ is the spatially varying local field and $g(t)$ determines the time dependent longitudinal field, and $g(t)h_i$ denotes the total applied longitudinal field. The site dependent local fields $h_i$ programmed at each qubit are set to a uniform positive coefficient of $4$ (which is the maximum local field that can be programmed on these three devices). 

We perform the magnetic hysteresis experiments on three D-Wave QPU's - these devices are summarized, including their exact chip ids, in Table~\ref{table:hardware_summary}. These QPUs are described by coupler connectivity graphs known as Pegasus~\cite{dattani2019pegasussecondconnectivitygraph, boothby2020nextgenerationtopologydwavequantum} and Zephyr~\cite{zephyr}. Oftentimes, we report or describe simulations in terms of hardware normalized control parameters, such as $s$ used in eq.~\eqref{equation:QA_Hamiltonian_h_gain}, but in all such cases there is an underlying physical energy scale given for example by the schedule in Figure~\ref{fig:Hysteresis}-b. 

\begin{table}[ht!]
    \begin{center}
        \begin{tabular}{|l||l|l|l|l|}
            \hline
            D-Wave QPU Chip & Graph name & Qubits & Couplers & Maximum h-gain field strength \\
            \hline
            \hline
            \texttt{Advantage\_system4.1} & Pegasus $P_{16}$ & 5627 & 40279 & $\pm 1.75$ \\
            \hline
            \texttt{Advantage\_system6.4} & Pegasus $P_{16}$ & 5612 & 40088 & $\pm 4$ \\
            \hline
            \texttt{Advantage2\_prototype2.6} & Zephyr $Z_{6, 4}$ & 1248 & 10827 & $\pm 3$ \\
            \hline
        \end{tabular}
    \end{center}
    \caption{D-Wave QPU summary. }
    \label{table:hardware_summary}
\end{table}

Figure~\ref{fig:Hysteresis}-a.i) and a-ii) illustrate examples of the time dependent control schedules for the D-Wave quantum annealing hardware that define the hysteresis sweep protocol. Left: the global longitudinal field control, which defines a time-dependent multiplier on all programmed local fields. Right: The anneal schedule, which defines the relative proportion of the driving Transverse field and the classical Hamiltonian energy scales at each point in time. We fix the proportion of transverse field to be constant during the simulation, and the longitudinal field sweeps are what carry out the hysteresis protocol. Because we can not monitor the state of the simulation while it is occurring, we progress through this protocol while stopping at intermediate points and measuring the state of all qubits, and then re-starting the simulation process from the beginning for a new anneal-readout cycle. In that sense, this protocol progresses in time, but the actual order in which the measurements are made does not matter. At each slice of the anneal protocol we generate many samples from independent anneal-readout cycles. The red region denotes the initial polarization phase, which is necessary because the system begins in the unpolarized state where the longitudinal field set to $0$ (in all experimental plots we do not show data from this red region). Next, the cyan and blue regions denote the complete hysteresis sweeps that drive the system through full magnetization polarity reversal.

We use four different total simulation times for different aspects of the analysis shown in this study. These simulation times are as follows: $11.2$ microseconds (h-gain sweep time of $10 \mu s$, with $\approx 400$ readout points, and $2000$ samples measured per point), $50 \mu s$ (h-gain sweep time of $48.8 \mu s$, with $\approx 400$ readout points, and $2000$ samples measured per point), $500 \mu s$ (h-gain sweep time of $498.8 \mu s$, with $\approx 2000$ readout points, and $400$ samples measured per point), and $1999.2 \mu s$ (h-gain sweep time of $1998 \mu s$, with $\approx 2000$ readout points, and $400$ samples measured per point). Typically most of the simulations were performed with the annealing time of $11.2 \mu s$, and some were simulated additionally with the other three annealing times. Note that these total simulation times are the \emph{final} simulation time -- the protocol progresses in time. Varying the duration of the annealing time (note that these are not actually performing annealing, so one may consider this simply the total physical simulation time) not only allows us to observe changes in the out of equilibrium dynamics of the measured hysteresis cycles, but in particular it also naturally gives experiments with different longitudinal field ramp rates.

In this study we examine two different types of ferromagnetic Ising models. The first is 1D systems with periodic boundary conditions. The second is a 2D square grid with open boundary conditions. The key feature of these models, for the purposes of hardware implementation, is that they are sparse, meaning that we can find a subgraph isomorphism mapping from these logical problems directly onto the hardware graph (meaning one spin is mapped to one qubit on the hardware), where the edges of the hardware graph are defined by the physical $J_{i, j}$ couplers. We use the graph isomorphism finder called the Glasgow solver \cite{mccreesh2020glasgow} to find hardware isomorphism embeddings for all Ising model instances used in this study, using iterated embedding attempts. For the 2D square grid, we use a $25\times 25$ spin grid on all 3 D-Wave QPU's. For the 1D ring with periodic boundary conditions, we embed a $100$ spin model onto the hardware graphs of all devices, and we also embed the largest ring we could find (using a reasonable amount of compute) on each of the hardware graphs. For \texttt{Advantage\_system4.1}, this is $4906$ spins, for \texttt{Advantage\_system6.4} this is $4885$ spins, for \texttt{Advantage2\_prototype2.6} this is $1131$ spins. Lastly, we embed $4$ spin rings onto the hardware graphs to study very small ferromagnetic system size cases. Because the amount of hardware used for the $4$ spin case is very small, we can embed many independent copies of the $4$ spin Ising model on the hardware using iterative subgraphs isomorphism embedding calls. Then, when running one anneal-readout cycle on the D-Wave hardware, we get many additional samples obtained. For \texttt{Advantage2\_prototype2.6}, $256$ disjoint instances can be embedded, on \texttt{Advantage\_system4.1} $1175$ instances be embedded, and on \texttt{Advantage\_system6.4} $1185$ instances can be embedded. All reported hysteresis data from these $4$ spin ferromagnetic models are averaged over all of these independent embeddings, with the exception of the central comparison between simulation and D-Wave experiment shown in Fig.~\ref{fig:4spin_ExpSim} where only $10$ of the disjoint embeddings were used. In addition to the advantage of having more samples, and therefore more robust statistics, using multiple embeddings can also result in averaging out any local hardware biases. This process is known as parallel quantum annealing, or tiling \cite{parallel_QA, pelofske2024short, PhysRevA.91.042314, Pelofske_PhysRevResearch_2023}. The purpose of simulating different system sizes is to determine to what extent the magnetic hysteresis we observe changes with respect to system size.

For the 1D ferromagnetic model implementation on D-Wave devices, we utilize two different simulation techniques. The first is the standard approach, where the models are implemented on the hardware directly as ferromagnetic couplers. However, with the aim of differentiating between at least some of the potential noise sources on the hardware, we also aim to implement 1D antiferromagnetic rings, which are transformed into ferromagnetic rings via a combination of gauge transformations\footnote{This process is also known as spin reversal transforms} (post qubit-readout) along with site-dependent alternating single site local fields (which are then uniformly acted on by the global longitudinal field). The gauge transformation modification of the longitudinal field is applied by programming $+h_i$ for every odd site on the ring, and $-h_i$ for every even site on the ring. The gauge transformation method is motivated by small hardware biases which can make the simulations of ferromagnetic and antiferromagnetic models different, thus there is a natural question of whether the ferromagnetic hardware simulations are significantly different from the antiferromagnetic hardware simulations. The use of these staggered local fields are required to be used for the gauge transformation method because otherwise, under the standard uniformly applied longitudinal field, the state we would end up polarizing towards would not be an antiferromagnetic ground state but rather a ferromagnetic ground state (in other words without the staggered field, post-spin reversal transform there would be no magnetic hysteresis observed). Moreover, this gauge transformation method is a slight departure from the uniform longitudinal field used in ref.~\cite{Pelofske_arxiv_2025}. We will refer to this gauge transformation case specifically when reporting data as \emph{AFM gauge transformation}, and in all other cases ferromagnetic coupling on the hardware is used. The \emph{AFM gauge transformation} method is used specifically for the 1D ferromagnetic system cases for $100$ spin and larger spin ($4906$, $4885$, $1131$) systems (not, for example, the 2D grid cases).

As in ref.~\cite{Pelofske_arxiv_2025}, when plotting the average magnetization observable $M_z$, we report the sign reversed quantity (relative to what is actually measured on the hardware) in order to compensate for the sign of the D-Wave hardware Hamiltonian eq.~\eqref{equation:QA_Hamiltonian_h_gain}.

\section{Derivation of first-order piecewise-constant propagator}
\label{apd:QuantEvol}

Here, we describe the quantum origin of magnetic memory and hysteresis in one-dimensional systems. 
Our approach reproduces experimental results by combining various conceptual features, and can be generalized to higher dimensional systems.  
The one-dimensional TFIM supports spin flips via the transverse field; we explicitly model these spin flip operations and the non-equilibrium response of the system to a time varying applied field. Our method accounts for both non-equilibrium driving and relaxation processes across distinct timescales.

Here we briefly review the Landau-Zener (LZ) model, which describes transitions between instantaneous eigenstates at avoided level crossings in asymptotic long time limit. Spin flip excitations in the presence of a ramped longitudinal field can be understood in terms of the LZ model. We examine this simple case to motivate our first order treatment of the quantum time evolution. 
In the LZ-model the system evolution is described by a linearly time dependent Hamiltonian
\begin{align}
    i\frac{d\psi}{dt} &={\mathcal{H}}(t)\psi, \quad \mathcal{H}(t) = \hat{A} + \hat{B}t
\end{align}
with $\hat{B}$ a diagonal $2^N\times 2^N$ matrix, and in the diabatic basis $\hat{A}$ is a $2^N\times 2^N$ matrix with off-diagonal elements. 
For two-level systems, diabatic excitations correspond to tunneling to an excited state when the eigenstates are close in energy. The probability to remain in the ground state (adiabatic states) depends on the energy gap at the avoided crossing (determined by off-diagonal elements, i.e., transverse field) and the rate at which the system is being swept by the time dependent component of the Hamiltonian. For a two-level system with Hamiltonian 
$\mathcal{H}(t)=\begin{pmatrix}
    b_1t  & a \\ a &b_2t
\end{pmatrix}$
the transition probability depends on the off-diagonal terms as well as the ramp rate. 
While the usual derivation relies on semiclassical methods, such as the Wentzel–Kramers–Brillouin (WKB) approximation or the Dykhne–Davis–Pechukas saddle-point method \cite{Benderskii_JExpTheorPhy_2003,dykhne1962adiabatic,Davis_JChemPhys_1976}, we provide a derivation of the transition probability using the path integral formalism and contour integration around a complex-time branch point in the Supplementary Information~\ref{section:appendix_derivation_of_LZ_transition_probability}. The transition probability for such a two-level system is $p= e^{-\frac{2\pi a^2}{\hbar\vert b_2-b_1\vert}}$. 

For an ideal two-level system in the TFIM without qubit interactions, $J=0$ the probability of a diabatic evolution   (p) or adiabatic transition, i.e., spin flip, (q) is
\begin{align}
p&=e^{-\frac{\pi\Gamma^2}{2\hbar \dot{h}}}    
, \qquad 
q =1-e^{-\frac{ \pi\Gamma^2}{2\hbar \dot{h}}}    .
\label{eqn:transition_prob}
\end{align} 
Let us now consider the TFIM Hamiltonian with interacting spins, under a longitudinal field sweep:
\begin{align}
    \mathcal{H}(t)=-\sum_{\langle i,j\rangle}J_{ij}\sigma^z_i\sigma^z_j -\sum_i\Gamma\sigma^x_i -h(0) \sum_i \sigma^z_i-\sum_i \dot{h} t  \sigma^z_i,
    \label{eqn:LZ_Ham}
\end{align}
where $\dot{h}$ is the longitudinal field ramp rate, i.e., the slope of the longitudinal field sweep (which is proportional to the maximum applied longitudinal field and inverse of the total annealing time).
To illustrate, we consider the TFIM for $N=3$ spins. Its Hamiltonian matrix is given:
\scriptsize
\begin{align}
    \mathcal{H}_{N=3}&=
    \begin{blockarray}{cccccccccc}
 & & \ket{\uparrow\uparrow\uparrow} & \ket{\downarrow\uparrow\uparrow} & \ket{\uparrow\downarrow\uparrow} & \ket{\uparrow\uparrow\downarrow} & \ket{\downarrow\downarrow\uparrow} & \ket{\downarrow\uparrow\downarrow} & \ket{\uparrow\downarrow\downarrow} & \ket{\downarrow\downarrow\downarrow} \\
\begin{block}{cc(cccccccc)}
  \ket{\uparrow\uparrow\uparrow} & & -3 J-3 h(t) & -\Gamma & -\Gamma & -\Gamma & 0 & 0 & 0 & 0  \\
   \ket{\downarrow\uparrow\uparrow} & & -\Gamma &  J-h(t)  & 0 & 0 & -\Gamma & -\Gamma & 0 & 0 \\
  \ket{\uparrow\downarrow\uparrow} & &-\Gamma & 0 &  J-h(t)  & 0 & -\Gamma& 0 & -\Gamma & 0 \\
  \ket{\uparrow\uparrow\downarrow} & & -\Gamma & 0 & 0 &  J-h(t)  & 0 & -\Gamma & -\Gamma & 0 \\
  \ket{\downarrow\downarrow\uparrow} & & 0 & -\Gamma & -\Gamma & 0 &  J+h(t)  & 0 & 0 & -\Gamma \\
  \ket{\downarrow\uparrow\downarrow} & & 0 & -\Gamma & 0 & -\Gamma & 0 &  J+h(t)  & 0 & -\Gamma \\
  \ket{\uparrow\downarrow\downarrow} & & 0 & 0 & -\Gamma & -\Gamma & 0 & 0 &  J+h(t)  & -\Gamma \\
  \ket{\downarrow\downarrow\downarrow} & &  0 & 0 & 0 & 0 & -\Gamma & -\Gamma & -\Gamma & - 3 J+3h(t)  \\
\end{block}
\end{blockarray}
\label{eqn:LZ-example}
\end{align}
\normalsize

For complex systems, such as multistate systems relevant to the $N$-qubit ring, the LZ-model is only valid exactly if several conditions are satisfied \cite{Sinitsyn_PRL_2018,Li_PRA_2017,Malla_PRB_2021}. 
Here, we specify the three main conditions: the time dependent Hamiltonian can be written as a time invariant component and a time dependent contribution with a linear time dependence,  that we can write the Hamiltonian in the diabatic basis such that there are no transitions between degenerate states,  and that the system satisfies integrability conditions including finding a set of nontrivial commuting Hamiltonians. The first and second conditions are satisfied by the TFIM with linear ramps, although the third condition is not guaranteed. 
We can find approximate conditions\cite{Sinitsyn_JPhysA_2017} (i) that closed paths in the diabatic level diagram should enclose zero areas and (ii) if diabatic levels intersect without coupling within a projected Hamiltonian to the subspace of intersecting levels then there is a regime of finite coupling wherein the corresponding adiabatic levels of the full-Hamiltonian are degenerate.
The TFIM typically violates these, except in restricted subspaces (e.g., single-spin flip states). Hence, the LZ model does not apply exactly to one-dimensional systems under a hysteresis protocol. This motivated finding a regime where the LZ model can approximate the quantum state transitions in a multi-state systems, as discussed in the main text.

Such an approximation can be found in the independent level crossing approximation, when the quantum evolution of a multilevel state can be approximated as a series of sequential and separable independent two-level crossings. When these conditions are satisfied  the systems satisfies the the independent crossing approximation~\cite{Sinitsyn_PRB_2015}.
There are hierarchy constraints such that if there are multiple eigenstate crossings as a function of applied field, such as multiple crossings for a single prepared state, each crossing needs to be separable. The instantaneous  diabatic transition probability in a two-level system is exponentially suppressed by the energy gap; avoided crossings between levels which diverge are well modeled by a finite time window. 
For example, consider two levels that have one avoided crossing at $t_0$, around $t_0$ there is a finite time window $\tau_{\mathrm{LZ}}$ that is sufficient to approximate the transition probability. This is valid when the transition probability is exponentially suppressed away from $t_0$ such that at time $t_0+\frac{1}{2}\tau_{\mathrm{LZ}}$ the instantaneous transition probability is negligible. 
Simply put, the transition probabilities for a given eigenstate at any point in time is negligible except for a single transition, and it is clear in which order the system will transition between states both in terms of sequence (time hierarchy) and eigenvalue (energy hierarchy).

In the case of the one-dimensional TFIM, 
crossings are separated by an energy $J=1$ and the system has a time hierarchy $\Delta t=\frac{J}{\dot{h}}$. Thus the independent level crossing approximation will not hold at some fast sweep rate. 
As the independent level crossing approximation does not capture interaction between excitations,  dynamics can be approximated with a mixed semiclassical and low order exact quantum evolution framework. 
In the TFIM, semiclassical dynamic phase effects do not affect the transition probabilities, as will be discussed below, thus incorporating the semiclassical dynamics do not violate additional constrains of the LZ-model. 
In the simple case of multiple well-separated levels that satisfy integrability conditions, the evolution can be approximated as 
\begin{align}
    \mathcal{T} = \hat{T}_\text{sc}^{t_\infty,t_n}S^{n,n-1}\cdots \hat{T}_\text{sc}^{t_2,t_1}S^{0,1}\hat{T}_\text{sc}^{t_1,t_{-\infty}}
    \label{eqn:MixedModel}
\end{align}
with $\hat{T}_\text{sc}$ is the operator that implements and interleaves the semiclassical evolution  and $S^{a,b}$ scattering matrices which reproduce the LZ-model transition probabilities.

Within the LZ framework alone, spin reversal following adiabatic evolution is exponentially suppressed due to large energy gaps away from the avoided crossing.  Thus the independent level crossing approximation suppresses relaxation mechanisms essential to the emergence of memory in hysteresis.  Diabatic transitions leave the magnetic system in an excited state due to the Zeeman energy under a reversed field. The independent level crossing approximation does not account for transitions between separated eigenstates. As such relaxation mechanisms need to be incorporated in the model.

Large multistate systems can be difficult to simulate, and as discussed above the the independent level crossing approximation fails to capture relaxation. Further, the open quantum system effects and quench dynamics introduce dynamical effects that can be modeled classically. This motivates incorporating the mixed semi-classical model discussed in the main text.
We argue that first order Magnus expansion, i.e, a first-order piecewise-constant propagator, is justified in a regime that nearly satisfies the independent level crossing. This low order propagator captures the relevant quantum excitations, but neglects relaxation mechanisms that can be influenced by classical affects. Thus, a low-order propagator is needed to incorporate classical dynamics.

A fully analytic treatment based on successive Landau–Zener (LZ) scatterers would require that each avoided crossing be strictly isolated in time so that the associated diabatic amplitudes evolve independently.  In the $N$-qubit ring the ideal of an isolated two-level avoided crossing is only approximate, the smallest instantaneous energy gap between eigenstates directly coupled via single-spin-flip operations, $\Delta E^{\text{min}}_{ij\in 2^N}(t_c)$, is tiny such that other Hamming-adjacent levels can lie within the same Landau–Zener window. In this case their transition amplitudes overlap and interfere, forcing a genuine multilevel treatment rather than a sequence of independent two-level scatterers.

A model sufficient to capture the relevant dynamics needs to incorporate the quantum excitations, higher order interference and relaxation, and classical dynamics incorporating open system and quantum quench effects.
 To reproduce LZ-model dynamics while incorporating multiple level interactions, beyond the idealized LZ chain, we can model dynamics with a short time first-order piecewise-constant propagator of the full Hamiltonian $\mathcal{H}(t)$.  During a short slice $\Delta t \ll \tau_{\mathrm{LZ}}=\Delta E^{\text{min}}/\dot{h}$ we apply the propagator $\exp[-i\mathcal{H}(t_n)\Delta t/\hbar]$.  
 Near the avoided crossing in finite time windows there is sufficient local entanglement such that quantum transitions are enabled. 
 When the crossings are well separated this step reproduces the single–crossing LZ exponent to leading order, and  the probability of a quantum transition is well approximated by a LZ-model transition probability.  Thus, near every avoided crossing the dynamics factorizes as a single complex branch point controls the probability of leaving the adiabatic state. When levels overlap the same step retains every off–diagonal matrix element, thereby capturing multilevel interference. 
 This motivates the proposed interleaved quantum and semi-classical dynamics presented in the main text. The evolution of well separated levels with relaxation mechanisms and open-system processes is defined by the time evolution
\begin{align}
    \mathcal{T} = \hat{T}_\text{sc}^{t_\infty,t_n}U(t_n,t_{n-1})\cdots \hat{T}_\text{sc}^{t_2,t_1}U(t_0,t_1)\hat{T}_\text{sc}^{t_1,t_{-\infty}},
    \label{eqn:MixedModel_2}
\end{align}
where we interleave quantum slices with a semiclassical drift–diffusion update for the domain-wall densities and first-order piecewise-constant propagator implemented with unitary operator $U(t_i,t_j)$.

\section{Derivation of semiclassical kinetics}
\label{apd:SemiClassEvol}

Diabatic evolution can leave the system trapped in excited states as higher-order processes like domain wall annihilation, as well as non-unitary evolution due to open system effects, are not captured in our low-order quantum evolution and coherent spin–flip dynamics. We can account for domain wall motion, interactions, and irreversible evolution through a kinetic equation for the kink
(domain–wall) density. Near the coercive field the switching time diverges, a critical-like slowing, in this regime quantum coherence and classical kinetics coexist, motivating a semiclassical treatment.

As discussed in the main text, the semiclassical approximation is valid when the system's wavefunction can be expressed as $\Psi=A(x,t)e^{\frac{-i}{\hbar}R(x,t)}$, allowing the independent treatment of phase and amplitude.
We write the wavefunction in a continuum coordinate $x$ along the chain (coarse–graining
over lattice spacing $a$) as
\[
\Psi(x,t)=A(x,t)\exp\!\left[-\frac{i}{\hbar}R(x,t)\right],
\]
where $A(x,t)$ is a slowly varying envelope and $R(x,t)$ the rapidly varying action (phase). 
We assume a hybrid scale separation, the spatial phase gradient is small (long–wavelength / heavy–mass limit) while the temporal phase winds rapidly (allowing phase averaging between crossings). Concretely,
\begin{align}
\tilde\epsilon_x = k\,\ell_A = \frac{|\partial_x R|\,\ell_A}{\hbar} \ll 1,
\qquad
\epsilon_t = \frac{\tau_\phi}{\tau_A}
           = \frac{\hbar}{|\partial_t R|\,\tau_A} \ll 1,
\end{align}
where $\ell_A$ and $\tau_A$ are the characteristic spatial and temporal envelope scales, 
$k=\partial_x R/\hbar$ is the (small) kink wavenumber, 
$\tau_\phi = 2\pi\hbar/|\partial_t R|$ the (short) dynamical phase winding time, and 
$\lambda_{\mathrm{dB}} = 2\pi/k$ the (large) kink de Broglie wavelength. 
Thus the spatial condition can equivalently be written as
\begin{align}
k\ell_A \ll 1 , \qquad \frac{|\partial_x R|}{\hbar} \ll \frac{1}{\ell_A},
\end{align}
while the temporal WKB–like condition reads
\begin{align}
\frac{|\partial_t A|}{A} \ll \frac{|\partial_t R|}{\hbar}.
\end{align}
This ordering arises because $J \gg \Gamma$ (heavy kinks with effective mass 
 giving small $k$ for a given drift velocity) and the longitudinal field ramp rate satisfies $\dot h \ll J^{2}/\hbar$, so intrinsic precession (energy) frequencies greatly exceed the sweep rate. To leading order in $(\tilde\epsilon_x,\epsilon_t)$ the phase satisfies a Hamilton–Jacobi equation with nearly position–independent spatial gradient, and the next order yields a continuity (kinetic) equation for the kink density

The transition between eigenbasis in the LZ-model does not depend on the dynamics phase factor,  $\hbar\Delta\phi_{ij}=\int^t \Delta E_{ij}=\int^t h_z(t)(N_i-N_j)$. Near avoided crossings, this reduces to the two-level LZ result. In open systems, dephasing washes out accumulated phase.  Similarly,  rapidly oscillating coherent phases are canceled out from the dynamic phase. Hence, only the near-degeneracy region contributes meaningfully to transitions as expected from the independent level crossing approximation. Finally, the D-Wave hardware sets the TFIM energy scale,
and the frequency of quantum oscillations  are much faster than the ramp rate of the longitudinal field, decoupling  the dynamics.

Under these conditions the amplitude evolves as
\begin{align}
    \frac{\partial A(x,t)}{ \partial t} &= -\frac{1}{M}\frac{\partial A(x,t)}{ \partial x}\frac{ \partial R(x,t)}{\partial x}-\frac{1}{2M}A\frac{\partial^2R(x,t)}{\partial x^2}
    \nonumber\\
    &= -v_0\frac{\partial A(x,t)}{ \partial x} ,
\end{align}
where $\partial_xR(x,t)$ is the canonical momentum and $v_0 = (1/M)\,\partial_x R$. In a uniform ring under a homogeneous longitudinal field and non-interacting domain walls, the canonical momentum and the domain wall velocity $v_0$ is approximately spatially uniform.  

Higher–order corrections (kink dispersion, interference between successive crossings,
multi–kink coherences) enter at $\mathcal O(\tilde\epsilon_x^2)$, $\mathcal O(\epsilon_t)$,
or $\mathcal O((\Gamma/J)^2)$ and are neglected here.

Since $A(x,t)$ represents the envelope of a domain wall at $x$, its spatial derivative reflects domain wall density variation. We construct a kinetic equation for domain walls (kinks)  using measured domain wall densities, $n$. Assuming the semiclassical approximation is valid, projecting to the diabatic basis, given $\Psi=A(x,t)e^{\frac{-i}{\hbar}R(x,t)}=\sum c_i\ket{\psi_i}$, the density of domain walls is $n\propto |A(x,t)|^2\delta_{DW}=\sum n_i=\sum |c_i|^2\delta_{DW}$, $\delta_{DW}$ denotes the domain–wall kernel (Dirac delta) that localizes a kink.
In this framework, we approximate the kink  density as $ n(x,t) \propto |A(x,t)|^2 $, corresponding to a statistical mixture over diabatic configurations weighted by $ |c_i(t)|^2 $.
 Writing $n=|A|^{2}$ gives the kinetic equation
\begin{align}
\partial_t n + v_0 \partial_x n = 0.
\end{align}
Experimentally measured $n(x,t)$ thus updates the envelope $A$, while coherent LZ rules govern the discrete spin–flip (gap–crossing) events.

Assuming that $A(x,t)$ is a smooth, real-valued function (consistent with the phase decoupling discussed above) and approximately uniform due to the near-uniform distribution of domain walls, the amplitude dynamics can be written  in terms of the density of kinks. Identifying the local density as $n(x,t) \propto |A(x,t)|^2 $, and $ n(x,t) > 0 $, we obtain $\partial_t |A(x,t)|^2=\sum_i \partial_t n_i(x,t)$. This implies the amplitude evolves according to $\partial_t A(x,t)=\frac{1}{2\sqrt{n(x,t)}}\partial_t n(x,t)$. Hence, the short-time dynamics of $  A(x,t) $ are governed by the redistribution of domain walls, as captured by the semiclassical kinetic equation.

\section{Mean Field Approximation}
\label{apd:appendix_mean_field_approximation}
In this section we derive the mean field approximation generalizing the derivation of Banerjee \textit{et. al.}\cite{Banerjee_PRE_1995} for time varying longitudinal fields.  We can write the full Hamiltonian coupled to the environment as $H_T=H_s+V+H_B$ where $V$ is the coupling between the bath and the spin system that induces spin flips, and $H_B$ is the bath Hamiltonian. We consider a system that is able to equilibrate with the bath in a fast time scale such that we can treat the longitudinal field as static. 
For the mean field approximation coupled with a bath we start with the spin system Hamiltonian 
\begin{align}
    H_s &=-\frac{1}{2}\sum_{\langle i,j\rangle } J_{ij}\sigma^z_i\sigma^z_j-\sum_i h(t)\sigma^z_i -\sum_i \Gamma\sigma^x_i
    \\
    H_{MFA} &\approx \sum_i-\tilde{h}(t)\sigma^z_i-\Gamma\sigma^x_i
\end{align}
with the mean field $\tilde{h}(t)=h(t)+ J_{ij}\langle\sigma^z\rangle = h(t)+ cJ\langle \sigma^x\rangle$ with $c$ the coordination number. With a one-dimensional system we have $\tilde{h}(t)=h(t)+J\langle\sigma^z\rangle$ and $H_s=-\sum_{i<j}\sigma^z_i\sigma^z_j-\sum_i h(t)\sigma^z_i-\Gamma\sigma^x_i$.   We can rotate the system about the $\hat{y}$-axis with $S_y=\exp{\frac{-i}{2}\sigma^y\arctan(\Gamma/\tilde{h}(t))}$, e.g., $S_y=\exp{-\frac{i}{2}\arctan\left(\frac{\Gamma}{J\langle\sigma^z\rangle+h(t)}\right)\sigma^y}$. Now the rotated Hamiltonian is diagonalized in the $\hat{z}$-basis,
\begin{align}
    \tilde{H}_s=-h^\prime_0\sum_i \sigma^z_i  ,
\end{align}
with $h_0 =\sqrt{\tilde{h}(t)^2+\Gamma^2}$ and $h^\prime_0 = \text{sign}(\tilde{h}(t))h_0$. This is an oscillating Hamiltonian.

We can obtain the rotated coupling $\tilde{V}=g\hat{b}\left(\sigma^x+\sigma^y\right)$ from an original coupling 
\begin{align}
V=g\hat{b}\left(\frac{1}{h_0}\left(|J\langle\sigma^z \rangle +h(t)|\sigma^x-\Gamma\sigma^z\right)+\sigma^y\right) ,
\label{eqn:OriginalBathCoupling}
\end{align}
with $g$ a scalar coupling strength, and $h_0=\sqrt{\left(J\langle\sigma^z \rangle +h(t)\right)^2+\Gamma^2}$. 


Now we can write the equation of motion for the density for states, we have the density of states in the rotated frame $\tilde{\rho}(t)=S_y^{-1}\rho(t)S_y$. We find the reduced reduced density of states by taking the trace over the bath states, $\rho_s(t)=\text{Tr}_b[\tilde{\rho}(t)]$.

Now using the interaction picture to account for the interaction with the bath, the density of state can be rewritten as
\begin{subequations}\begin{align}
\rho_s(t) &= e^{-i\tilde{H}_s t}\text{Tr}_b\left(\hat{\mathcal{T}}\exp{\left( -i\int^t_0 dt^\prime [\tilde{V}_I(t^\prime),\circ]\right)}\rho(0)\right)e^{i \tilde{H}_s t}
\\
&\approx  e^{-i \tilde{H}_s t}\left(\hat{\mathcal{T}}\exp{\left( -i\int^t_0 dt^\prime \langle[\tilde{V}_I(t^\prime),\circ] \rangle  -i\int^t_0 \int_0^t dt^\prime  dt^{\prime\prime} \langle[\tilde{V}_I(t^\prime),[\tilde{V}_I(t^{\prime\prime}),\circ]] \rangle  \right)}\rho_s(0)\right)e^{i \tilde{H}_s t}
\label{eqn:rho_interactionpic}
\end{align}
\end{subequations}
where the second line is keeping terms to second order and we have assumed the density of states begins as separable state $\rho(0)=\rho_s\otimes\rho_b$. Here the interaction potential between system and bath in the interaction picture is given as $V_I(t)=e^{i(H_s+H_b)t} V e^{-i (H_s + H_b)t}$.  Assuming the bath coupling is invariant to time translation and that the correlations are non-trivial only in short time ,  the first moment of the coupling vanishes and  we can write eq.~\eqref{eqn:rho_interactionpic} as
\begin{align}
    \rho_s(t)&=e^{-i \tilde{H}_s t}\left(\hat{\mathcal{T}}\exp{\left(  -i\int^\infty_0 dt^\prime  \langle[\tilde{V}_I(t^\prime),[\tilde{V}_I(0),\circ]] \rangle  \right)}\rho_s(0)\right)e^{i \tilde{H}_s t}  .
\end{align}

Keeping terms to second order we can write the equation of motion for the reduced density of states in the rotated frame as 
\begin{align}
    \frac{\partial \rho_s(t)}{\partial t} &= -i\left[\tilde{H}_s,\rho_s(t)\right] - e^{-i \tilde{H}_st} \int_0^\infty d t^\prime \langle [\tilde{V}_I(t^\prime),[\tilde{V}_I(0),\circ]]) \rangle e^{i \tilde{H}_s t} \rho_s(t).
\end{align}
Now we can examine the magnetization in terms of the reduced and rotated density of states
\begin{align}
m^\mu(t) &=\text{Tr}\left(\rho(t)\sigma^\mu\right)
\nonumber \\
&= \text{Tr}_s\left(\rho_s(t)S_y^{-1}\sigma^\mu S_y\right)
\nonumber \\
&= \Tr_s \left(\tilde{\rho}_s e^{iH_s t}S_y^{-1}\sigma^\mu S_y e^{-iH_s t}\right)
\end{align}
with $\tilde{\rho}_s$ in the interaction picture. Similarly we can find the a dynamical equation of the magnetization as,
\begin{align}
    \frac{\partial m^\mu(t)}{\partial t} &= \text{Tr}_s\left(\frac{\partial \rho_s(t)}{\partial t}S_y^{-1}\sigma^\mu S_y\right)+\text{Tr}_s\left( \rho_s(t) \partial_t \left(S_y^{-1}\sigma^\mu S_y\right)\right)    .
    \label{eqn:Mmu_rho}
\end{align}

We can write the  bath correlation in terms of a phenomenological relaxation rate $\lambda=\int_{-\infty}^\infty d\tau e^{ih\tau}(\langle \hat{b}(\tau)\hat{b}(0)\rangle+\langle \hat{b}(0)\hat{b}(\tau)\rangle)$. Now we can determine the magnetization and the equation of motion. 
First we  define 
\begin{align}
    g(h(t)):=\frac{i}{2}\frac{\Gamma\dot{h}(t)}{\Gamma^2+\tilde{h}(t)^2}
\end{align}
with $\dot{h}(t)$ the time derivative of $h(t)$, i.e., $\dot{h}$. We take the derivative of the rotation ($S_y^{-1}\sigma^\mu S_y$) in eq.~\eqref{eqn:Mmu_rho}. We determine the equation of motion for the magnetization components as, 
\begin{subequations}
\begin{align}
    \frac{d m^z}{dt}  = & 2 h^\prime_0\left(-m^y\sin{\left(\arctan{\left(\frac{\Gamma}{\tilde{h}}\right)}\right)}\right) +\lambda\left( 2\left(-m^z +\tanh\left(\beta h^\prime_0\right)\right)\cos{\left(\arctan{\left(\frac{\Gamma}{\tilde{h}}\right)}\right)} +\langle\sigma^x\rangle\cos{(h^\prime_0 t)}\sin{\left(\arctan{\left(\frac{\Gamma}{\tilde{h}}\right)}\right)} 
    \right. \nonumber\\ &\left.
    +\langle\sigma^y\rangle\sin{(h^\prime_0 t)}\sin{\left(\arctan{\left(\frac{\Gamma}{\tilde{h}}\right)}\right)}\right) + 2i g(h(t))\left(2 m^x \cos{\left(\arctan{\left(\frac{\Gamma}{\tilde{h}}\right)}\right)}+m^z \sin{\left(\arctan{\left(\frac{\Gamma}{\tilde{h}}\right)}\right)}\right) ,
    \\
    \frac{d m^x}{dt} = & 2h^\prime_0\left(m^y\cos{\left(\arctan{\left(\frac{\Gamma}{\tilde{h}}\right)}\right)}\right)+\lambda\left( 2\left(-m^z +\tanh\left(\beta h^\prime_0\right)\right)\sin{\left(\arctan{\left(\frac{\Gamma}{\tilde{h}}\right)}\right)} 
    -\langle\sigma^x\rangle\cos{(h^\prime_0t)}\cos{\left(\arctan{\left(\frac{\Gamma}{\tilde{h}}\right)}\right)} 
    \right.\nonumber\\ &\left.
    -\langle\sigma^y\rangle\sin{(h^\prime_0t)}\cos{\left(\arctan{\left(\frac{\Gamma}{\tilde{h}}\right)}\right)}\right) +2i g(h(t))\left(-m^z\cos{\left(\arctan{\left(\frac{\Gamma}{\tilde{h}}\right)}\right)} +m^x\sin{\left(\arctan{\left(\frac{\Gamma}{\tilde{h}}\right)}\right)}\right) ,
    \\
    \frac{d m^y}{dt}  = & 2 h^\prime_0(-m^x)+\lambda\left(-\langle\sigma^y\rangle\cos{(h^\prime_0t)}+\langle\sigma^x\rangle\sin{(h^\prime_0 t)}\right)+2 g(h(t))\text{Tr}_s(\rho_s) .
\end{align} 
\end{subequations}
Following the notation in Banerjee \textit{et. al.}\cite{Banerjee_PRE_1995}, we distinguish $m^\mu=\Tr\left(\rho(t)\sigma^\mu\right) = \Tr_s\left(\rho_s\sigma^\mu\right)$ and the expectation value in the interaction picture $\langle\sigma^\mu\rangle=\Tr_s\left(\tilde{\rho}_s\sigma^\mu\right)=\Tr_s\left(e^{i H_s t }\rho_s(t)e^{-iH_s t}\sigma^\mu\right)$. In the small $\Gamma$ limit we obtain eq.~\eqref{eqn:MF-LLG} in the main text. 

The expectation value is determined similarly 
\begin{subequations}\begin{align}
\langle\sigma^z\rangle = m^z\cos{\left(\arctan{\left(\frac{\Gamma}{\tilde{h}}\right)}\right)} + m^x\sin{\left(\arctan{\left(\frac{\Gamma}{\tilde{h}}\right)}\right)}
\\
\langle\sigma^x\rangle = m^x\cos{(h^\prime_0 t)}\cos{\left(\arctan{\left(\frac{\Gamma}{\tilde{h}}\right)}\right)} + m^y\sin{(h^\prime_0 t)}
\\
\langle\sigma^y\rangle = m^y\cos{(h^\prime_0 t)} -m^z\sin{(h^\prime_0 t)} \sin{\left(\arctan{\left(\frac{\Gamma}{\tilde{h}}\right)}\right)} 
\end{align}
\label{eqn:MFA_expectation}
\end{subequations}

The mean field theory gives a characteristic scaling of the hysteresis loop area with respect to the strength of the transverse field. In the case of hysteresis driven by oscillating transverse field $A\propto \Gamma^\alpha$ with $\alpha=2$ \cite{Banerjee_PRE_1995}. With oscillating longitudinal field we can identify a similar scaling law. We can generalize from the a sinusoidal oscillating longitudinal field $h_z(t)=h_1\sin(\omega t)$, the hysteresis loop area is then, 
$A=\oint \vec{m}\cdot d\vec{h}$. Substituting in the magnetization
\begin{subequations}
\begin{align}
    A&=\int_0^T \vec{m}\cdot d\vec{h}(t) 
   \nonumber \\ 
    &=-\int_0^T h_z(t)\frac{\partial m^z}{\partial t}dt 
    \nonumber\\
    &= -h_1\int_0^T \sin{(\omega t)}\left(-2m^y\Gamma +\lambda\left(2\left(-m^z +\tanh\left(\beta h^\prime_0\right)\right) \frac{|\tilde{h}(t)|}{h_0} +\langle\sigma^x\rangle\cos{(h^\prime_0 t)}\frac{\tilde{h}(t)}{|\tilde{h}(t)|}\frac{\Gamma }{h_0} +\langle\sigma^y\rangle\sin{(h^\prime_0 t)}\frac{\tilde{h}(t)}{|\tilde{h}(t)|}\frac{\Gamma }{h_0} \right) 
    \right.\nonumber\\  & 
    \qquad\qquad\qquad \left. +2ig(h(t))\left( m^x \frac{|\tilde{h}(t)|}{h_0}+ m^z \frac{\tilde{h}(t)}{|\tilde{h}(t)|}\frac{\Gamma }{h_0} \right) \right)dt
    \label{eqn:AreaScaling_Lambda}
    \\
    &\approx 2h_1\Gamma\int_0^T \sin{(\omega t)} m^y +  m^x |\tilde{h}(t)| \frac{h_1 \omega\sin{(\omega t)}\cos(\omega t)}{\left(\Gamma^2+\tilde{h}^2(t)\right)^{\frac{3}{2}}} d t +h_1 \Gamma^2 \int_0^T m^z \frac{\tilde{h}(t)}{|\tilde{h}(t)|} \frac{h_1 \omega\sin{(\omega t)}\cos(\omega t)}{\left(\Gamma^2+\tilde{h}^2(t)\right)^{\frac{3}{2}}}  d t 
    \label{eqn:AreaScaling}
\end{align}
\end{subequations}
wherein in eq.~\eqref{eqn:AreaScaling} we treat the heat bath fluctuations, $\lambda$, as small. In most longitudinal hysteresis experiments the longitudinal field applied is larger than the transverse field,  $h_\text{max}>\Gamma$. Examining eqs.~\eqref{eqn:AreaScaling_Lambda} and \eqref{eqn:AreaScaling},  we can see that to leading order $A\propto \Gamma^2$. We fit the area  as $A=a\Gamma^\alpha+c$, as discussed in the main text we see $\alpha\sim 2$. 
\begin{figure}[hbt!]
    \centering
    \includegraphics[width=0.4\linewidth]{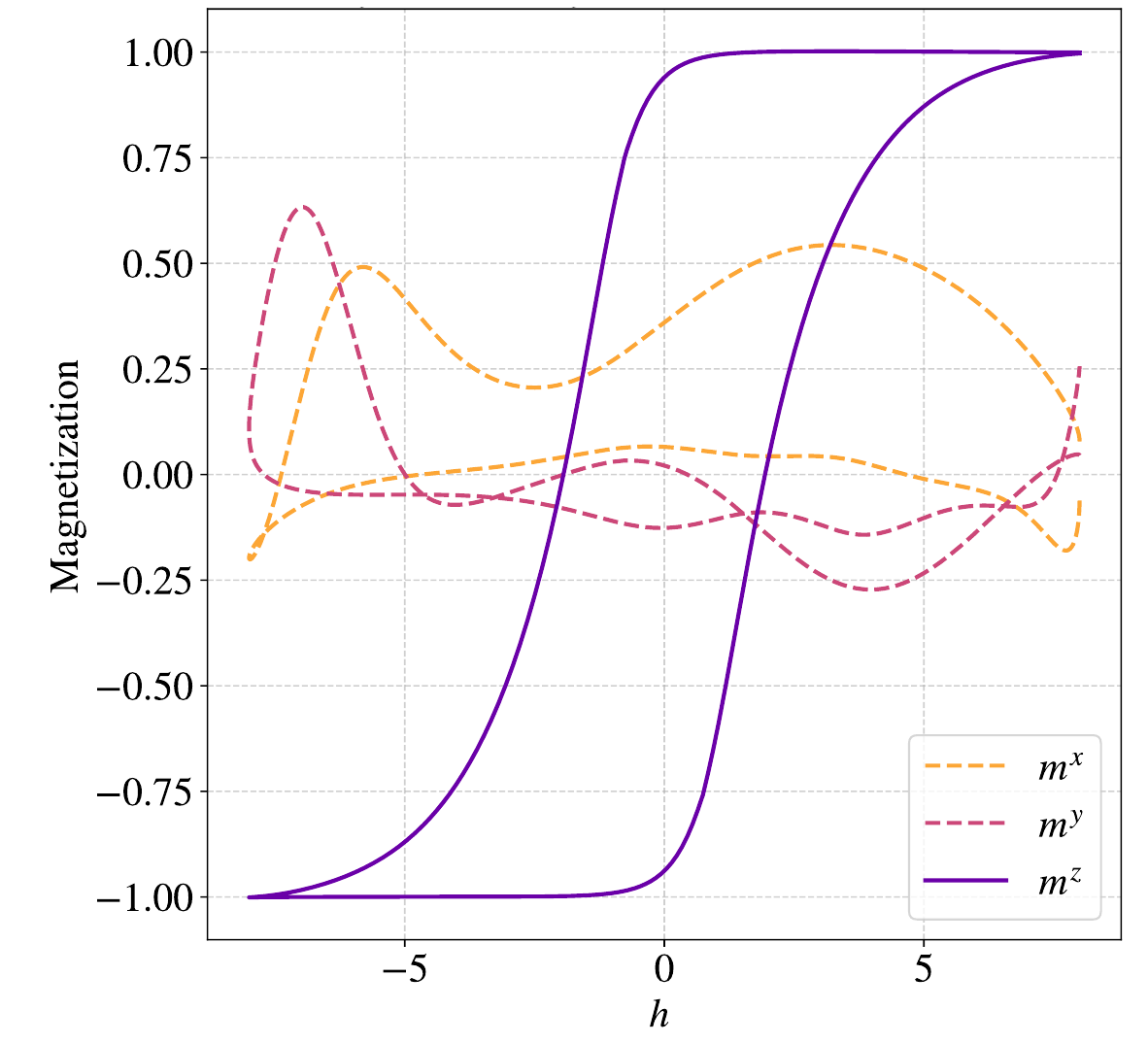}
    \caption{ Three components of magnetization determined by the mean field approximation, eq.~\eqref{eqn:MF-LLG}, are plotted versus applied field during a hysteresis protocol under a sinusoidal longitudinal field.  }
    \label{fig:MFA}
\end{figure}

In Fig.\ref{fig:MFA} we plot hysteresis determined by the mean field dynamics, eq.~\eqref{eqn:MF-LLG} under a sinusoidally varying field with hysteresis plots corresponding to $t\in [ \frac{\pi}{2},\frac{5\pi }{2} ]$ with ramp up to maximum applied field at for times $t\in [0,\frac{\pi}{2}]$.

\bibliography{references}

\clearpage
\newpage

\newcounter{supsection}
\renewcommand{\thesupsection}{S\arabic{supsection}}

\newcommand{\suppsection}[1]{
  \refstepcounter{supsection}
  \section*{Supplementary Information \thesupsection. #1}
  \addcontentsline{toc}{section}{Supplementary Information \thesupsection. #1}
}

\renewcommand{\thefigure}{S\arabic{figure}}
\setcounter{figure}{0}
\renewcommand{\thetable}{S\arabic{table}}
\setcounter{table}{0}
\renewcommand{\theequation}{S\arabic{equation}}
\setcounter{equation}{0}
\suppsection{Weak \texorpdfstring{$\Gamma$}{} Hysteresis Simulations}
\label{section:appendix_Weak_Gamma_hysteresis_simulations}

\begin{figure}[ht!]
    \centering
    \includegraphics[width=0.45\linewidth]{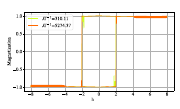}
    \caption{Hysteresis loops of four ferromagnetically coupled qubits in a one-dimensional ring (with periodic boundary conditions) are simulated with a total anneal time of $5E4 \mu s$, step size of  $\Delta t = 10 ps$.}
    \label{fig:WeakGamma_hyst}
\end{figure}
In Fig.~\ref{fig:WeakGamma_hyst} the hysteresis loop is shown determined by similar implementation as in Fig.~\ref{fig:LZ-hyst}. Simulations implemented  a combined first-order piecewise-constant propagator and semiclassical dynamics, as described in the main text. Hysteresis loops for $\Spause  > 0.5$ are simulated with   parameters $T_\text{total}=5E4 \mu s$, $\Delta t = 10 ps$ and $N =4$, for both $\Spause\in\{0.6,0.7\}$ $(J\Gamma^{-1}\in\{310.11,5274.37\})$. These simulation parameters produced closed hysteresis loops with full magnetization reversal, similar to as seen in experimental data in Fig.~\ref{fig:Hysteresis}. Here, full magnetization reversal are observed in simulations with small time steps compared with simulations shown in the main text. For these weak transverse field simulation the hysteresis loops do not qualitatively look similar to the experimental data. This would suggest that when the transverse field is sufficiently weak the independent level crossing approximation, and low order Magnus expansion, does not adequately capture the dynamics magnetization reversal process. In the case of simulations, the avoided crossing is small thus the independent level crossing approximation is less valid and diabatic transitions have comparably larger probability over a wider range of longitudinal field values, as diabatic transitions occur more broadly in $h(t)$. In addition, as $B(s)\gg A(s)$ the eigenvalues diverge rapidly away from the avoided crossings, thus simulation requires small timesteps.  In experiments, fluctuations can assist magnetization reversal as the transverse field is small compared to the hardware device temperature. Thus within this weak transverse field limit we don't expect the simulations to capture the magnetization reversal seen in experiment. 

\suppsection{Defect Density as a Function of Longitudinal Field Ramp Rate}
\label{section:appendix_defect_density_as_a_function_of_longitudinal_field_ramp_rate}

\begin{figure}[ht!]
    \centering
    \includegraphics[width=\linewidth]{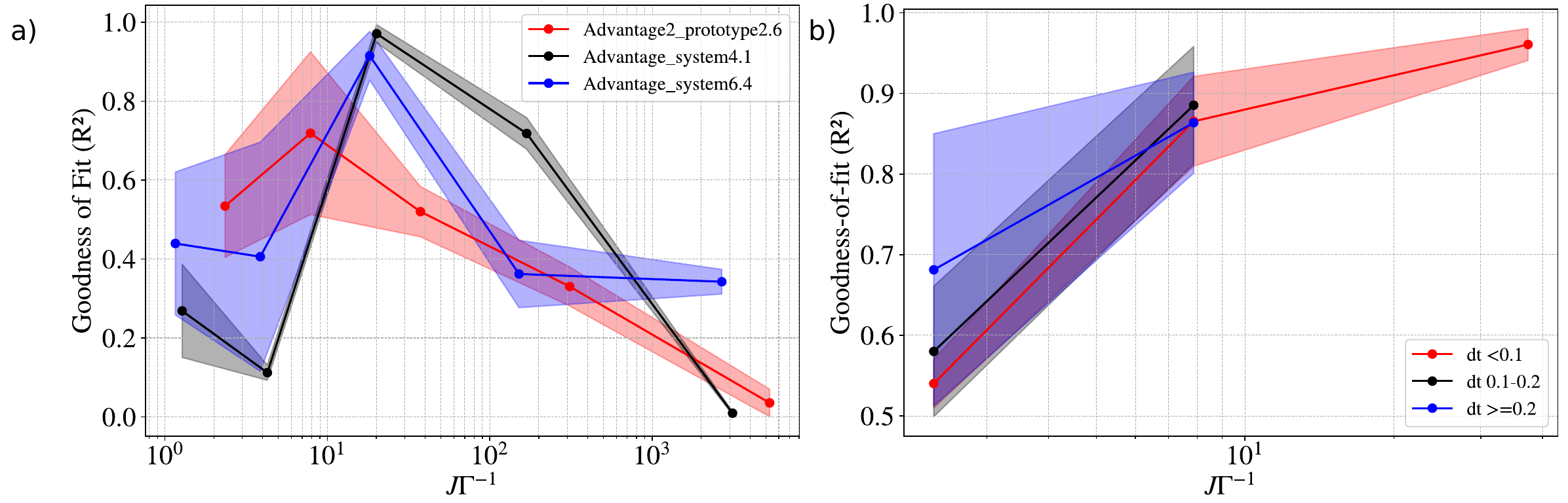}
    \caption{ Linear fit $R^2$ value for D-Wave QPU experimental (a) and simulated (b) hysteresis curves, as a function of $J\Gamma^{-1}$. Shaded region indicates standard deviation. (a) Different hardware devices shown in different colors, indicating device specific performance. (b) Different step sizes shown in different colors, showing similar performance across step sizes. For weak $\Gamma$ and step sizes $dt\geq 0.1$, there are insufficient data points to obtain a linear fit.  }
    \label{fig:R2s}
\end{figure}

Linear fit $R^2$ values for all experimental and simulated hysteresis curves are shown in Fig.~\ref{fig:R2s}. The shaded region indicates standard deviation over annealing time and system sizes. $R^2$ values are for the linear fitting of $\ln(1-n_d)$ inverse to ramp rate, $\dot{h}$, a proxy that suggests domain wall densities are governed by an independently level crossing approximation in the LZ-model. 
We see that each device has a regime where the dynamics and domain wall density does scale inversely with the longitudinal field ramp rate. We note at very small $\Gamma$ the fit is bad and the dynamics do not seem governed strictly by the LZ mechanism. In this regime, the avoided crossing gap is small and domain wall nucleation is suppressed, thus the magnetization reversal may be governed by relaxation mechanisms from excited states. In (b), simulations are shown for different simulation step size. We see the defect density is not strongly influenced by step size, but instead on sweep time and sweep rate. There are insufficient data points for $J\Gamma^{-1}>10$ for large step sizes as simulation are sensitive to step size in the weak field limit and do not form closed hysteresis loops.
We observe that under strong transverse fields, the linear fit is worse in simulation, thus scaling does not as closely resemble scaling expected by the LZ-model. This is a result of the anneal time sensitivity in the strong transverse field regime, as well as the domain wall motion and probability of domain wall annihilation in small systems. 

The summarized experimental data that is shown in Fig.~\ref{fig:R2s} includes all 1D ferromagnetic Ising model hysteresis cycle simulations for the spin systems with $\geq 100$ qubits, at $s$ pause values $\in \{0.3, 0.4, 0.5, 0.6, 0.7\}$, using all 4 annealing times, including the $11.2\mu s$ AFM gauge transformed simulations -- however, we have excluded the data from $s=0.3$ with $11.2\mu s$ simulation times run on \texttt{Advantage\_system4.1} and \texttt{Advantage\_system6.4} due to anomalous, and hardware specific, results which we will investigate in future work. This data (Fig.~\ref{fig:R2s}) does not include the $4$ spin ferromagnetic ring data.

\suppsection{Eigenvalues Crossings via Exact Diagonalization}
\label{section:appendix_eigenvalue_crossing}
\begin{figure}[ht]
    \centering
    \includegraphics[width=.6\linewidth]{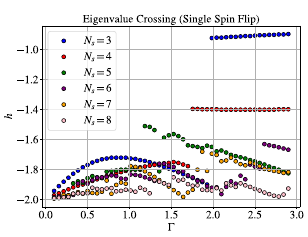}
    \caption{ Scattered points correspond to the longitudinal field values where the avoided  crossing to first excited state (spin flip state) is minimized during a backward sweep, as a function of system sizes and transverse field strength, $\Gamma$. }
    \label{fig:Crossings}
\end{figure}

The longitudinal field values corresponding to the minimum energy gap at the avoided crossing between an spin-flip adjacent eigenstates. From an initial all aligned state there is a single spin flipped state that determines the longitudinal field where we expect domains wall to nucleate. In Fig.~\ref{fig:Crossings}, this field is plotted versus the transverse field for systems of size $3\leq N \leq 8$. This was determined by numerical diagonalization for matrices of the type in eq.~\eqref{eqn:LZ_Ham}, e.g, eq.~\eqref{eqn:LZ-example}. A longitudinal field value $h=-2J$ corresponds to the exact value at which spin flip would reduce the energy from the aligned state during a backward field sweep. We can see that the avoided crossing occurs near $|h|=2J$, except for large values of $\Gamma$ where level splitting begins to interfere across multiple levels. We see that an increase number of spins does not change the general field value of the minimum avoided crossing, but it does introduce local variation such that the longitudinal field value does not vary smoothly with the transverse field.  
Discontinuities correspond to changes in the splitting of eigenvalues as a function of $\Gamma$. This is a result of more interaction between levels and more paths through the diabatic levels. For $\Gamma$ values investigated in this work, we expect the longitudinal field corresponding to the avoided crossing with the first spin flip state in one-dimensional systems to be $|h|\sim 2J$.

\suppsection{Derivation of Landau-Zener Model Transition Probability}
\label{section:appendix_derivation_of_LZ_transition_probability}
For a two-level system with Hamiltonian   $\mathcal{H}(t)=\begin{pmatrix}
    b_1t  & a \\ a &b_2t
\end{pmatrix}$ the eigenvalues are $E_\pm=\frac{(b_1+b_2)t}{2}\pm \sqrt{\frac{((b_1-b_2)t)^2}{4}+a^2}$, which  are degenerate at an imaginary time $t=\pm \frac{i2a}{b_2-b_1}$.
The transition probability, $P$, between basis states (in the diabatic basis)  can be determined by the path integral
\begin{align}
    \bra{\uparrow} \mathcal{T}e^{-i\int^\infty_{-\infty}\mathcal{H}(\tau)d\tau}\ket{\uparrow} &= \bra{\psi_+(\infty)}e^{-i\int^\infty_{-\infty}\mathcal{H}(\tau)d\tau}\ket{\psi_-(-\infty)}  
    \\
    P &=\left| \bra{\psi_+(\infty)} e^{-i\int_\mathcal{C} E_{\pm }(\tau)d\tau}\ket{\psi_-(-\infty)}  \right|^2, 
\end{align} 
with contour integral in complex time, thus transitions can be understood through open system effects. Integrating around a complex-time degeneracy gives a non-zero phase, e.g., at $t = \frac{i2a}{b_2 - b_1}$, the contour integral contribution around the branch point 
yields $p= e^{-\frac{2\pi a^2}{\hbar\vert b_2-b_1\vert}}$.

\suppsection{Truncated Density Matrix Renormalization Group}
\label{section:appendix_truncated_DMRG}

\begin{figure}[ht]
    \centering
    \includegraphics[width=.6\linewidth]{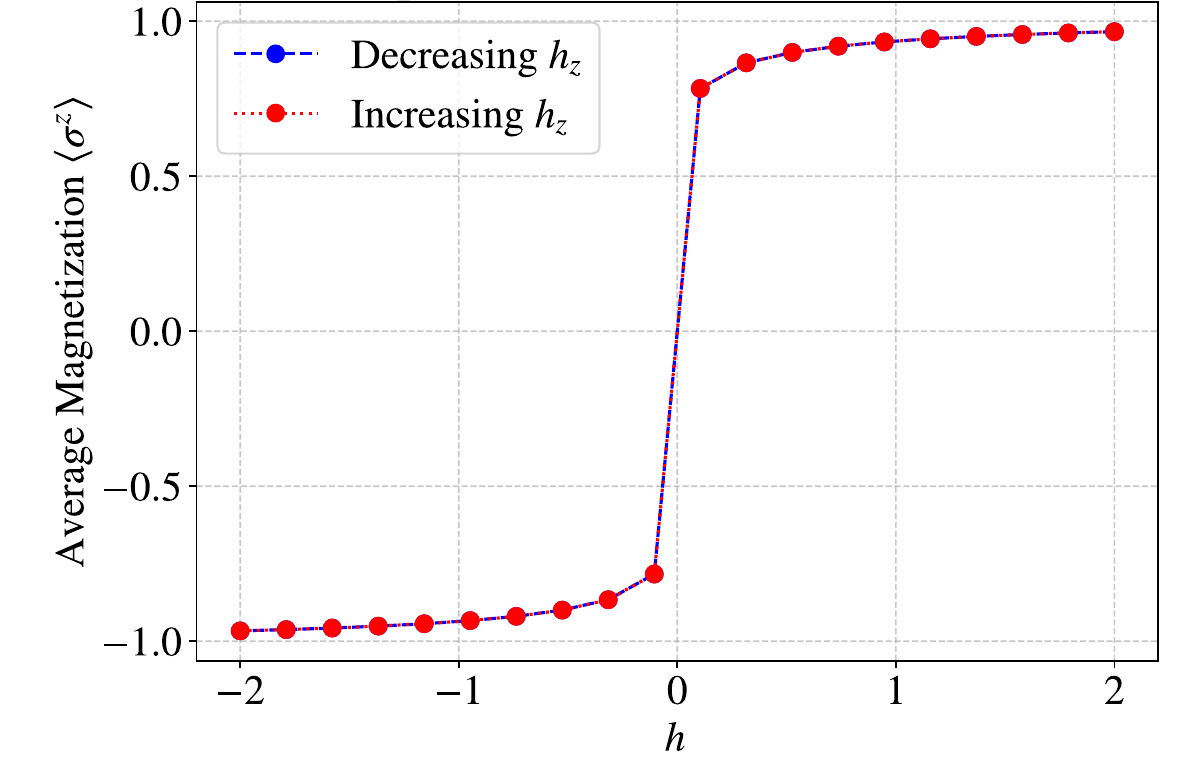}
    \caption{ Magnetization versus longitudinal field during a hysteresis protocol implemented with DMRG. The backward sweep is shown in blue (decreasing longitudinal field) and the forward sweep is shown in red (increasing longitudinal field).}
    \label{fig:dmrg}
\end{figure}

We simulated the magnetic hysteresis with matrix product states using a density matrix renormalization group approach. To introduce quench like dynamics we truncated the density matrix renormalization group (DMRG) equilibration. We observed that the system undergoes magnetization reversal without hysteresis. 

We simulated magnetic hysteresis in the transverse field Ising model (TFIM) using matrix product states (MPS) and a density matrix renormalization group (DMRG) approach, as implemented in the TeNPy library~\cite{Hauschild_2018}. The TFIM Hamiltonian includes both a transverse field  which induces quantum fluctuations, and a tunable longitudinal field $h(t)$ , which biases spin alignment along the $\hat{z}$-direction. We constructed a custom MPS model with explicitly defined Pauli operators and simulated a one-dimensional open chain of spin-$\frac{1}{2}$ sites. The DMRG method was used to variationally minimize the energy within the MPS ansatz, yielding approximate ground states at each value of the external longitudinal field.

To mimic quench-like dynamics and capture possible non-equilibrium features, we truncated the DMRG procedure by limiting the number of sweeps to a single iteration per field step. This under-converged DMRG prevents the system from fully relaxing to its ground state at each step, creating a computational analog of slow or finite-rate driving. We swept the longitudinal field in a backward and then forward sweep, recording the average magnetization at each point. This protocol is designed to probe for hysteresis-like memory effects, where the spin response would depend on the direction of the field sweep.

In the resulting simulations shown in Fig.~\ref{fig:dmrg}, the magnetization exhibited sharp transitions around $h=0$, consistent with a collective reorientation of spins and the response was  identical for both forward and reverse sweeps, indicating an absence of hysteresis under these conditions. This result underscores a key feature of DMRG, even when convergence is limited, the algorithm still tends to track low-energy (near-equilibrium) states rather than explore diabatic evolution or excited-state pathways. Local transitions and excited sates are not captured by static or weakly non-equilibrium DMRG. Thus, while DMRG captures the magnetization reversal, it did not demonstrate hysteresis.

\suppsection{Simulation Without Relaxation via Semiclassical Motion of Domain Walls}
\label{section:appendix_simulation_without_relaxation_semiclassical_motion_of_domain_walls}
\begin{figure}[ht]
    \centering
    \includegraphics[width=0.45\linewidth]{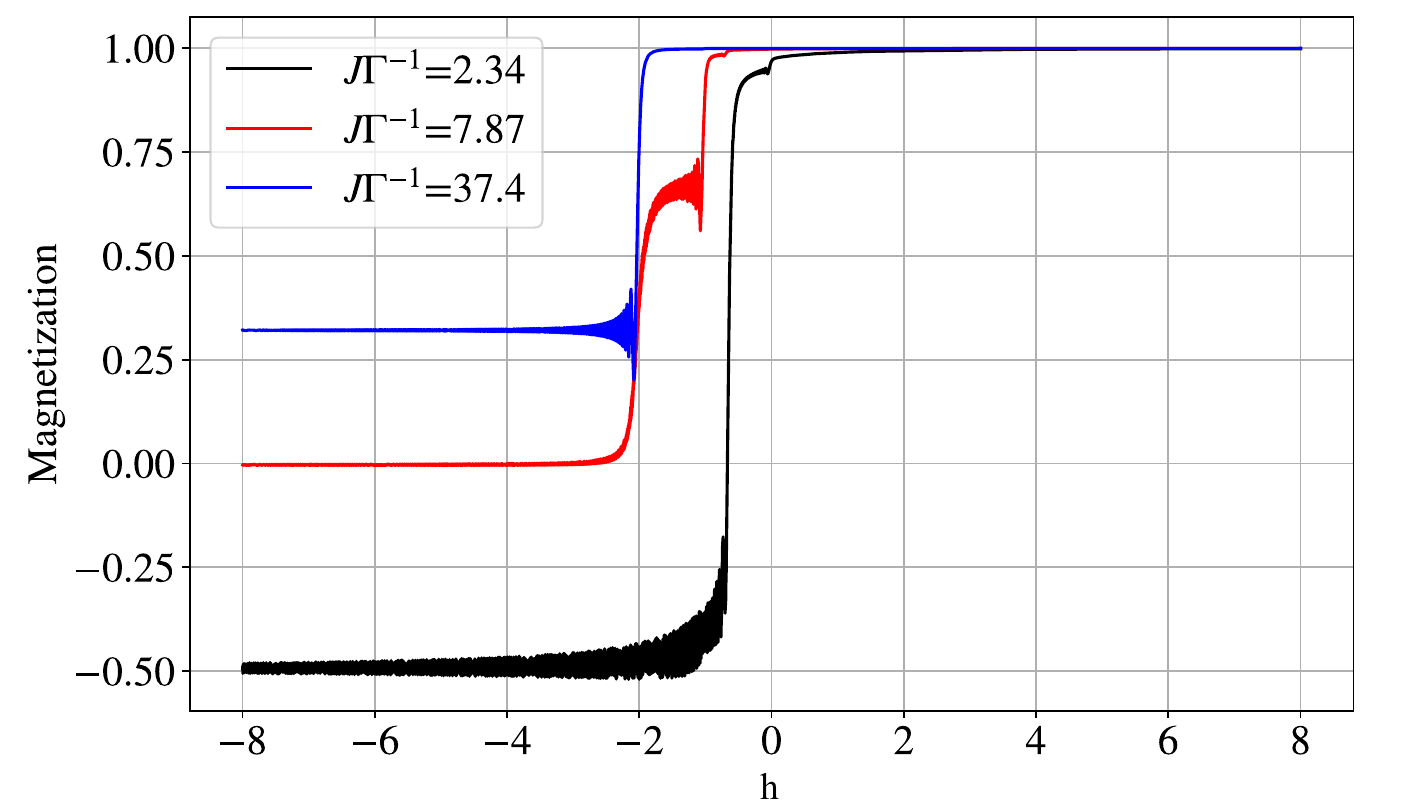}
    \caption{Magnetization reversal modeled with successive first-order piecewise-constant propagation but without semiclassical evolution of the domain walls. During the backward sweep the system does not fully reverse magnetization, though the final magnetization does depend on the strength of the transverse field. }
    \label{fig:LZ_transition}
\end{figure}
Figure~\ref{fig:LZ_transition} plots magnetization during a backwards magnetic field sweep, representing one half of the hysteresis sweep. This system is simulated without the semiclassical evolution of the domain walls, and thus represents the system dynamics driven entirely by first-order piecewise-constant propagation applied successively. It is apparent that the system does undergo some magnetization reversal, and the initiation of magnetization reversal is similar to that observed in the experiments and the numerical simulations that do incorporate domain wall kinetics. The value of applied field at which magnetization reversal begins is similar to experimental and numerical simulation data in the main text. Further, the initiation of magnetization reversals has similar characteristics including a combination of Barkhausen jumps as well as rapid oscillation of the magnetization once domain walls are nucleated that corresponds to some magnetization fluctuations. Notably the system does not fully reverse it's magnetization and saturate aligned with the applied field. Without the semiclassical motion of domain walls the low-order Magnus expansion is unable to grow domains and fully reverse magnetization.

\suppsection{Interaction Picture}
\label{section:appendix_interaction_picture}
We can study the dynamics fo the TFIM driven under time varying longitudinal filed in the interaction picture. 
This serves as a comparison for systems which display hysteresis like behavior under equilibrium driving.
Via the Jordan-Wigner transformation it is well known that there is a duality between the high and low transverse field regimes, at low field the system is ferromagnetically ordered and the transverse field acts to introduce spin flips. At high transverse field the system is paramagnetic, consisting of domain walls. As hysteresis involves varying the longitudinal field neither of these approximations remains valid throughout the entire sweep and the standard Jordan-Wigner transformation is not applicable, this is discussed below describing time periodic states. As such, we have to employ other methods to study the state evolution under time varying fields.

Here we use the interaction picture to account for the noncommuting terms of the Hamiltonian and study the time evolution of a initial state, and operators. Conventionally we would not expect hysteresis with uncoupled spins, $J=0$. Experimental data has indicated that as we reduce the strength of the $z-$basis Hamiltonian, $B(s)\rightarrow 0$, there is no hysteresis~\cite{Pelofske_arxiv_2025}. This suggest we should  treat the coupling term as the interaction term. 
We treat the interaction terms as the Ising coupling, $J\sigma^z\sigma^z$. Without spin-spin coupling the ground-state, $\ket{GS}$, of the non-interacting system, $H_0$, under longitudinal and transverse fields will oscillate in time,
    \begin{align}
        \ket{GS} &= \bigotimes^N(a(t) \ket{0}+b(t) \ket{1}) , \quad \frac{h(t)}{\Gamma} = \frac{a(t)^2-b(t)^2}{2a(t)b(t)} 
    \label{eqn:InteractionGS}
    \end{align}
with the second equation being constraints on the coefficients.


Incorporating the interaction term $J\neq 0$ gives us the full TFIM Hamiltonian. Often in the interaction picture  the interaction term is chosen to be  the time varying field; as the $\hat{z}$-basis Hamiltonian is well behaved and serves as suitable non-interacting Hamiltonian. 
Similarly, the TFIM without the longitudinal field could be the non-interacting Hamiltonian, but the TFIM without longitudinal field is a little less well behaved. As such our choice is, 
\begin{subequations}\begin{align}
    \mathcal{H} &= H_0+H_1 \nonumber
    \\
    H_0 &= A(s)\Gamma\sigma^x+B(s)h(t)\sigma^z 
    \\
    H_1 &= B(s)J_{ij}\sigma^z_i\sigma^z_j .
\end{align}
\end{subequations}
During the time evolution of the longitudinal field we hold $s$ constant, as such we absorb the $A(s)$ and $B(s)$ into the field terms and the coupling constant. We now define the operators in the interaction picture as the interaction terms:
\begin{align}
    H_1^\text{int} &=e^{i\int H_0(t^\prime) dt^\prime}H_1 e^{-i\int H_0(t^\prime)dt^\prime}
\end{align}
\begin{align}
\sigma^{z,\text{int}}(t)&=\hat{\mathbf{T}}e^{i(\Gamma\sigma^x \Delta t+\int^t h(t^\prime)dt^\prime\sigma^z)}\sigma^z(0)\hat{\mathbf{T}}e^{-i(\Gamma\sigma^x \Delta t+\int^t h(t^\prime)dt^\prime\sigma^z)}
\nonumber\\
&=\sigma^z \left[\cos{(\theta(t))}+\cos{(a(t))}\left[1-\cos{(\theta(t))}\right]\right]+\sigma^y\sin{(\theta(t))}\sin{(a(t))} .
\end{align}
Here we have $a(t)=2\sqrt{(\Gamma\Delta t)^2+(\int^t h(t^\prime)dt^\prime)^2}$ and $\theta(t)=\arctan\left(\frac{\Gamma\Delta t}{\int^t h(t^\prime)dt^\prime}\right)$. Thus the expectation value  $\langle\sigma^{z,\text{int}}\rangle$ oscillates in time as a different rate then the applied field, giving rise to an apparent hysteresis.

\begin{figure}[h!]
    \centering
    \includegraphics[width=0.8\linewidth]{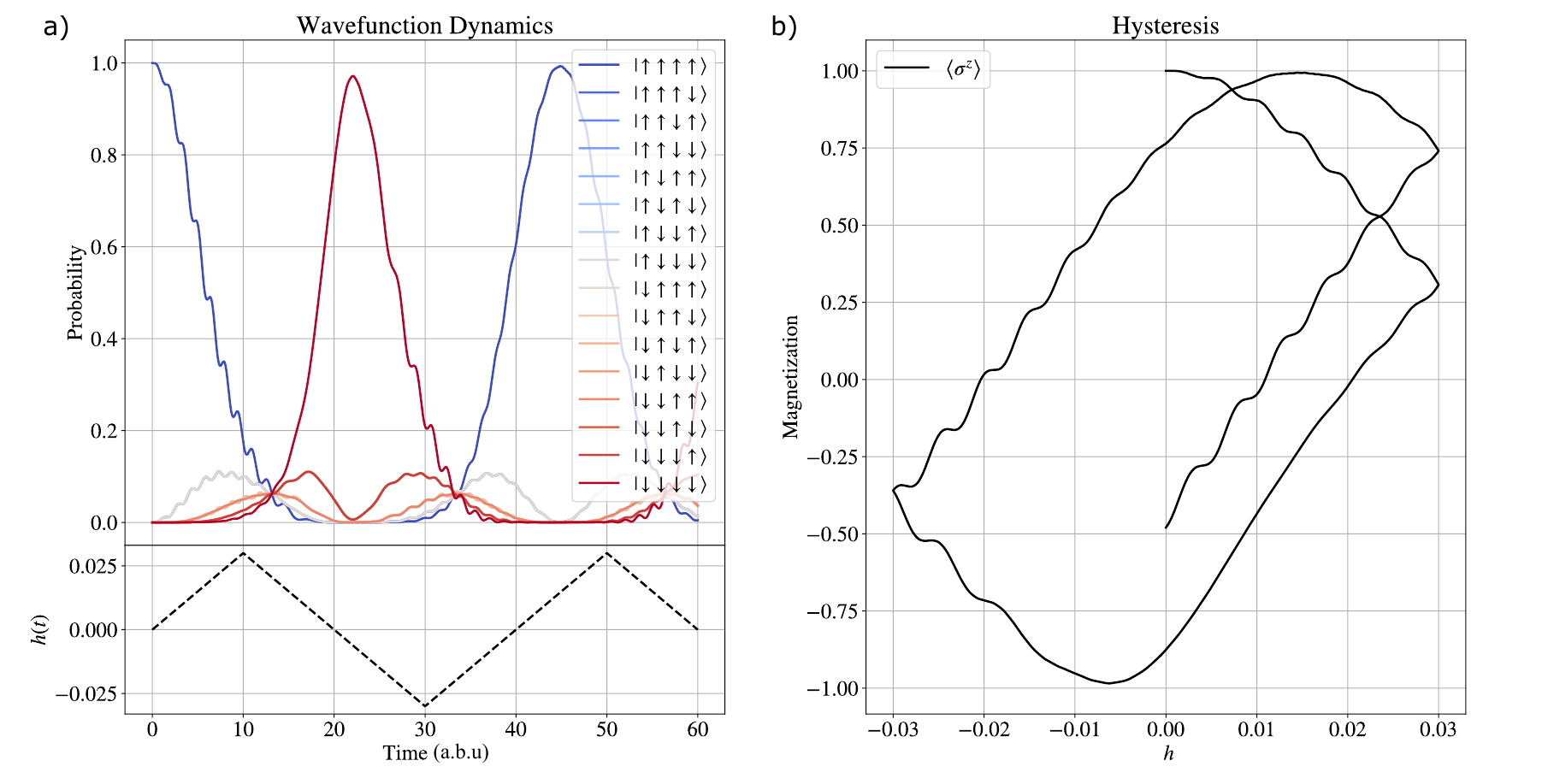}
    \caption{(a) Density of states during hysteresis protocol determined by the interaction picture model, (below inset) the corresponding applied field is shown in black dashed line. (b) Magnetization during hysteresis protocol determined by the interaction picture. The system is initialized in a fully magnetized state ($m^z=1$) at a longitudinal field of $h=0$. Density of states plots (a) correspond to projecting the density of states and measurements in the interaction picture. Magnetization (b) is measured for the corresponding density of states $\langle\sigma^{z,\text{int}}\rangle$.}
    \label{fig:InteractionPic_StateEvolution}
\end{figure}

We can investigate the time evolution of state, in the interaction picture as
\begin{align}
   \psi_I(t)=\hat{\mathbf{T}}e^{i(\Gamma\sigma^xt+\int_0^t h(t^\prime)dt^\prime\sigma^z)}\psi_s(t) ,
\end{align}
with $\psi_s(t)$ a basis in the Schroding picture. 
Similarly, the density of state in the interaction picture can be studied. We can set an initial state at initial time $t_0=0$ and study the evolution of the density of states in terms of the non-interacting Hamiltonian,
 \begin{align}
     \rho_I(t_0) &=\rho_s(t_0)
     \\
     \rho_I(t) &=\hat{\mathbf{T}}e^{i(\Gamma\Delta t\sigma^x+\int_0^th(t^\prime)dt^\prime\sigma^z)}\rho_s(t)e^{-i(\Gamma\Delta t\sigma^x+\int_0^th(t^\prime)dt^\prime\sigma^z)}
 \end{align}
Now, we can observe the evolution of the basis of states and perform measurements of magnetization, $\langle\sigma^{z,\text{}int}\rangle$, in the interaction picture.  
In Fig.~\ref{fig:InteractionPic_StateEvolution}, we plot the evolution of states and measured magnetization during hysteresis sweeps. We study the evolution in terms of the $\hat{z}$-basis states, and initialize the system at $t=0$ in an all spin-up state. 

The simulation consists of four qubits, it is worth noticing the magnetization takes multiple states during the sweep beyond four distinct magnetization values, as could be expected from classical system of spins, each with two orientations.

Hysteresis is due to the phase accumulation as we sweep $h(t)$. 
This is distinct from hysteresis observed in experiments, in particular the shape of the hysteresis and the lack of fluctuations, the lack of non-monotonicity, and saturation at large applied field values. Instead, in the interaction picture the system oscillates between  magnetized states  while we sweep the applied field, these competing timescales produce dynamical hysteresis like behavior. 

\suppsection{Periodic states}
\label{section:appendix_Floquet_states}

Here, we briefly review time periodic states so as to demonstrate the challenge of incorporating longitudinal fields in time periodic TFIM. This is motivated by previous works that have identified exact solutions for a time-dependent TFIM~\cite{Suzuki_Springer_2013,Shukla_EDP_2020}. 

We consider the coherent dynamics of a one-dimensional quantum many-body system described by the transverse-field Ising model (TFIM) with open boundary conditions, defined by the Hamiltonian:
\begin{equation}
\mathcal{H}(t) = -J\sum_{i=1}^{L-1}\sigma_i^z \sigma_{i+1}^z - \Gamma(t)\sum_{i=1}^{L}\sigma_i^x,
\end{equation}
where $\sigma_i^{x,z}$ are Pauli matrices acting at site $i$, $J$ is the nearest-neighbor interaction strength, $\Gamma(t)$ is a time-dependent transverse field, and $L$ denotes the number of spins.

The TFIM in the absence of the longitudinal field can be exactly solved by first mapping the spin operators to spinless fermionic operators via bosonization followed by a Jordan-Wigner (JW) transformation \cite{LIEB1961407,PFEUTY197079}. Introducing fermionic operators $ c_i $ and $ c_i^\dagger $, we first represent spins through bosonic ladder operators and then map these to fermions. The Jordan-Wigner transformation explicitly reads:
\begin{equation}
\sigma_i^z = -\prod_{j<i}(1 - 2 c_j^\dagger c_j) (c_i^\dagger + c_i), \quad \sigma_i^x = 1 - 2 c_i^\dagger c_i.
\end{equation}

After applying the JW transformation, the Hamiltonian becomes quadratic in fermionic operators:
\begin{equation}
\mathcal{H}(t) = -J\sum_{i=1}^{L-1}(c_i^\dagger - c_i)(c_{i+1}^\dagger + c_{i+1}) - \Gamma(t)\sum_{i=1}^{L}(1 - 2 c_i^\dagger c_i).
\label{eqn:JW_standard}\end{equation}
This approach works also if $\Gamma$ is time-dependent, and in particular if it is periodic as we will assume soon.

\subsection{Momentum-Space Representation}

Assuming translational invariance and periodic boundary conditions for the fermions, we introduce the Fourier transforms:
\begin{equation}
c_j = \frac{1}{\sqrt{L}}\sum_k e^{i k j} c_k,\quad c_j^\dagger = \frac{1}{\sqrt{L}}\sum_k e^{-i k j} c_k^\dagger.
\end{equation}

The Hamiltonian thus decomposes into independent $2\times 2$ momentum-space sub-blocks:
\begin{equation}
\mathcal{H}(t) = \sum_{k>0}\Psi_k^\dagger \mathcal{H}_k(t)\Psi_k,
\end{equation}
with the Nambu-spinor notation $\Psi_k = (c_k, c_{-k}^\dagger)^T$, and the Bogoliubov-de Gennes (BdG) Hamiltonian:
\begin{equation}
\mathcal{H}_k(t) = [\Gamma(t)-J\cos k]\,\sigma^z + J\sin k\,\sigma^y.
\end{equation}

Each mode $ k $ thus represents an independent two-level system with instantaneous eigen-energies:
\begin{equation}
E_k(t) = \sqrt{[\Gamma(t)-J\cos k]^2 + (J\sin k)^2}.
\end{equation}

\subsection{Floquet Theory and Diagonalization}

When the transverse field is periodic, $\Gamma(t+\tau) = \Gamma(t)$, the dynamics of the system is best described using Floquet theory. Floquet states $\{|\Phi_\alpha(t)\rangle\}$ and quasi-energies $\{\mu_\alpha\}$ are defined through the time evolution operator over one period:
\begin{equation}
U(\tau,0) = \mathcal{T}\exp\left[-i\int_0^\tau H(t')\,dt'\right],
\end{equation}
where $\mathcal{T}$ denotes time ordering. The Floquet eigenvalue equation is then:
\begin{equation}
U(t,0)|\Phi_\alpha(t)\rangle = e^{-i\mu_\alpha t}|\Phi_\alpha(t)\rangle,
\end{equation}
with periodic Floquet modes defined as $|\Phi_\alpha(t+\tau)\rangle=|\Phi_\alpha(t)\rangle$, and for $0\leq t\leq \tau$.

Expanding the initial state in the Floquet basis, we have:
\begin{equation}
|\Psi(t)\rangle = \sum_{\alpha}R_\alpha e^{-i\mu_\alpha t}|\Phi_\alpha(t\mod \tau)\rangle,\quad R_\alpha=\langle\Phi_\alpha(0)|\Psi(0)\rangle.
\end{equation}
where $t\in [0,\infty)$.

\subsection{Nonzero longitudinal field}

Returning to eq.~\eqref{eqn:JW_standard}, we can incorporate a longitudinal field
\begin{align}
\mathcal{H}(t) = -J\sum_{i=1}^{L-1}(c_i^\dagger - c_i)(c_{i+1}^\dagger + c_{i+1}) - \Gamma(t)\sum_{i=1}^{L}(1 - 2 c_i^\dagger c_i) 
+h(t)\Pi_{j<i}(-1)^{c_j^\dagger c_j}(c_i+c_i^\dagger).
\label{eqn:JW_h(t)}\end{align}
Immediately we can see the longitudinal field introduces a nonlocal term that counts the kinks and flips the sign along the one-dimensional system. During the hysteresis protocol the longitudinal field takes large and small values, and thus the this non-local term cannot be treated perturbatively. 
\begin{align}
\mathcal{H}(t) = \sum_{k>0}\Psi_k^\dagger \mathcal{H}_k(t)\Psi_k 
+h(t)\Pi_{j<l}\left(1-2\left(\frac{1}{L}\sum_{\Delta k,k} e^{ij \Delta k} c_k^\dagger c_{\Delta k}\right)\right)\left(\frac{1}{\sqrt{L}}\sum_k e^{i k l} c_k+\frac{1}{\sqrt{L}}\sum_k e^{-i k l} c^\dagger_k\right).
\label{eqn:JW_h(t)_FT}
\end{align}
It is apparent we cannot diagonalize and identify coherent Floquet states via the JW transform in the presence of non-trivial time varying longitudinal field.

\begin{figure}[ht!]
    \centering
    \includegraphics[width=0.99\linewidth]{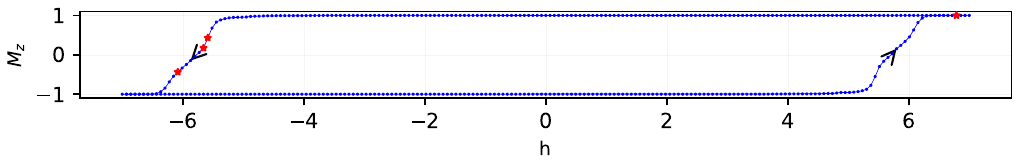}\\
    \includegraphics[width=0.24\linewidth]{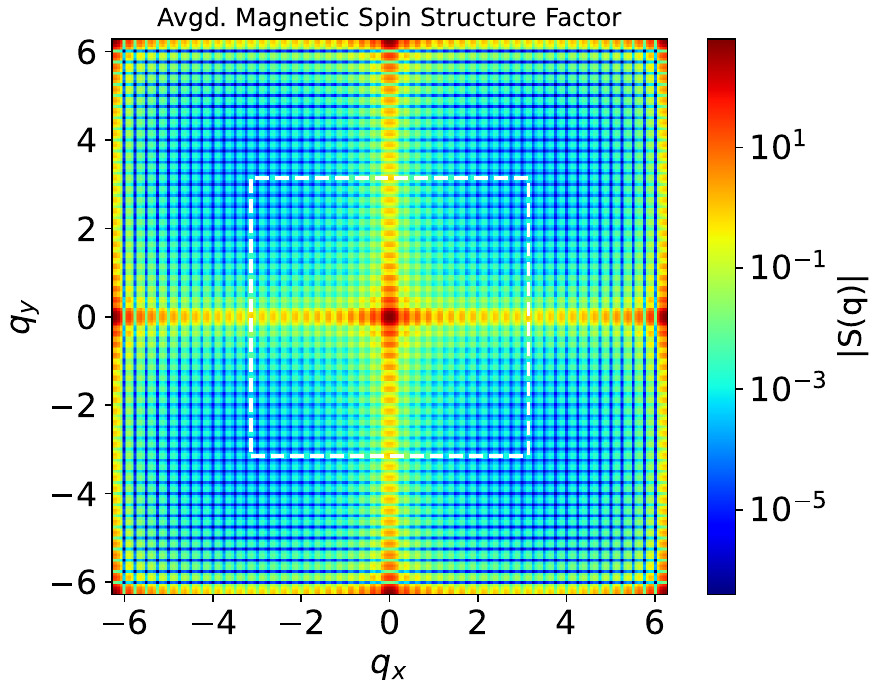}
    \includegraphics[width=0.24\linewidth]{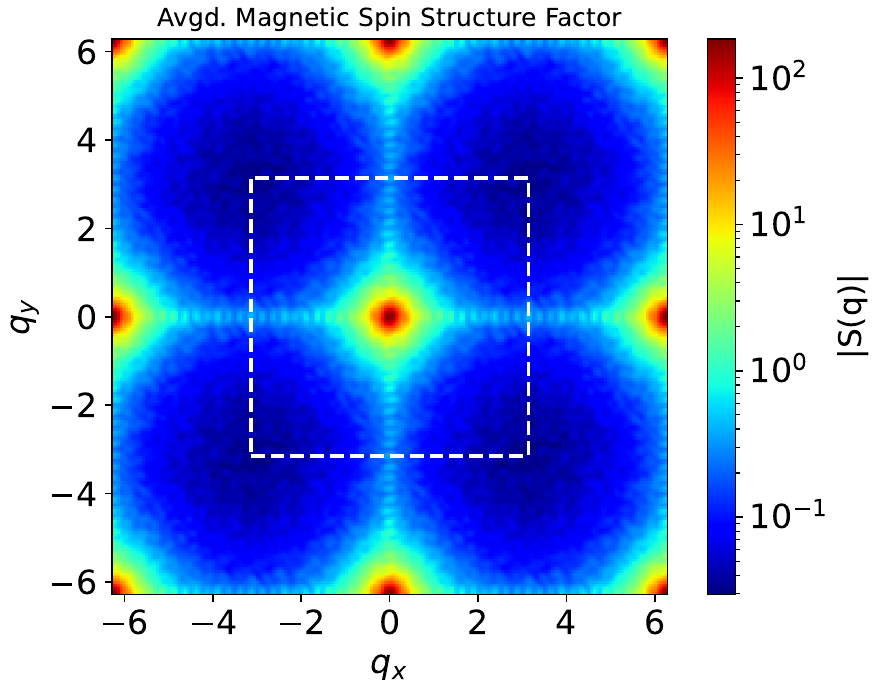}
    \includegraphics[width=0.24\linewidth]{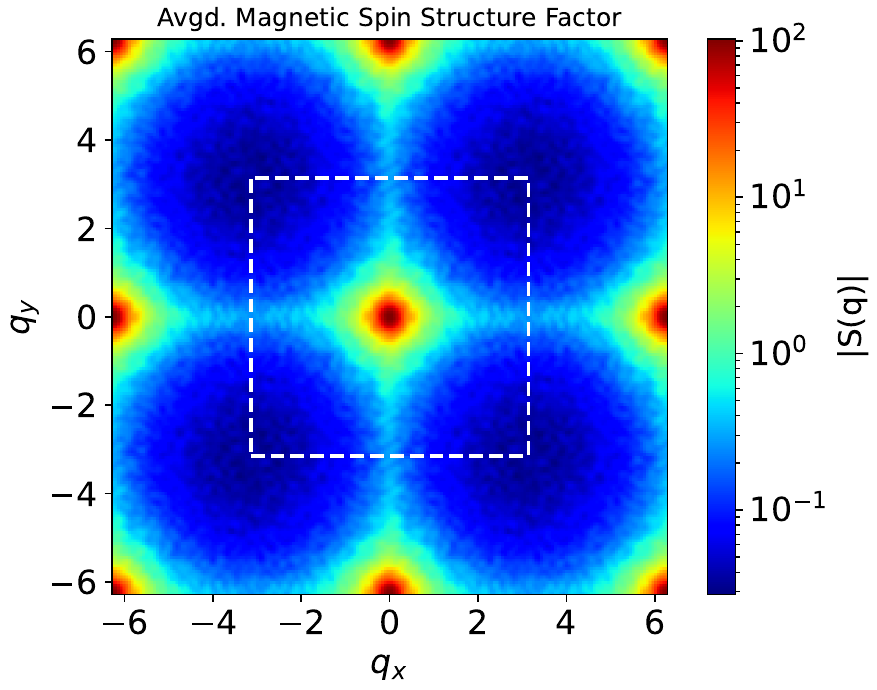}
    \includegraphics[width=0.24\linewidth]{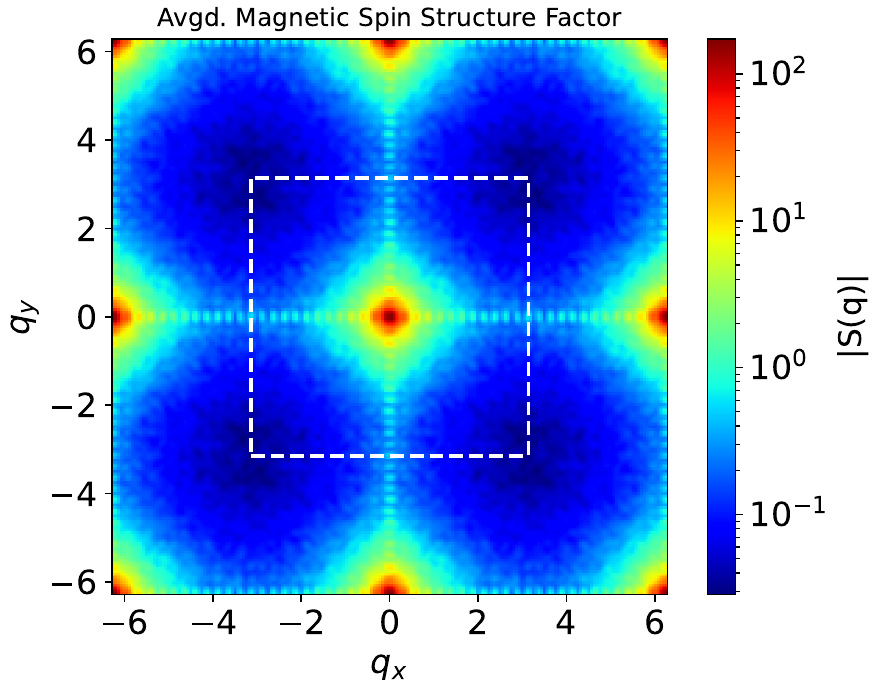}
    \vspace{0.01em}
    \tikz{\draw[dashed, thick] (0,0) -- (18,0);}
    \vspace{0.01em}
    \includegraphics[width=0.99\linewidth]{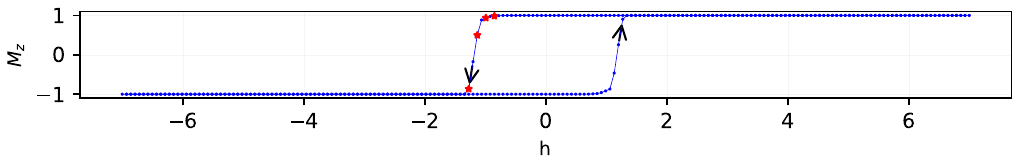}\\
    \includegraphics[width=0.24\linewidth]{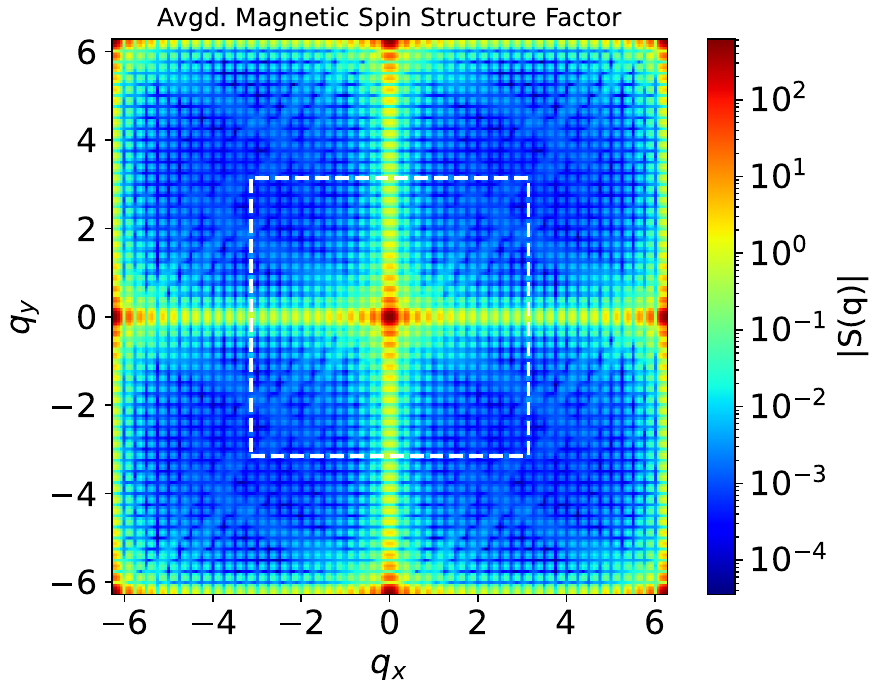}
    \includegraphics[width=0.24\linewidth]{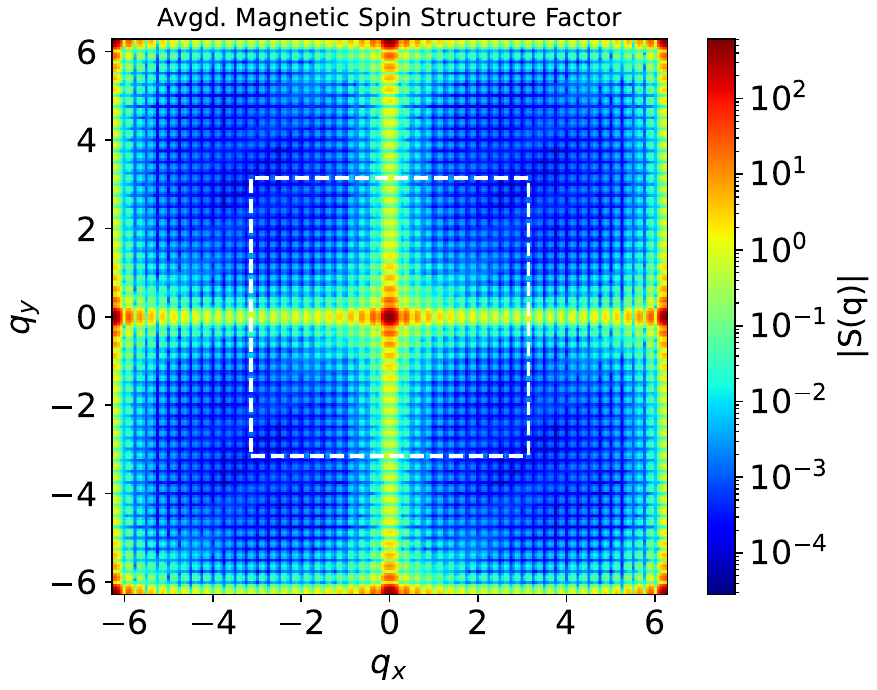}
    \includegraphics[width=0.24\linewidth]{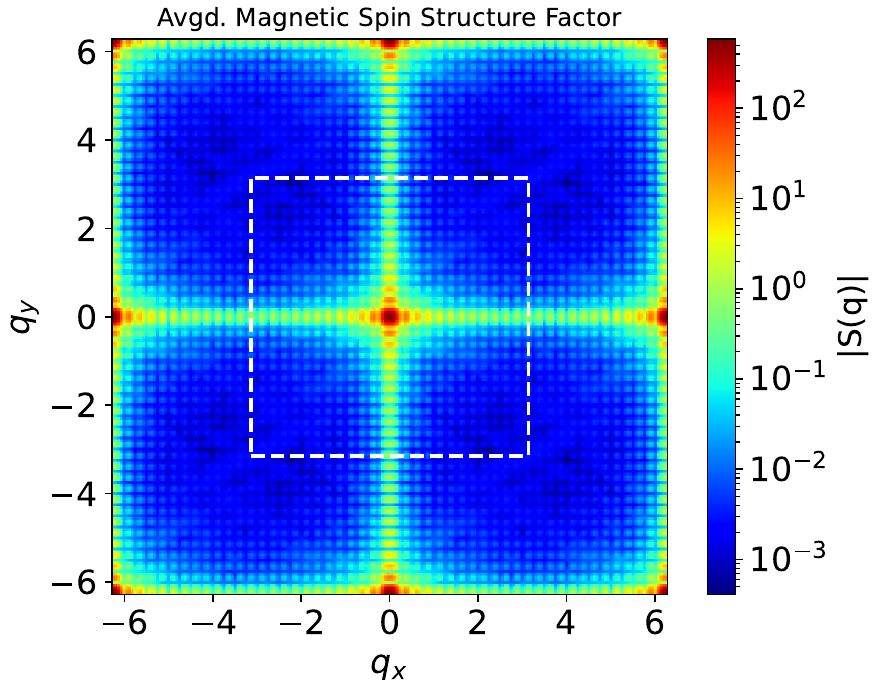}
    \includegraphics[width=0.24\linewidth]{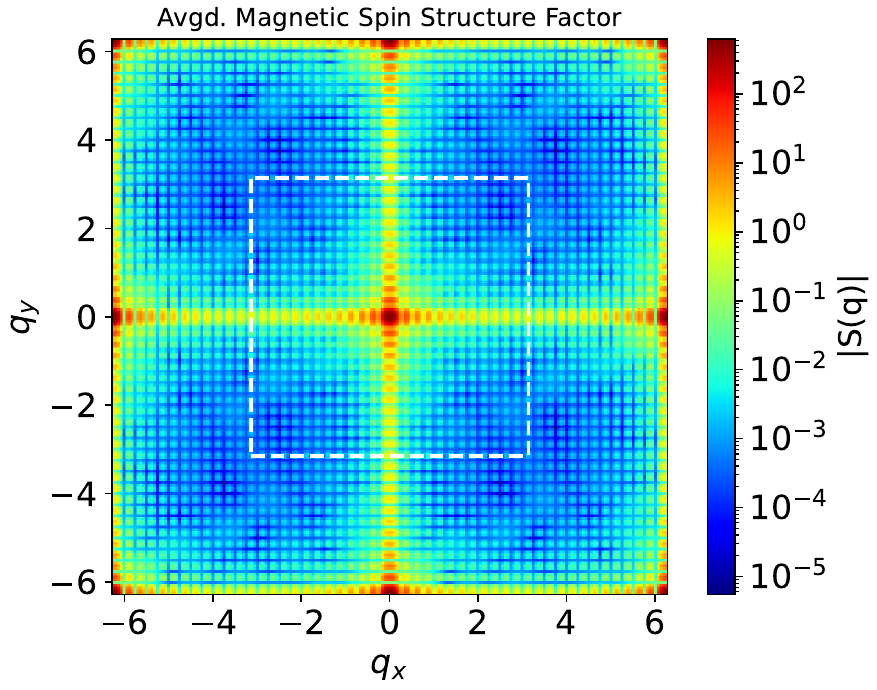}
    \caption{Magnetic spin structure factor of two-dimensional square lattices of $625$ ($25\times 25$) qubits at specific points along the hysteresis cycles (the heatmaps and the red points along the hysteresis curves have the same order from left to right). Data obtained from \texttt{Advantage2\_prototype2.6} with $s=0.7$ (top) and $s=0.3$ (bottom). All $\abs{S(q)}$ SSF heatmap plots use log-scale, and represent an average over $100$ spin configurations sampled on the D-Wave hardware. The dashed white box outlines the first Brillouin zone. Arrows along the hysteresis loop denotes both the direction of the longitudinal field sweep and the time progression of the protocol. }
    \label{fig:2D_ferromagnetic_hysteresis_zephyr_SSF}
\end{figure}

\begin{figure}[ht!]
    \centering
    \includegraphics[width=0.99\linewidth]{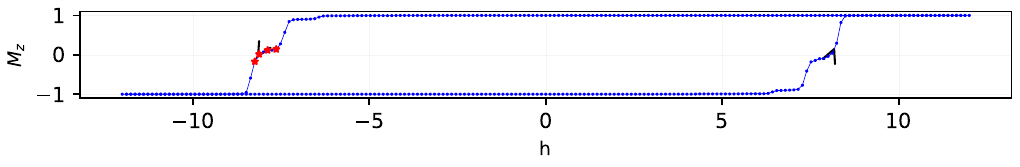}\\
    \includegraphics[width=0.24\linewidth]{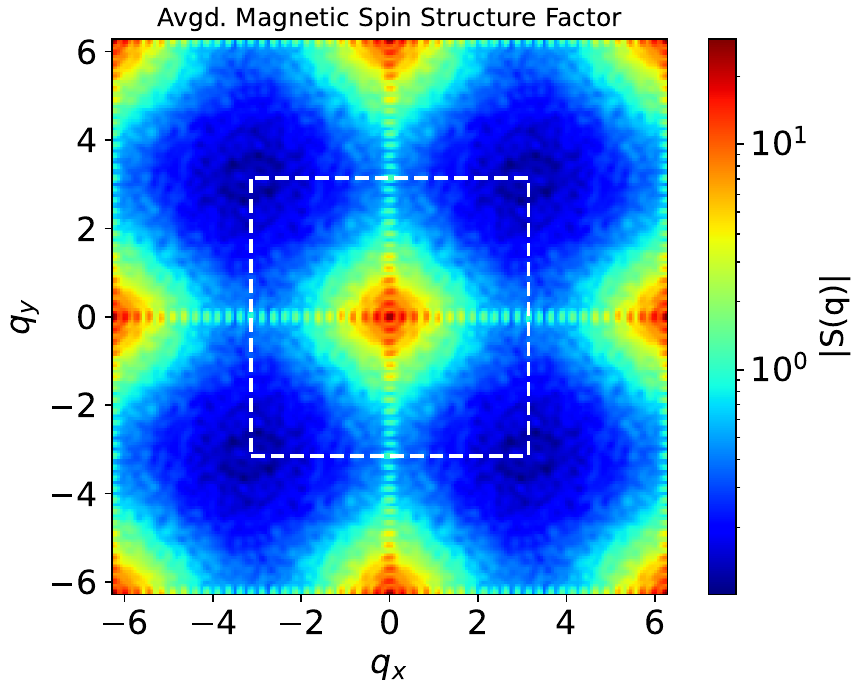}
    \includegraphics[width=0.24\linewidth]{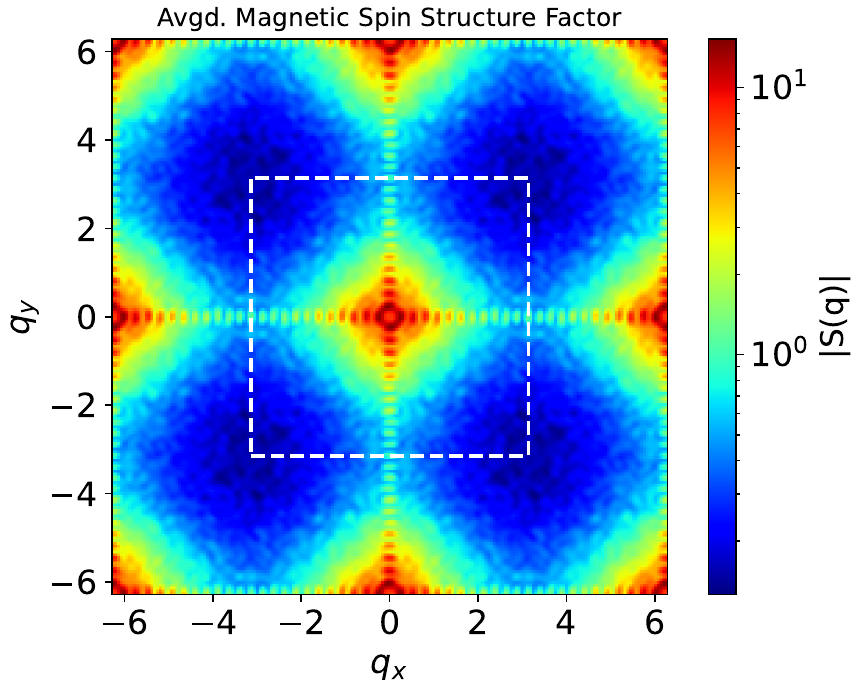}
    \includegraphics[width=0.24\linewidth]{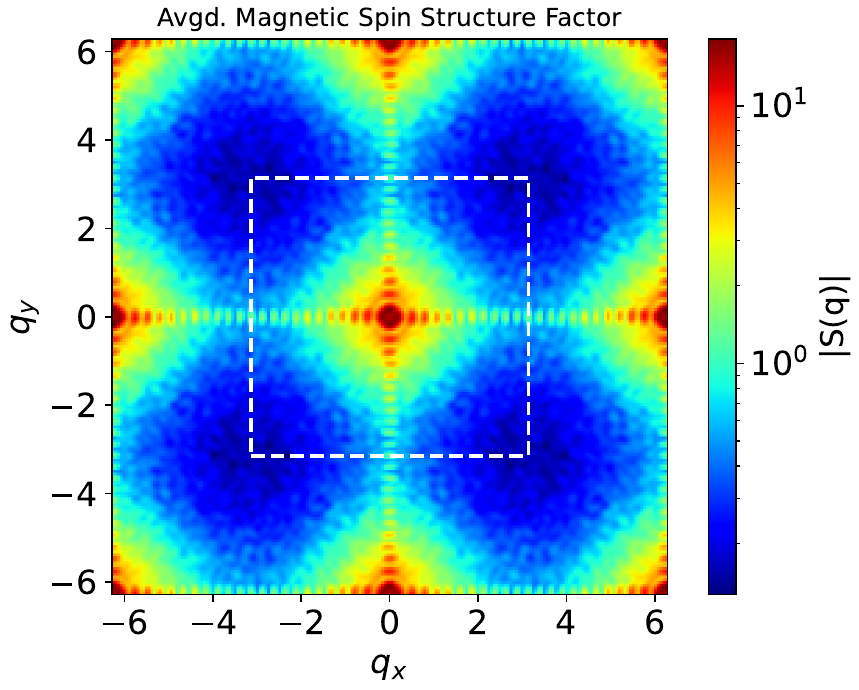}
    \includegraphics[width=0.24\linewidth]{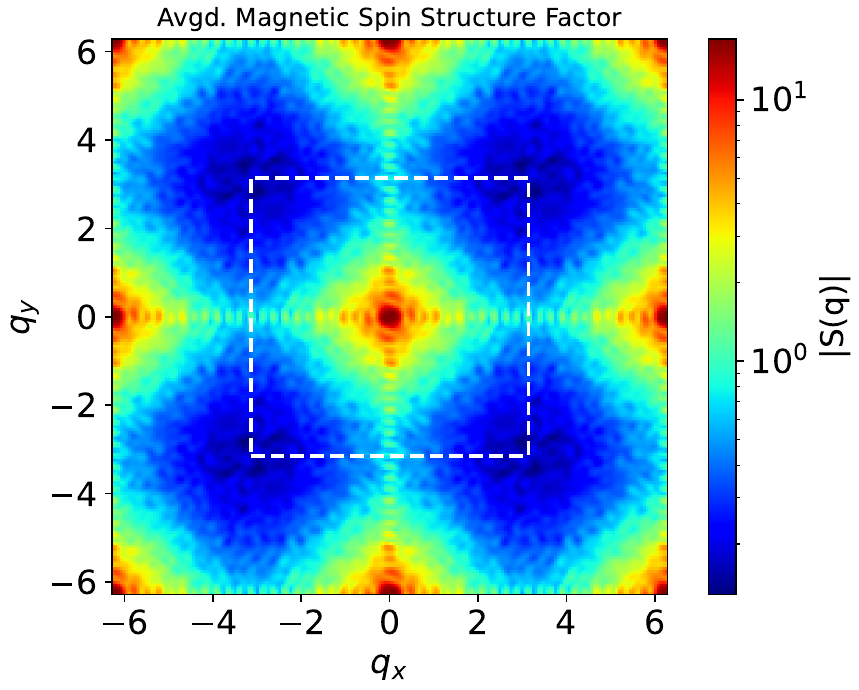}
    \caption{2D $625$ ($25$ by $25$) averaged spin magnetic spin structure factor at selected points in the non-saturated region of the hysteresis loops on \texttt{Advantage\_system4.1} with $s=0.7$. The red points on the hysteresis loop (top) denotes the sampleset from which the $100$-configuration averaged SSF heatmaps are shown (bottom), where the hysteresis cycles denote average magnetization computed from all $2000$ measured samples from the QPU at each $h$ value. Dashed white box outlines the first Brillouin zone. Arrows along the hysteresis cycle (top) denotes both the direction of the longitudinal field sweep and the time progression of the protocol. Log-scale $\abs{S(q)}$ heatmap.  }
    \label{fig:2D_ferromagnetic_hysteresis_Pegasus4.1_SSF}
\end{figure}

\suppsection{Spin Structure Factor of 2D Ferromagnets During Hysteresis Cycles}
\label{section:appendix_2D_spin_structure_factor}

The magnetic hysteresis protocol run on D-Wave QPU's allows the measurement of spin configurations, and one of the tools that we can use to visualize the dynamics occurring within the hysteresis cycles is spin structure factor (SSF). The magnetic spin structure factor is defined as
\begin{align}
{S(q) =  \sum_{i, j} e^{i q \cdot (r_i - r_j)}  \sigma_z^i \sigma_z^j },
\end{align}
where $r_i$, $r_j$ are position vectors, and the magnetic spins are $S_i$, $S_j$ which have values $\pm 1$. We set the lattice spacing to $1$. Here we average over $S(q)$ from the first $100$ sampled configurations on the D-Wave hardware to mitigate finite system size and finite sampling effects, and we then plot $\abs{S(q)}$ as heatmaps on a $200\times 200$ uniformly spaced grid over $(-2\pi,2\pi)$. 

SSF for $s=0.3$ and $s=0.7$ on the \texttt{Advantage2\_prototype2.6} device are shown in Fig.~\ref{fig:2D_ferromagnetic_hysteresis_zephyr_SSF}, top and bottom, respectively. Four SSF corresponding to demagnetized configurations are shown beneath the corresponding hysteresis loop. The corresponding location of the applied longitudinal field is marked on the hysteresis loop (left to right). SSFs demonstrate ferromagnetic ordering, bright Bragg peaks appear predominantly at the center of the first Brillouin zone. The averaged ordering from the demagnetized samples is weakly ferromagnetic as the peak intensity decreases, but there is a lack of ordering beyond ferromagnetic ordering. The train of intensity along the central $x-$ and $y-$axis results from the finite square lattice and open boundary conditions~\cite{Pelofske_arxiv_2025}. We see similar trends in $s=0.3$ and $s=0.7$, where the system is predominantly ferromagnetic.  

Fig.~\ref{fig:2D_ferromagnetic_hysteresis_Pegasus4.1_SSF} shows averaged spin structure factor heatmaps from one of the demagnetized regions of the low-transverse field hysteresis cycles ($s=0.7$) run on \texttt{Advantage\_system4.1}. 

Fig.~\ref{fig:2D_ferromagnetic_hysteresis_zephyr_SSF} and Fig.~\ref{fig:2D_ferromagnetic_hysteresis_Pegasus4.1_SSF} all show that any correlations that arose due to magnetic domains are averaged out over the $100$ distinct sampled configurations, but we can see some broadening of the central Bragg peak especially in Fig.~\ref{fig:2D_ferromagnetic_hysteresis_Pegasus4.1_SSF} suggesting the existence of more magnetic domain formation in those experiments.

\begin{figure}[ht!]
    \centering
    \includegraphics[width=0.32\linewidth]{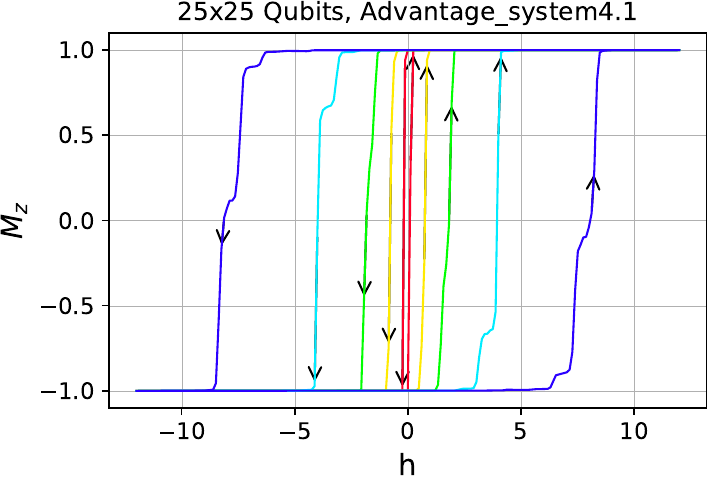}
    \includegraphics[width=0.32\linewidth]{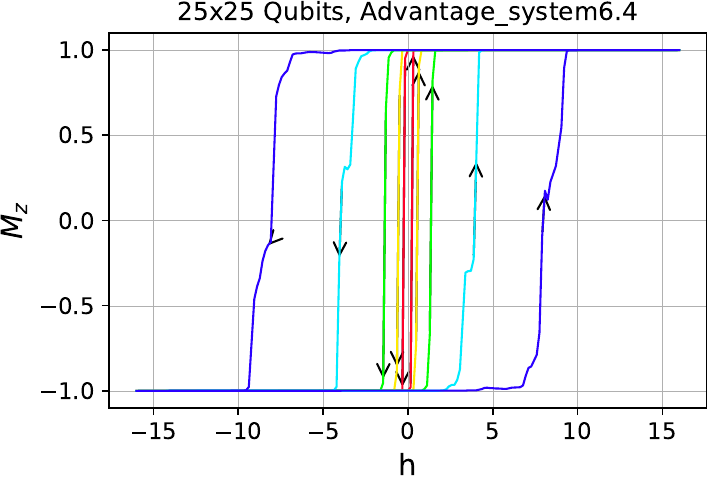}
    \includegraphics[width=0.32\linewidth]{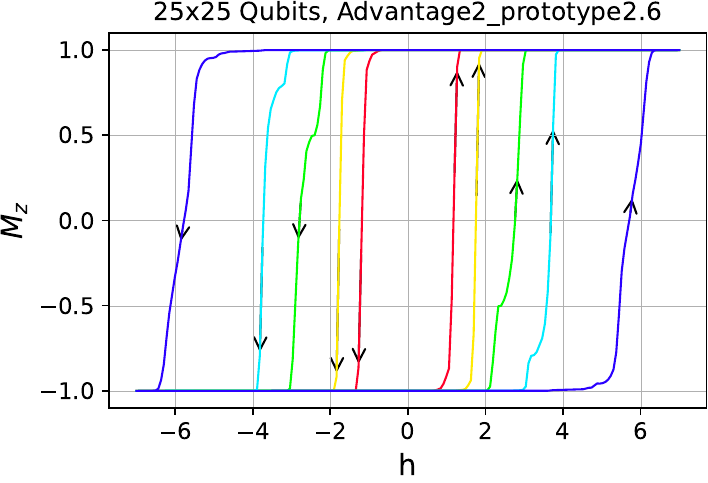}\\
    \hspace{0.02\linewidth}
    \includegraphics[width=0.29\linewidth]{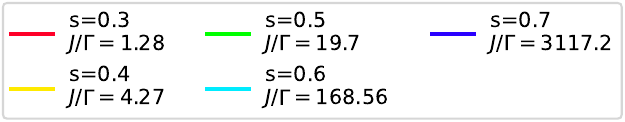}
    \hspace{0.02\linewidth}
    \includegraphics[width=0.29\linewidth]{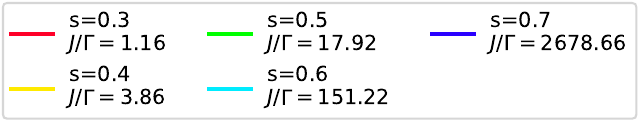}
    \hspace{0.02\linewidth}
    \includegraphics[width=0.29\linewidth]{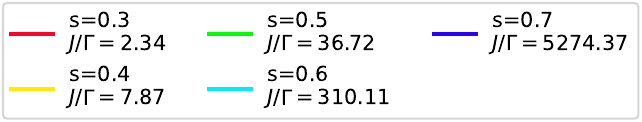}
    \caption{2D ferromagnetic hysteresis curves, run on the three different D-Wave QPU's. The system sizes are all $25\times 25$ square grids (open boundary conditions). Average net magnetization $M_z$ (variance not shown) as a function of the applied longitudinal field $h$ (x-axis). These simulations used $11.2 \mu s$ annealing times. The overlayed black arrows denote simultaneously the longitudinal field sweep direction and the time progression of the protocol. For all $s$ values, the equivalent physical ratio $\Gamma/J$ is given in the legend, and for all $s$ values full saturation is achieved.  }
    \label{fig:2D_ferromagnetic_hysteresis_loops}
\end{figure}

\begin{figure}[ht!]
    \centering
    \includegraphics[width=0.32\linewidth]{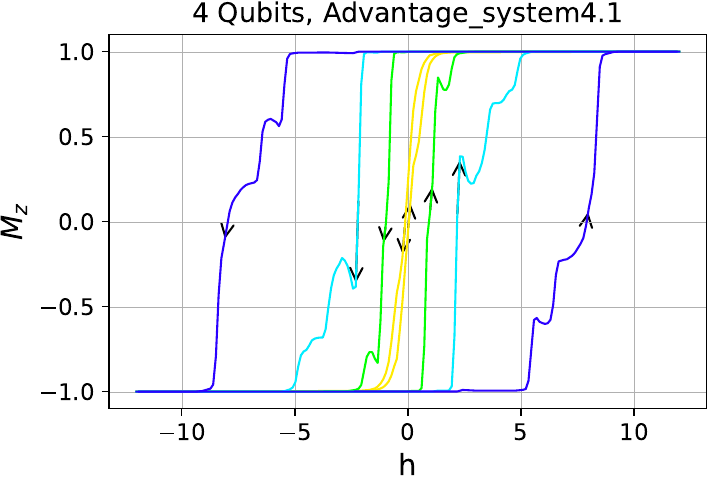}
    \includegraphics[width=0.32\linewidth]{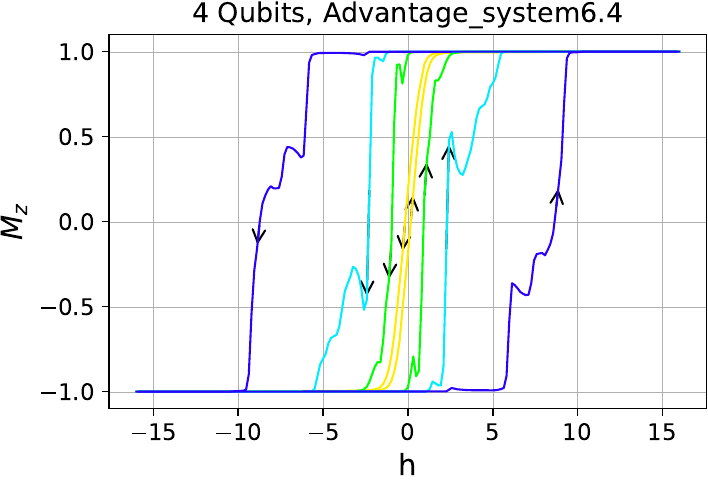}
    \includegraphics[width=0.32\linewidth]{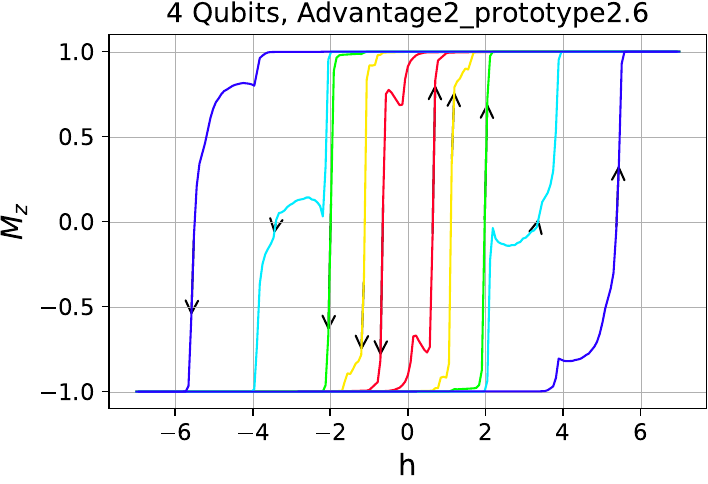}\\
    \hspace{0.02\linewidth}
    \includegraphics[width=0.29\linewidth]{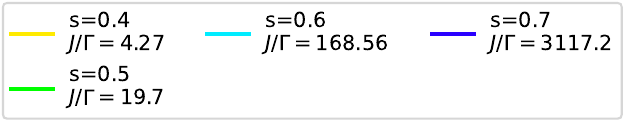}
    \hspace{0.02\linewidth}
    \includegraphics[width=0.29\linewidth]{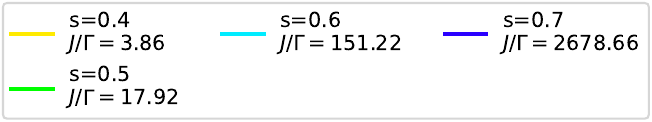}
    \hspace{0.02\linewidth}
    \includegraphics[width=0.29\linewidth]{figures/legends/legend_Advantage2_prototype2.6.pdf}
    \caption{1D 4-spin ferromagnetic Ising models, with periodic boundary conditions, closed hysteresis loops run on the three different D-Wave QPU's. Average net magnetization $M_z$ as a function of the applied longitudinal field $h$ (x-axis). The overlayed black arrows denote simultaneously the longitudinal field sweep direction and the time progression of the protocol. These simulations used $11.2 \mu$ s annealing times. For all $s$ values, the equivalent physical ratio $\Gamma/J$ is given in the legend, and for all $s$ values full saturation is achieved.  }
    \label{fig:4_spin_ferromagnetic_hysteresis_loops}
\end{figure}

\begin{figure}[ht!]
    \centering
    \includegraphics[width=0.32\linewidth]{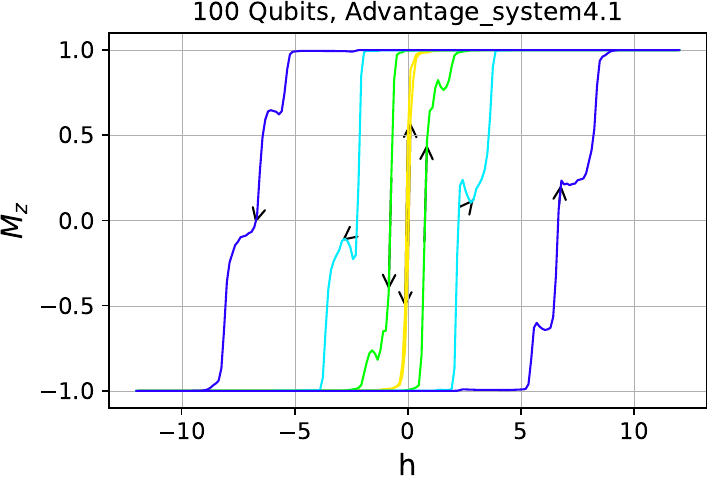}
    \includegraphics[width=0.32\linewidth]{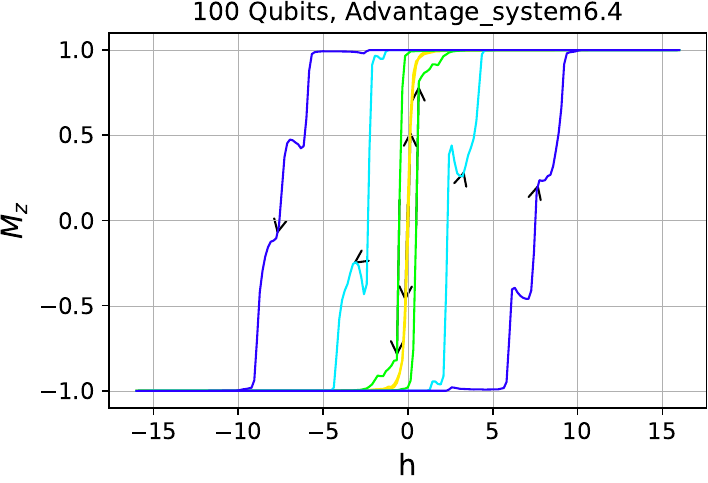}
    \includegraphics[width=0.32\linewidth]{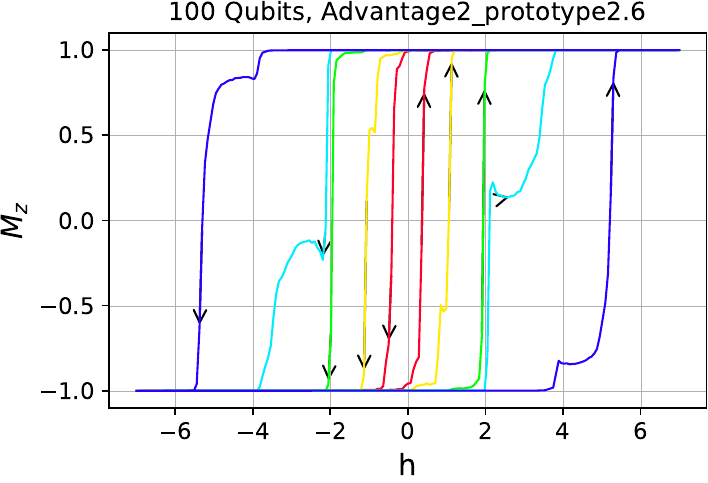}\\
    \includegraphics[width=0.32\linewidth]{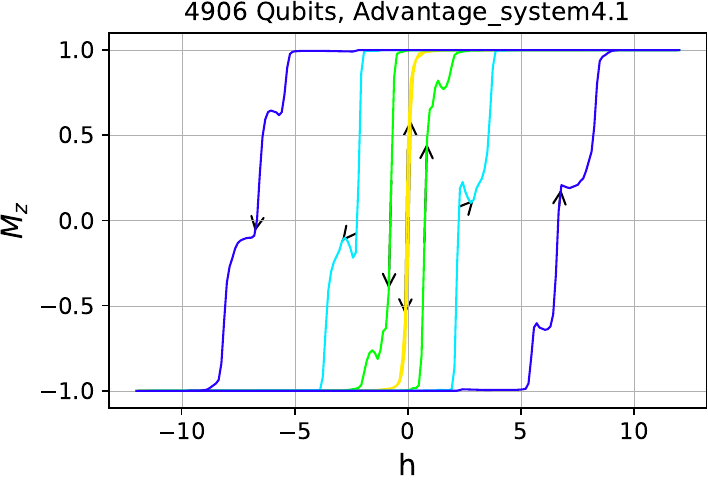}
    \includegraphics[width=0.32\linewidth]{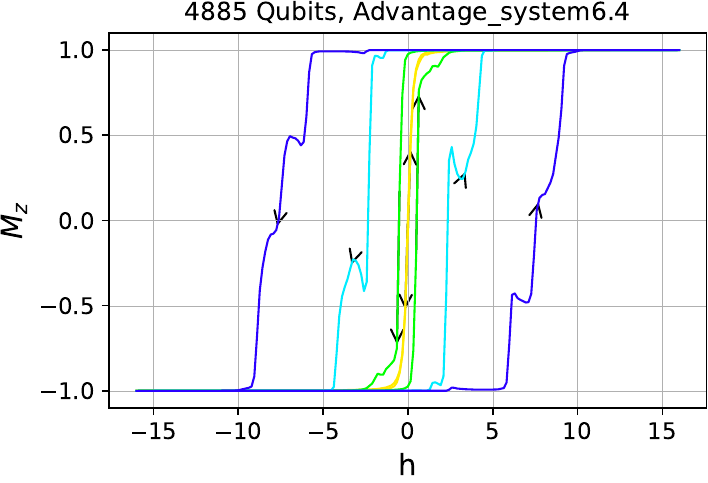}
    \includegraphics[width=0.32\linewidth]{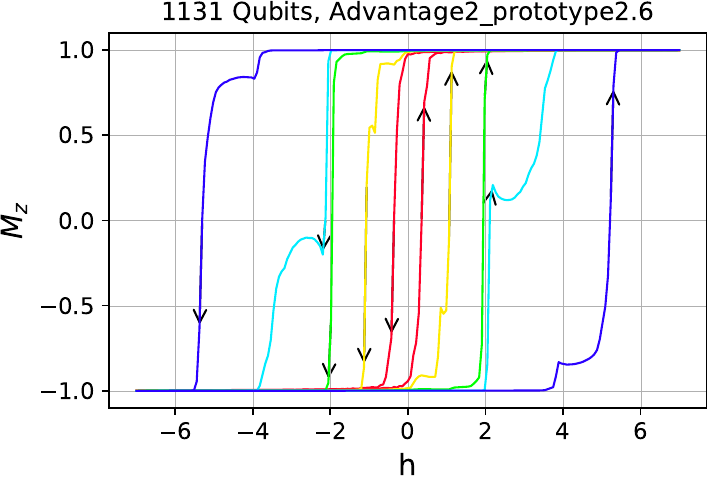}\\
    \hspace{0.02\linewidth}
    \includegraphics[width=0.29\linewidth]{figures/legends/legend_Advantage_system4.1_v2.pdf}
    \hspace{0.02\linewidth}
    \includegraphics[width=0.29\linewidth]{figures/legends/legend_Advantage_system6.4_v2.pdf}
    \hspace{0.02\linewidth}
    \includegraphics[width=0.29\linewidth]{figures/legends/legend_Advantage2_prototype2.6.pdf}
    \caption{1D ferromagnetic hysteresis (periodic boundary conditions) curve simulations, implemented on hardware as AFM gauge transformations, run on the three different D-Wave QPU's. Average net magnetization $M_z$ as a function of the applied longitudinal field $H$ (x-axis). These simulations all used $11.2 \mu$ s annealing times. The number of spins in the 1D chain is notated in the plot titles. We observe very little difference between the $100$ spin case and the larger (up to $4906$ spin) instances. For all $s$ values, the equivalent physical ratio $\Gamma/J$ is given in the legend, and for all $s$ values full saturation is achieved. The overlayed black arrows denote simultaneously the longitudinal field sweep direction and the time progression of the protocol. }
    \label{fig:1D_AFM_gauge_transforms_hysteresis_loops}
\end{figure}

\begin{figure}[ht!]
    \centering
    \includegraphics[width=0.32\linewidth]{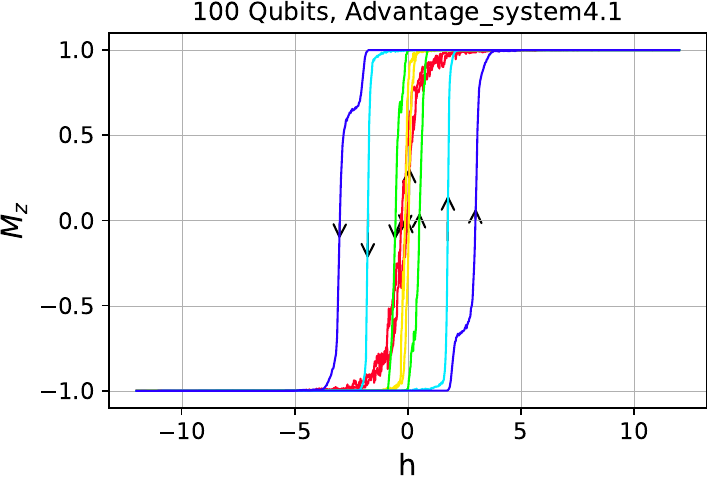}
    \includegraphics[width=0.32\linewidth]{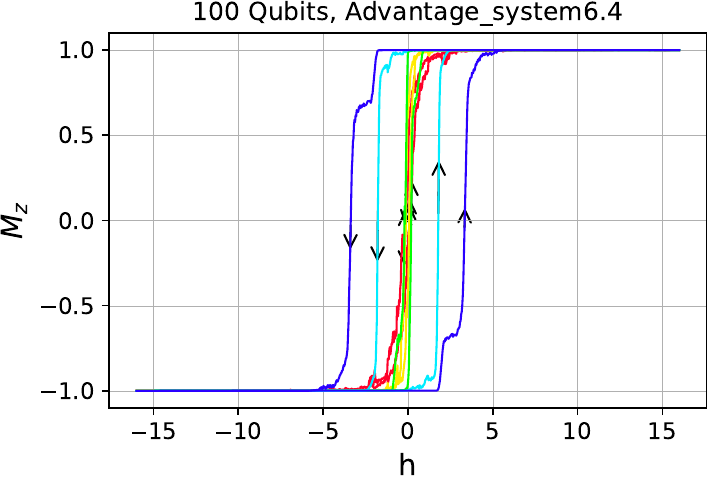}
    \includegraphics[width=0.32\linewidth]{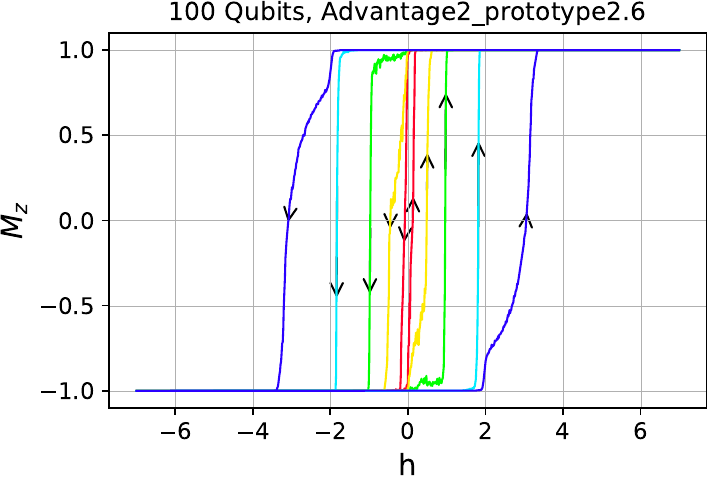}
    \includegraphics[width=0.32\linewidth]{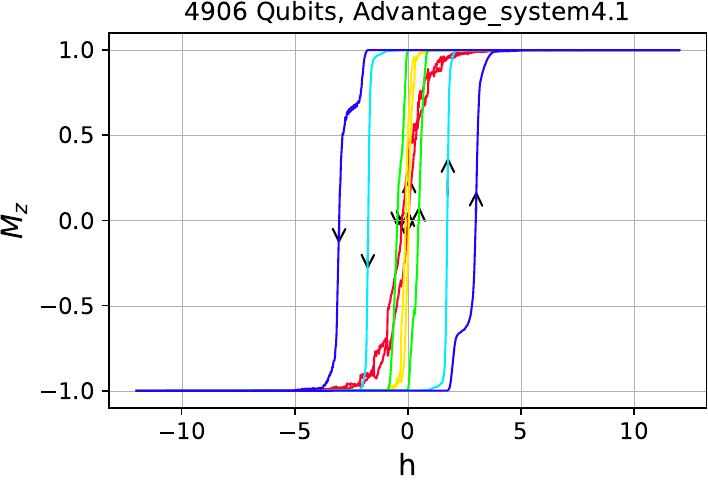}
    \includegraphics[width=0.32\linewidth]{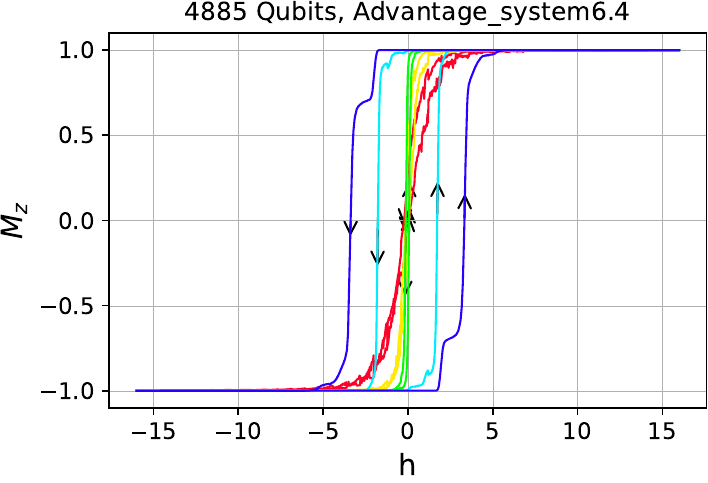}
    \includegraphics[width=0.32\linewidth]{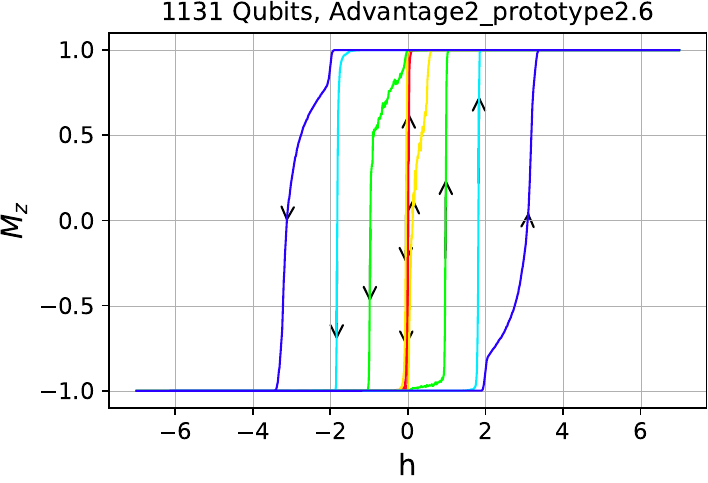}
    \vspace{-0.03cm}
    \tikz{\draw[dashed, thick] (0,0) -- (18,0);}
    \vspace{-0.03cm}
    \includegraphics[width=0.32\linewidth]{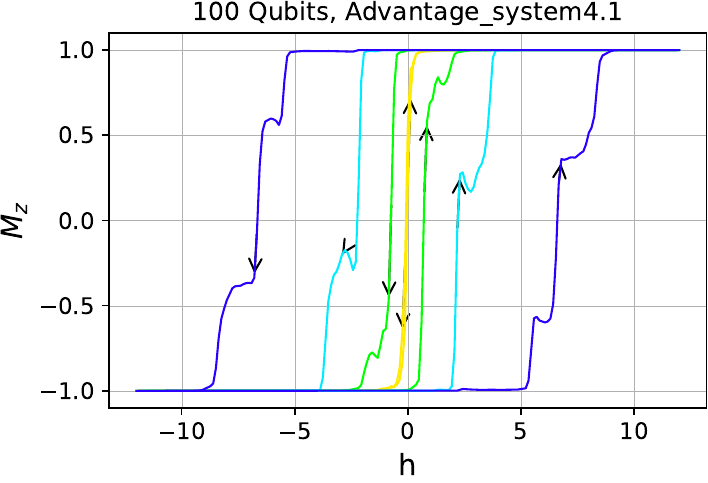}
    \includegraphics[width=0.32\linewidth]{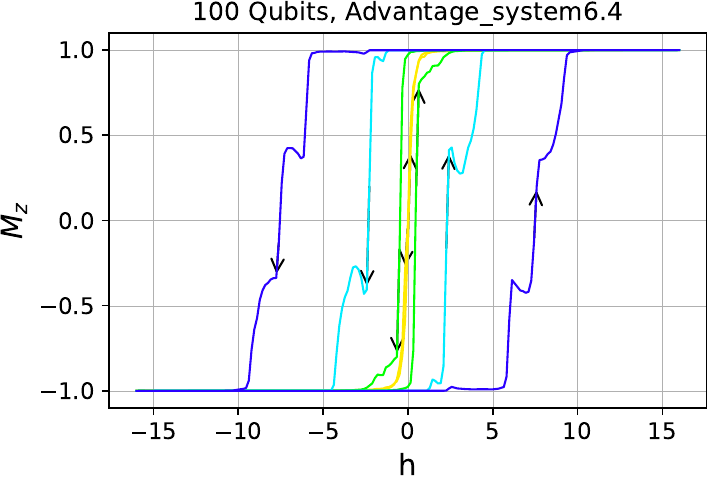}
    \includegraphics[width=0.32\linewidth]{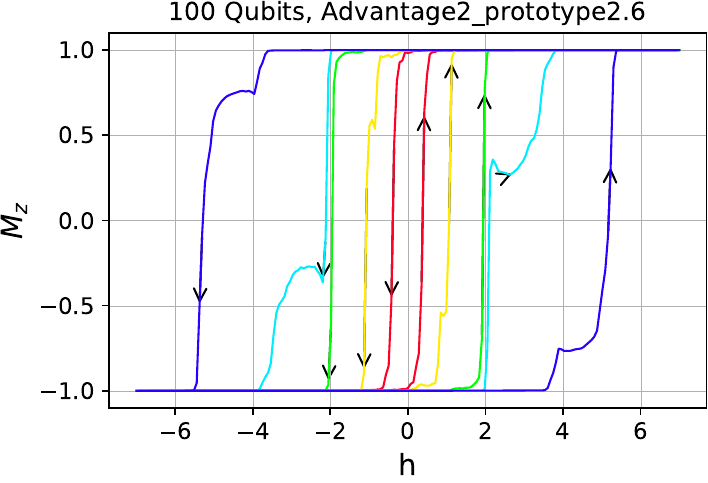}
    \includegraphics[width=0.32\linewidth]{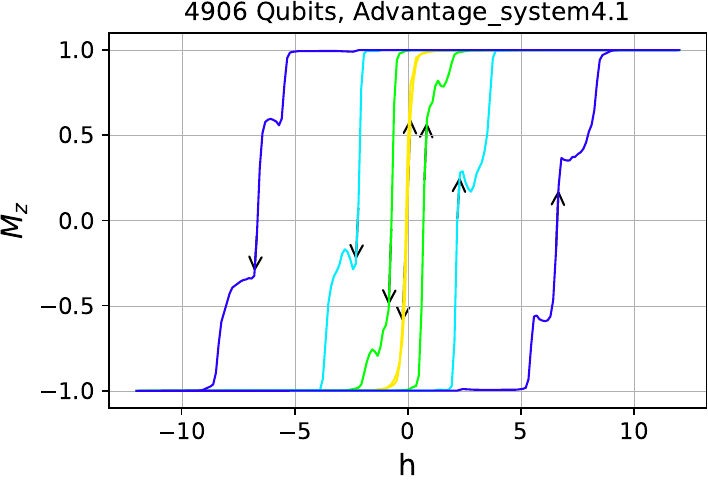}
    \includegraphics[width=0.32\linewidth]{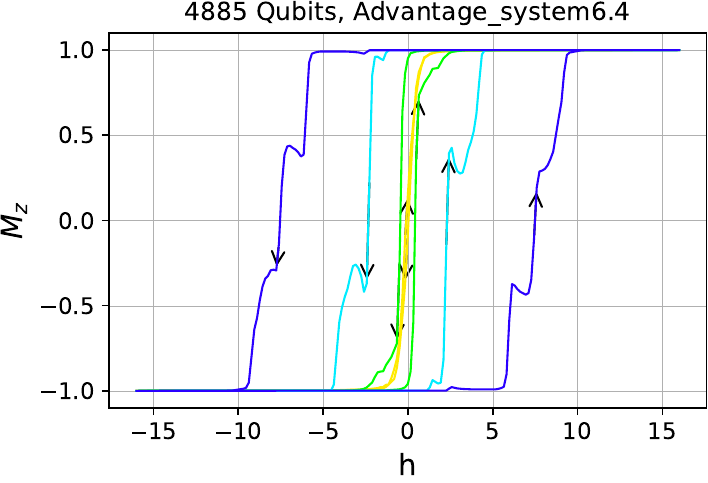}
    \includegraphics[width=0.32\linewidth]{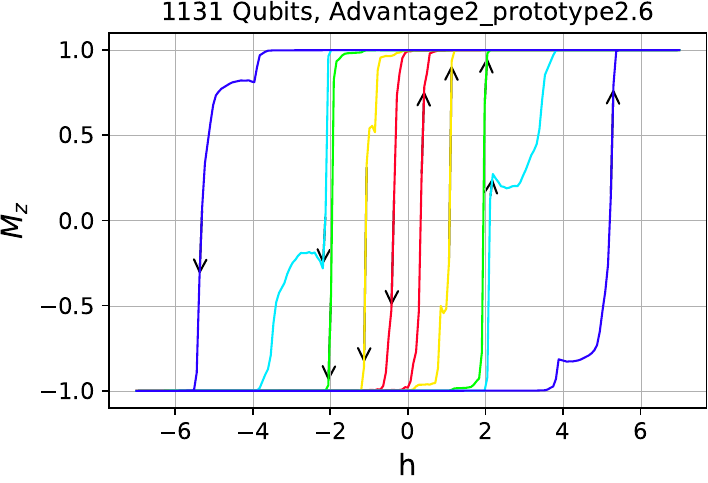}
    \hspace{0.02\linewidth}
    \includegraphics[width=0.29\linewidth]{figures/legends/legend_Advantage_system4.1.pdf}
    \hspace{0.02\linewidth}
    \includegraphics[width=0.29\linewidth]{figures/legends/legend_Advantage_system6.4.pdf}
    \hspace{0.02\linewidth}
    \includegraphics[width=0.29\linewidth]{figures/legends/legend_Advantage2_prototype2.6.pdf}
    \caption{1D ferromagnetic hysteresis (periodic boundary conditions) curve simulations run on the three different D-Wave QPU's. Average net magnetization $M_z$ as a function of the applied longitudinal field $h$ (x-axis). The plots in the top two rows used $1999.2 \mu$ s annealing time, and the bottom two rows used $11.2 \mu$ s annealing time. The number of spins in the 1D chain is notated in the plot titles. Simulations using $50 \mu $ seconds and $500 \mu$ seconds are not shown for brevity, but they show very similar characteristics to these plots (longer annealing times result in hysteresis loops with smaller areas, and fewer non-monotonic dips).  }
    \label{fig:1D_hysteresis_loops}
\end{figure}

\suppsection{Additional Experimental Data: Magnetic Hysteresis Loops on D-Wave Quantum Annealers}
\label{section:appendix_additional_experiment_DWave_magnetic_hysteresis_loops}

Fig.~\ref{fig:2D_ferromagnetic_hysteresis_loops} shows 2D ferromagnetic hysteresis experimental data run on the three D-Wave devices. These hysteresis simulations result in closed hysteresis loops with no non-monotonic dips during either forward or backward longitudinal field sweeps, and we always achieve maximum saturation at all values of $s$.

Fig.~\ref{fig:4_spin_ferromagnetic_hysteresis_loops} shows magnetic hysteresis loops from D-Wave processors on $4$ spin ferromagnetic models. Despite the very small system size, we observe very clear and consistent magnetic hysteresis for all $s$ pause values. In contrast, for much larger spin systems (Figures~\ref{fig:1D_hysteresis_loops} and \ref{fig:1D_AFM_gauge_transforms_hysteresis_loops}) show remarkably similar hysteresis curves despite the system sizes having a difference of over three orders of magnitude. The forward and backward hysteresis sweeps in Figure~\ref{fig:4_spin_ferromagnetic_hysteresis_loops} are symmetric, but the exact location of small non-monotonic dips (such as at $s=0.5$) are slightly different across the different devices, which could be either due to the different $J/ \Gamma$ ratios or small intrinsic hardware differences.

One of the key questions in regards to examining hardware level noise for the ferromagnetic compared to antiferromagnetic systems is the use of the AFM gauge transformation technique. Examining Fig.~\ref{fig:1D_AFM_gauge_transforms_hysteresis_loops}, we see very high agreement compared to Fig.~\ref{fig:1D_hysteresis_loops} (bottom two rows). This suggests that any hardware level noise differences (or biases) between ferromagnetic and antiferromagnetic coupling is a negligible contribution to these magnetic hysteresis simulations.

Comparing Fig.~\ref{fig:2D_ferromagnetic_hysteresis_loops} (2D simulations) and Figures~\ref{fig:4_spin_ferromagnetic_hysteresis_loops}, \ref{fig:1D_AFM_gauge_transforms_hysteresis_loops}, \ref{fig:1D_hysteresis_loops} (1D simulations), we observe that the 1D simulations have many more non-monotonicities compared to the 2D case. Longer annealing times shown in Fig.~\ref{fig:1D_hysteresis_loops} also reduce the number of non-monotonicities, along with a smaller hysteresis loop area (while still achieving full magnetization). 

We have omitted all $s=0.3$ 1D ferromagnetic data run on both of the Pegasus-chip device (\texttt{Advantage\_system4.1} and \texttt{Advantage\_system6.4}) at $11.2 \mu$ s annealing time (the fastest annealing time used in this study) from Figures~\ref{fig:4_spin_ferromagnetic_hysteresis_loops}, \ref{fig:1D_AFM_gauge_transforms_hysteresis_loops}, \ref{fig:1D_hysteresis_loops} because of anomalous (and hardware-specific) behavior that we will investigate in follow up study.

\end{document}